\documentclass[ieeetr]{sn-jnl}
\usepackage{graphicx}
\usepackage{hyperref}
\usepackage{xcolor}

\usepackage{xspace}

\usepackage{subcaption}

\usepackage{mymacros}

\begin{document}

\title{Parallel Adaptive Anisotropic Meshing on cc-NUMA Machines}

\author{\fnm{Christos} \sur{Tsolakis}}

\author{\fnm{Nikos} \sur{Chrisochoides}}
 \affil{CRTC, Department of Computer Science, Old Dominion University, Norfolk, VA, USA}

\abstract{
Efficient and robust anisotropic mesh adaptation is crucial for
Computational Fluid Dynamics (CFD) simulations. The CFD Vision
2030 Study highlights the pressing need for this technology, particularly for
simulations targeting supercomputers.
This work applies a fine-grained speculative approach to
anisotropic mesh operations. Our implementation exhibits more than
90\% parallel efficiency on a multi-core node. Additionally,
we evaluate our method within an adaptive pipeline
for a spectrum of publicly available test-cases that includes both
analytically derived and error-based fields. For all test-cases,
our results are in accordance with published results in the literature.
Support for CAD-based data is
introduced, and its effectiveness is demonstrated on one of NASA's High-Lift
prediction workshop cases.
}

\maketitle

\section{Introduction}\label{sec1}

The goal of mesh adaptation is to modify an existing mesh so that it can accurately
capture features of the underlying simulation.
Metric-based methods guide the
process of mesh adaptation through the use of metric fields.
Metric fields control the orientation and size of the
elements individually \textbf{for each} direction, thus enabling the creation of anisotropic meshes.
Metric-based methods have been utilized in a breadth of applications, including solving neutron transfer problems in nuclear physics \cite{goffin_minimising_2013}, efficiently applying diffusion tensor calculus on medical imaging data~\cite{Arsigny_LogEuclidean_2006} and topology optimization in structural mechanics~\cite{ejlebjerg_jensen_anisotropic_2016}. In the CFD field,  metric-based methods are capable of resolving
both inviscid \cite{loseille_achievement_2007} and viscous flows over complex
geometries \cite{michal_anisotropic_2018}.

The CFD 2030 Vision \cite{vision_cfd_2014} study identifies
efficient and robust mesh adaptation as one of the key areas that
need improvement to advance CFD simulations.
In \cite{park_unstructured_2016},  the authors present an overview
of mesh adaptation and offer suggestions for achieving the
objectives outlined in the CFD Vision study. One of the
recommendations is to optimize existing adaptive techniques to
function effectively on current and future systems.
An example of these architectures is \emph{Frontier},  the world's first public exascale machine \cite{oak_ridge_national_lab_frontier_2022} that can
provide the platform for simulation analyses at high fidelity. As of today \emph{Frontier} is composed out of
74 cabinets, each cabinet holds 64 blades, and each blade has two 64-core nodes for a total of  60,6208 CPU cores.

Handling these kind of hardware efficiently starts by using an efficient shared memory implementation of mesh adaptation operations. \cdt is a 
\emph{scalability-first} approach to mesh adaptation that implements 
a scalable software an initially incomplete functionality and the intention of completing functionality as it is needed.

\cdt uses the speculative method that explores concurrency at a
fine-grain (element) level by executing in parallel multiple meshing kernels
that attempt to capture their dependencies upon runtime. If any of the kernels fail to capture
its dependencies, due to conflicts with another kernel, it will release its
acquired dependencies and proceed to a different element.

In previous work \cite{tsolakis_parallel_2021} we compared 
an earlier version of the method of this paper against
state-of-the-art anisotropic mesh adaptation methods from both
the industry and academia that included both \emph{scalability-} and \emph{functionality-first} approaches. 
The analysis of the paper shows that 
\cdt provides comparable mesh quality to the other methods for the cases studied
in the paper. 

In this work, we describe the implementation of the speculative
fine-grained scheme for anisotropic mesh generation and
adaptivity of \cdt in detail and evaluate the generated meshes in a number of CFD applications.
In particular, the contributions of this work include:
\begin{itemize}

  \item Application of the speculative fine-grained scheme to a family of
        metric-aware mesh operations applicable to volume and  boundary adaptation
        (Section~\ref{sec:metric-aware}). As seen in previous work \cite{tsolakis_parallel_2021}
        this combination is highly effective.  As of today and to the best of the authors knowledge, it offers on shared memory machines 
        the highest strong scaling speedup among \emph{scalability-first} approaches and better weak scaling speedup 
        than state-of-the-art \emph{functionality-first} approaches. This is
        achieved via the low startup cost of our methods and the nature of the
        optimistic/speculative approach that exploits parallelism as soon as possible.
        Moreover, the ability to adapt both
        surface and volume elements at the same time yields better robustness and  offers more than 3 orders of
        magnitude improvement when compared to adapting only the volume.

  \item Extension of the above operations in order to support for geometrical
       (CAD-based) models (Section \ref{sec:introducing-geometry}). Access to a
        geometrical kernel allows to interrogate the underlying domain an recover
        information about its curvature and local feature size which could be absent in a
        coarse mesh representation which is often the starting point of the adaptation procedure. 
        Handling of CAD data gives access to well-known verification examples like the ONERA-M6
        wing and the JAXA highlift model.

  \item Evaluation of the method in an adaptive pipeline that includes a CFD
        solver and real-world geometry cases and comparison of the computed quantities with
        values present in the literature (Section \ref{sec:evaluation}).
        In contrast to our previous work that focused mainly on element quality
        measures, we compiled a suite of publicly available cases and bring our
        attention to the main goal of mesh adaptation: capturing features of the underlying
        simulation.  At the same time, we test how well \cdt operates as a part of an
        adaptive pipeline composed of
        open-source components.
\end{itemize}

In the next couple of sections, we review related methods in the literature
(Section \ref{sec:literature}). Appendix \ref{sec:background} provides a short
introduction on the use of metric spaces for mesh adaptation.

\section{Related work}
\label{sec:literature}

The method presented in this work belongs to the wider class of
\emph{Tightly Coupled} methods for parallel mesh generation \cite{chrisochoides_survey_2006,tsolakis_unified_2021}.
Tightly Coupled methods are characterized by intense
communication. Communication can be direct
through messages or indirect
through accessing regions of shared memory which due to false sharing and cache
invalidation creates overheads. Moreover, synchronization primitives and
constructs such as locks and barriers add to the overall overhead.

One of the first speculative tightly coupled methods for mesh generation is presented in
\cite{chrisochoides_parallel_2003,nave_guaranteed-quality_2004}. This method
is designed for distributed-memory architectures and it uses the Delaunay kernel \cite{bowyer_computing_1981,watson_computing_1981}
for introducing new points into the mesh. The main idea is to allow multiple cavity expansions (i.e. dependency resolutions) to take place in parallel,
however, in this approach, the data dependencies are
evaluated and acquired at run-time and not in any pre-processing step.  This
characteristic
gives this method the name \emph{speculative} or \emph{optimistic}.
If the dependency acquisition is successful the operation is applied otherwise,
the process ``rollbacks'' to the previous state by releasing any acquired data dependencies.
Although the amount of rollbacks
is low, the intense communication incurs many messages, resulting in sub-optimal results. Still, by acting directly upon touched data, this approach improves
cache utilization and allows tolerating more than $80\%$ of system latencies~\cite{nave_guaranteed-quality_2004}.
Later efforts extend this approach to a tightly-coupled Parallel Delaunay
Triangulation algorithm in ~\cite{foteinos_dynamic_2011} and an image-to-mesh
method ~\cite{foteinos_high_2014,feng_lapd_2016} optimized for Distributed Shared Memory machines (DSM).

A more complete review of parallel mesh generation methods categorized
by communication intensity appears in \cite{tsolakis_unified_2021}.
Due to length constraints, we limit our review section to parallel metric-based
mesh adaptation methods instead of parallel mesh generation methods in general.
All the presented methods utilize a mix of the same elementary mesh operations:
Point Insertion for refining larger than the target elements, edge collapse for
suppressing small edges and elements attached to them, Edge-Face swapping for
improving the quality of small clusters through topological operations  as well
as Vertex Smoothing to improve the quality of the elements attached to a
vertex. There are, however significant differences in the implementation of
each operation across the various mesh adaptation approaches/strategies which
we present below. Adopting the characterization of methods followed in \cite{tsolakis_unified_2021},
we organize the related work into two categories based on the workload decomposition method.

Data decomposition methods decompose the data that corresponds to
the given domain into datasets that can be safely accessed and updated.
This category includes \pragmatic as presented in \cite{gorman_thread-parallel_2015,pragmatic_web}
and \omegah~\cite{ibanez_mesh_2016}. Both methods are based
on coloring the dependency graph of the cavities of their mesh operations
in order to exploit parallelism. Once coloring is in place the methods
evaluate the quality of the elements in a cavity before and after applying the
mesh operation.  
The optimal configuration between the two is then picked and the mesh is updated.
\pragmatic defers the connectivity updates until the end of the mesh iteration to allow for updating the data structure in parallel.
\omegah on the other hand
is based on data structures designed specifically for mesh adaptation
\cite{ibanez_pumi_2016} and offers a unique approach where during each
adaptation pass the mesh becomes read-only and a new updated copy is derived.

Another method based on data decomposition is described in
\cite{de_cougny_parallel_1999-1,alauzet_parallel_2006}. This method deals with parallelism
by creating mesh-specific data structures \cite{fmdb_seol_2006} designed to
handle concurrent read and write access in a distributed memory environment resembling
distributed databases.
Moreover, it has been extended for metric-based adaptation in
\cite{sahni_parallel_2017}.  Each mesh operator pass is coupled with a global
synchronization step where all the processes commit and receive modifications
of the boundary elements.

Domain decomposition methods decompose the initial mesh into subdomains in
order to exploit parallelism. A few methods of this category are
\refine \cite{park_parallel_2008,park_refine_2020}
\epic \cite{Michal_Boeing_AnisotropicMeshAdaptation_2012}
\fefloa  \cite{loseille_unique_2017},
\avro \cite{caplan_parallel_2022, avro_gitlab},
and the method presented in \cite{digonnet_massively_2019}.

\refine, \epic, \avro and \cite{digonnet_massively_2019},
decompose the mesh using graph-based methods
and ``freeze'' the elements along subdomain boundaries during adaptation.
Mesh operations are applied only to internal elements. Several element migration passes allow the fixed elements to become internal and also give the opportunity
of load balancing.

\fefloa  \cite{loseille_unique_2017} approaches the discrete
domain decomposition problem using a hierarchical partitioning technique
combined with a region-grow greedy algorithm.
In particular, the initial mesh is partitioned based on a breadth-first approach.
The internal elements are adapted while the subdomain boundary elements remain fixed.
In the next iteration the fixed elements are partitioned and become the new subdomain while
their common boundaries become fixed etc.

When it comes to the repertoire of operations,
\refine and  \epic utilize edge-split, edge-collapse, edge-face swap, and vertex smoothing.
On the other hand, \fefloa, \avro and \cite{digonnet_massively_2019} 
use a cavity-based approach that allows using a single operator to express all the above mesh operations.
This feature not only
simplifies implementation and maintenance but in the case of \avro it enables performing
mesh adaption in higher dimensions \cite{caplan_four-dimensional_2020}.
When it comes to \cite{digonnet_massively_2019} the cavity-based criterion combines both element size and
element quality into a single criterion instead of utilizing separate operator passes.

Our method, \cdt, utilizes the same elementary operations but employs no domain decomposition or intensive data decomposition like the above
adaptive methods. Instead, it applies the speculative approach to
exploit parallelism and decompose data on the fly.
Also, in contrast to  previous speculative methods that arose from the telescopic approach
such as \cite{nave_guaranteed-quality_2004, foteinos_high_2014} the proposed approach includes
a variety of different mesh operations instead of just the Delaunay kernel.
To the authors' best knowledge, the method of this work is the first
tightly-coupled speculative fine-grained method for anisotropic 3D mesh adaptation for shared-memory architectures.

\section{Metric-based Adaptation within the \cdt library }
\label{sec:metric-aware}

\begin{figure}[!htpb]
	\centering
	\includegraphics[width=0.6\linewidth]{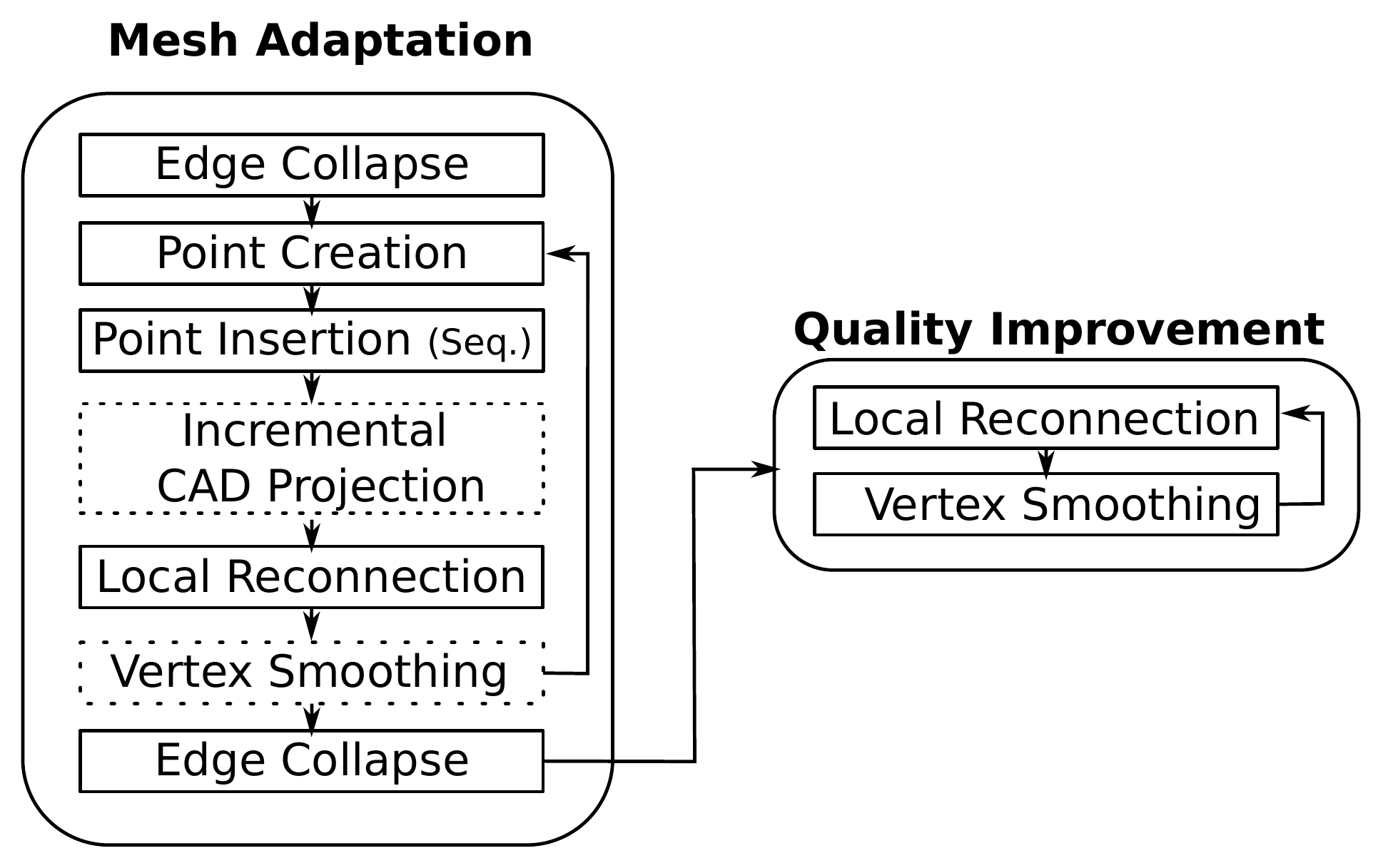}
	\caption{Pipeline of the presented approach. Dotted modules are utilized only when CAD data are
	available.}
	\label{fig:cad_pipeline}
\end{figure}

The metric-based approach of this work builds on top
of \cdt~\cite{drakopoulos_finite_2017}, a mesh generation toolkit developed at
the CRTC lab\footnote{\url{https://crtc.cs.odu.edu} (Accessed 2023-03-30).}
of Old Dominion University.
\cdt has  demonstrated significant improvements in end-user-productivity
compared to  state-of-the-art isotropic advancing front mesh generator~\cite{drakopoulos_fine-grained_2019}.
Its modular design  allowed  the addition of refinement
zones for the isotropic method that enable its use in Large Eddy Simulations as presented in~\cite{zhou_hybrid_2019}.
When faced with the challenge of  introducing metric-based adaptation capabilities,
we opted to decompose each mesh operation into
\emph{topological}  and \emph{geometrical} components.
Topological steps access and modify only the connectivity information (a 2-3
flip, for example, see Figure~\ref{fig:flips:flip23}) and as such, there is no
need for modifications for metric adaptation.
On the other hand, geometrical steps, such as evaluating a predicate that
decides whether a flip should be performed, will need to incorporate the metric
information.
Figure~\ref{fig:cad_pipeline} depicts the
metric-adaptive pipeline built and evaluated throughout this work.
In the rest of the section we iterate through the diagram discussing the most significant contributions required
to enable metric adaptation in \cdt and introduce support for handling
CAD-based information. Moreover, to improve the end-user-productivity of \cdt
the speculative fine-grained scheme
presented in~\cite{drakopoulos_fine-grained_2019} is 
adapted to handle all the mesh operations utilized in the pipeline. The next sections present technical details of each of the parallel operations.

\subsection{Parallel Speculative Local Reconnection in Metric Spaces}\label{sec:local_reconnection}

The fine-grained speculative scheme for local reconnection employed by \cdt
has been already presented in detail in~\cite{drakopoulos_fine-grained_2019}. We provide here only the contributions pertinent to  metric-based mesh adaptation.

\begin{figure}[!htbp]
	\centering
	\begin{subfigure}{0.42\textwidth}
		\includegraphics[width=\linewidth]{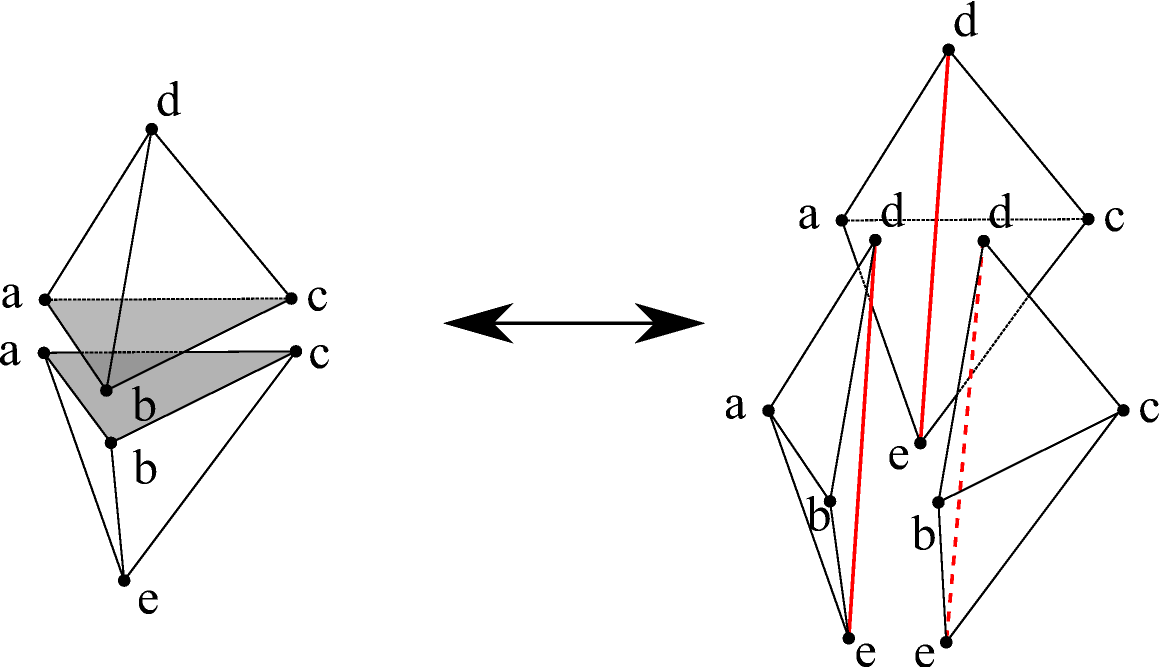}
		\caption{2-3 and 3-2 flips.}\label{fig:flips:flip23}
	\end{subfigure}
	\hspace{0.2cm}
	\begin{subfigure}{0.55\textwidth}
		\includegraphics[width=\linewidth]{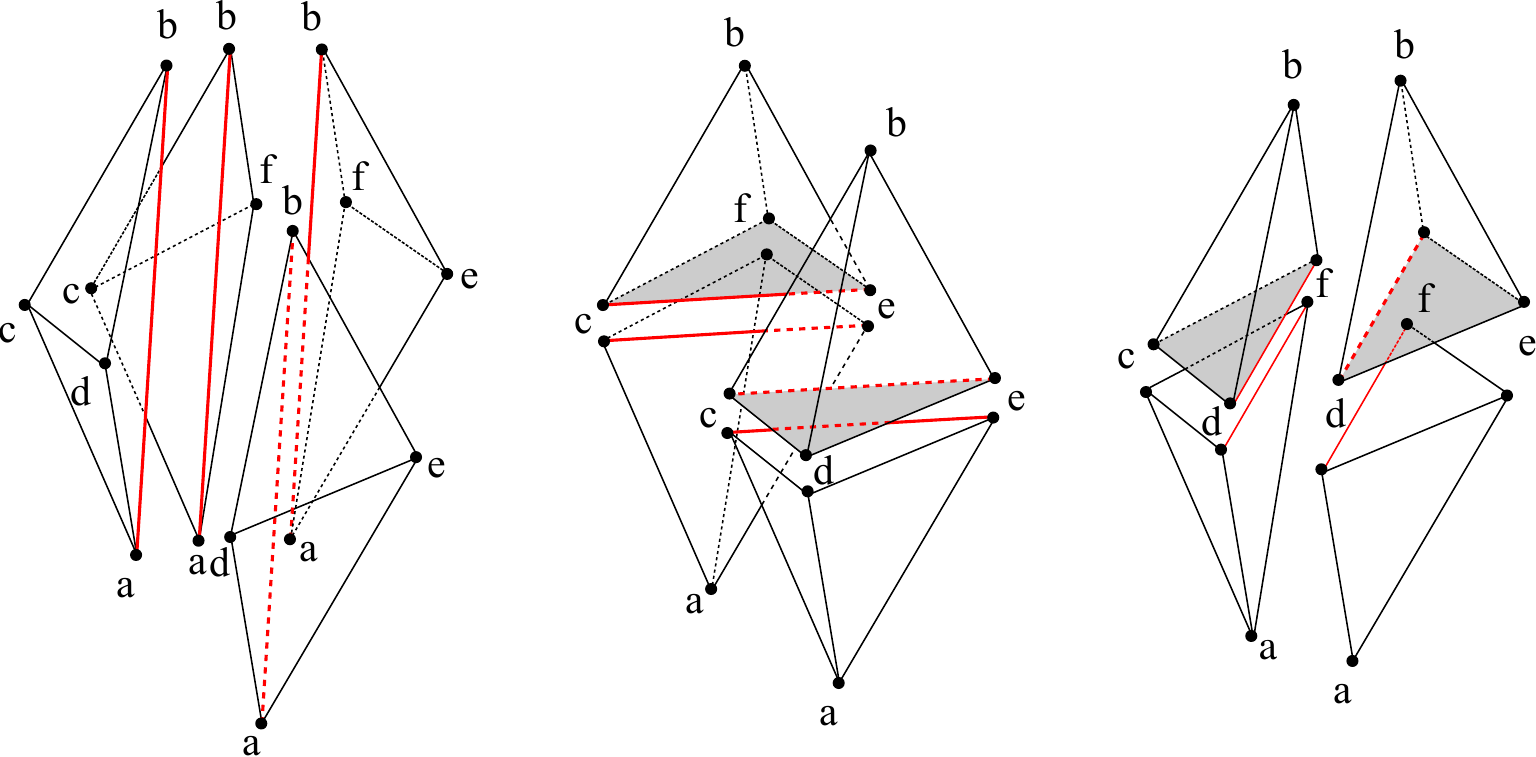}
		\caption{The three configurations of a 4-4 flip.}\label{fig:flips:flip44}
	\end{subfigure}
	\begin{subfigure}{0.4\textwidth}
		\includegraphics[width=\linewidth]{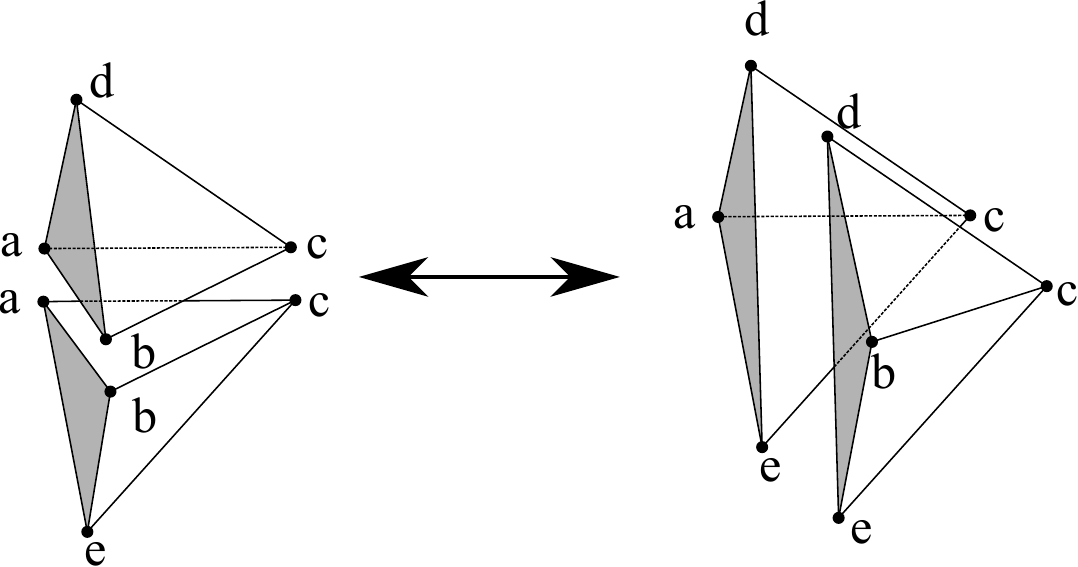}
		\caption{ 2-2 Flip, $abd$ and $abe$ are boundary faces.}
		\label{fig:flips:flip22}
	\end{subfigure}
	\caption{Topological Flips utilized by \cdt for local reconnection.}
	\label{fig:flips}
\end{figure}

A local reconnection pass
consists of four types of flips~\cite{lawson_software_1977} depicted in Figure~\ref{fig:flips}.
The flip operations are purely topological transformations, however, the
criteria they use are based on geometrical quantities, and as such, they need to
be substantially enhances in order to incorporate metric-based information. The first
criterion used in conjunction with a 2-3/3-2 flip is Delaunay-based: face
$abc$ will be replaced with edge $ed$ if $d$ is in the circumsphere of $abce$
(see~\autoref{fig:flips:flip23}). Extensions of the Delaunay criterion for
metric spaces has been suggested in reference~\cite{borouchaki_delaunay_1997} 
where the authors provide approximations
of the criterion based on the \emph{Delaunay measure} which is defined as follows: Let $K=(x_1,x_2,x_3,x_4)$ be a tetrahedron,
the Delaunay measure of a point $p$ with respect to $K$ is defined as:
\begin{equation}\label{eq:delaunay_measure}
	\alpha_\metric(p,K) = \frac{d_\metric(O_K,p)}{d_\metric(O_K,x_1)}
\end{equation}
where, $O_K$ is the circumcenter of $K$ evaluated in metric $\metric$.
If $\alpha_\metric(p,K) < 1$ then $p$ is in the circumsphere of $K$.
Notice that we did not specify $\metric$ explicitly. In fact, by incorporating
the metric from $1$, $2$ or even all $4$ points of $K$ one can get better
approximations of the Delaunay criterion~\cite{borouchaki_delaunay_1997}.
In this work, we adopt the criterion presented in ~\cite{dobrzynski_anisotropic_delaunay_adaptation_2008}, that uses not only
metric information of $K$ put also of the point $p$ itself:
\begin{equation}\label{eq:delaunay_criterion}
	\left\{
	\begin{array}{r}
		\alpha_{\metric_p}(p,K) < 1 \\
		\sum\limits_{i=1}^4\alpha_{\metric_{x_i}}(p,K)  + \alpha_{\metric_p}(p,K) < 5
	\end{array}
	\right.
\end{equation}

In practice, this criterion consists  of evaluating the traditional Delaunay
criterion in $5$ different spaces equipped with a different (but constant each time) metric
and averaging the results.
Similarly, in order to optimize the connectivity on the surface of the
mesh, a 3D in-circle test is coupled with a 2-2 Flip. In particular, a
surface edge $ab$ is flipped for $ed$  if $d$ is in the circumcircle of $abc$
(see Figure~\ref{fig:flips:flip22}).

The second criterion used is the maximization of the minimum  Laplacian edge
weight of an element $K$~\cite{barth_numerical_1991}. This criterion is combined
with the 2-3/3-2 and the 4-4 Flips (see Figure~\ref{fig:flips:flip23}
and~\ref{fig:flips:flip44}). In isotropic mesh generation it
consists in  evaluating the following quantity:
\begin{equation*}
Q(K) = \max_{i=1 \ldots 6} \frac{ \inner{n_{F_{i,1}},n_{F_{i,2}} }}{6\abs{K}}
\end{equation*}
where $F_{i,1}, F_{i,2}$ are the two faces attached to the $i$-th edge of a tetrahedron $K$, $n_{F_{i,1}}, n_{F_{i,2}}$
the respective face normals and $\abs{K}$ the volume of $K$. The algorithm performs
a flip only if the new configuration increases this quantity.  For the anisotropic
case, the formula is adapted using a metric tensor $\metric$
interpolated at the centroid of $K$. Moreover, in order to avoid the
numerically expensive evaluation of the normals, the formula is replaced with an equivalent formula presented in~\cite{marcum_advancing_layers_metric_2013}
that uses only inner products between the edge vectors of each face:
\begin{equation}
	Q_\metric(K) = \max_{i=1 \ldots 6}
	\frac{\inner{e_i,e_j}_\metric \cdot \inner{e_i,e_k}_\metric - \inner{e_i,e_i}_\metric \cdot \inner{e_j,e_k}_\metric}
	{6\abs{K}_\metric}
\end{equation}

\subsection{Parallel Point Creation for Anisotropic Mesh Adaptation}\label{sec:point_creation}
\footnotetext{Parallel implementation was developed in collaboration with Fotis Drakopoulos.}

The parallel implementation of the point creation kernel is structured in a similar fashion to the local
reconnection kernel.
Each
thread iterates a subset of   elements that do not satisfy the local length criteria in order to generate
\emph{candidate} points for insertion.
New candidate points are compared against existing mesh vertices and  other
candidates for proximity (in the metric space). This check allows to avoid the
creation of points too close to each other that will impact the quality and the
local size requirements. A similar approach named \emph{Anisotropic
	Filtering} is presented in~\cite{loseille_metric_orthogonal_2014}. Once a
point passes all proximity tests it is stored in a list assigned to the contained
element. Storing the candidate points in the contained element gives a
significant advantage; both the proximity checks and the subsequent point
insertion step (see Figure~\ref{fig:cad_pipeline}) can be performed in constant
time since the point location step of the direct insertion kernel will only require constant time to execute.

In contrast to local reconnection, the point creation step does not perform any
topological modification and thus no cavity locks are required. Moreover,
vertices are allocated into thread-local memory
pools~\cite{chernikov_experience_2007} and thus vertex allocation can be performed
concurrently. The only step that requires synchronization is
adding the candidate point to the  list of the contained element. Our
experiments showed that this lock is short-lived and the use of spinlocks
is a sufficiently efficient solution to handle concurrent accesses.

Our approach uses a centroid-based point-creation method.
This method will check the edge lengths of an element that does not meet the quality criteria
in the metric space, and if any of them does not satisfy the spacing
requirements, it will generate its centroid as an initial candidate point.
If the tetrahedron has a boundary face, or if any of its edges is  a ridge (i.e., lays
between two different surface markers) encroachment rules similar to those used
in Constrained Delaunay refinement~\cite{shewchuk_tetrahedral_1998} are
utilized. In particular,
the candidate point will be checked for encroachment
(in the metric space) against the boundary face, and if encroachment occurs,
the candidate is rejected and the centroid of the boundary face becomes the new
candidate. The same procedure is applied to the new point which is checked for
encroachment against any ridge edges, see Figure~\ref{alg:centroidalgo} in Appendix \ref{sec:supplementary_figures} for more
details.
Once a candidate is created, the metric is interpolated using formula~\eqref{eq:metric_interpolation}.
Inspired by~\cite{marcum_aligned_2014}, we store
alongside the metric value at a point $\metric(p)$ its logarithm
$\log(\metric(p))$. Although this requires more space, it reduces significantly
the time required for metric interpolation.

\subsection{Speculative Edge Collapse for Metric-based Adaptation}\label{sec:edge_collapse}
The goal of the edge collapse operation is to suppress edges with lengths smaller than a target value.
In this work, the edge collapse operation is utilized as a pre- and
post-refinement operation (see~\autoref{fig:cad_pipeline}). The pre-refinement
step removes short edges present in the input mesh. By default an edge
is considered short if it is smaller than $1/\sqrt{2}$ as measured by ~\eqref{eq:euclidean_length}).
 Depending on
the input mesh, the user may increase this value in order to create a coarser
initial mesh which can lead to a better quality of the final mesh as we demonstrated in \cite{tsolakis_parallel_2021}. The post-refinement use of
the operation allows the removal of any short edges created during refinement.

\begin{figure}[!tbph]
	\centering
	\includegraphics[width=\linewidth]{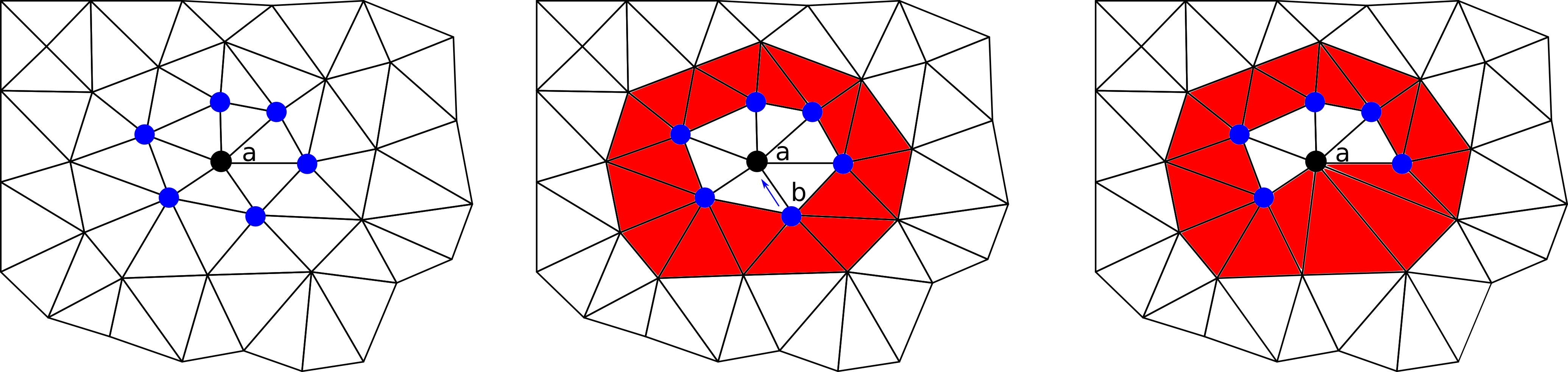}
	\caption{Steps of speculative edge collapse.}
	\label{fig:edge_collapse}
\end{figure}

The parallel implementation of the edge coarsening the algorithm iterates through the vertices allocated
by each thread and exploits parallelism utilizing a \texttt{\#pragma omp parallel for
schedule(guided)} OpenMP construct.
Each thread picks and locks (speculatively)
the vertex (\emph{a} in~\autoref{fig:edge_collapse})
corresponding to the iterator value. Then it speculatively locks its adjacent
vertices (blue in~\autoref{fig:edge_collapse}). If any of the locks fail,
the thread will release any acquired locks and it will proceed to the next vertex.
Notice that locking the vertices implicitly grants exclusive access to all
their adjacent tetrahedra (red elements of~\autoref{fig:edge_collapse}).
Once the required locks have been acquired, the edge lengths between the
vertex \emph{a} and the rest of the vertices of its cavity are evaluated. If an edge
with a length less than a user-defined value is found 
then the edge will be collapsed. Additional criteria such as not applying edge collapse
in cases that will increase the surface deviation (see
Section~\ref{sec:introducing-geometry}) are also applied for edges on the surface of
the mesh. Finally, the edge is collapsed by moving the second point to the
first. The \texttt{guided} scheduler was
selected because it performs on average better for cases that require
different levels of coarsening 
\cite{tsolakis_parallel_2021}. More sophisticated parallel iteration and tasking schemes
are out the scope of this work but have been explored in \cite{tsolakis_tasking_framework_2021}.

\subsection{Vertex Smoothing}
\label{sec:smoothing}

\begin{figure}[!htpb]
	\centering
	\includegraphics[width=0.3\linewidth]{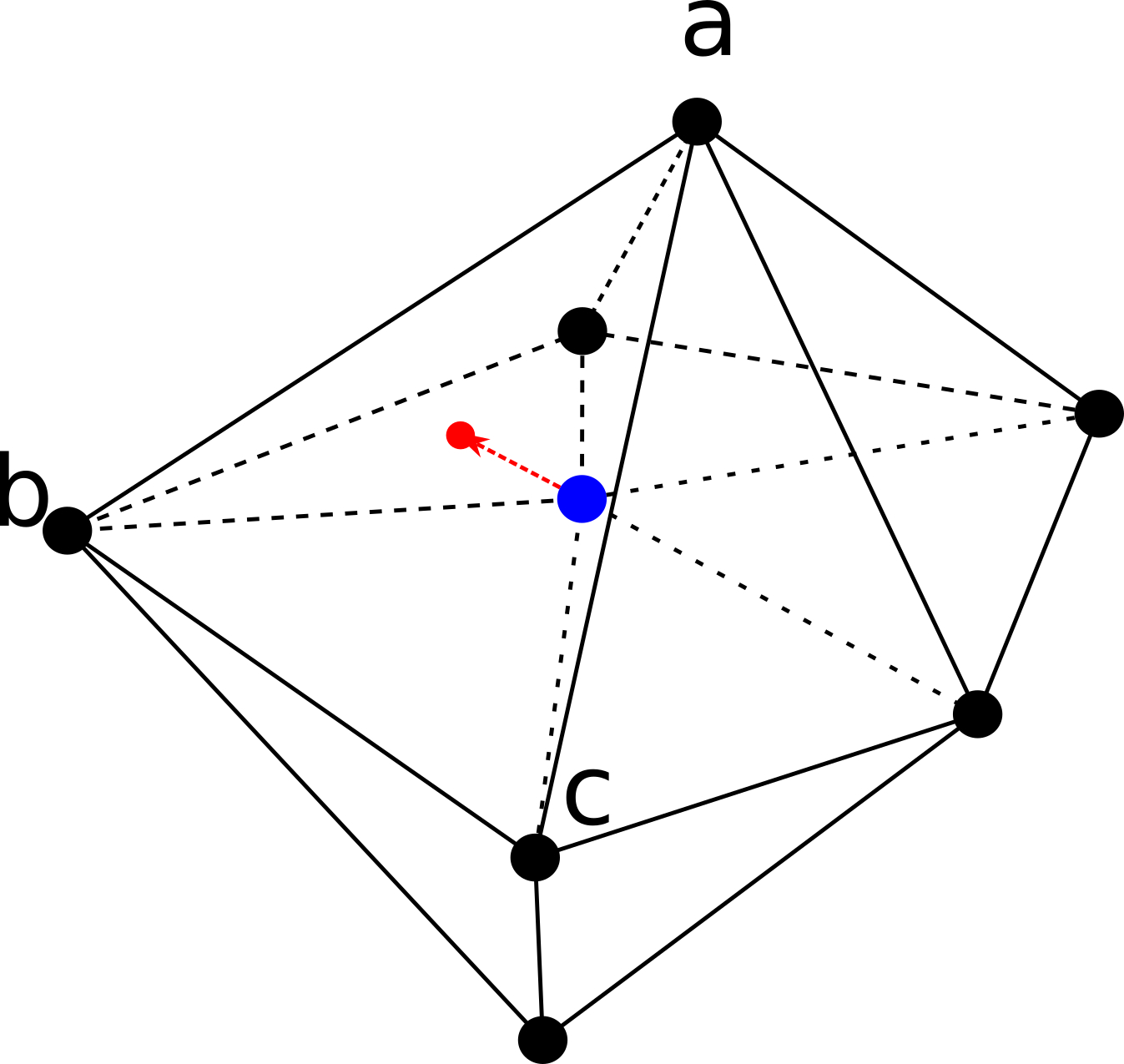}
	\hspace{1cm}
	\includegraphics[width=0.3\linewidth]{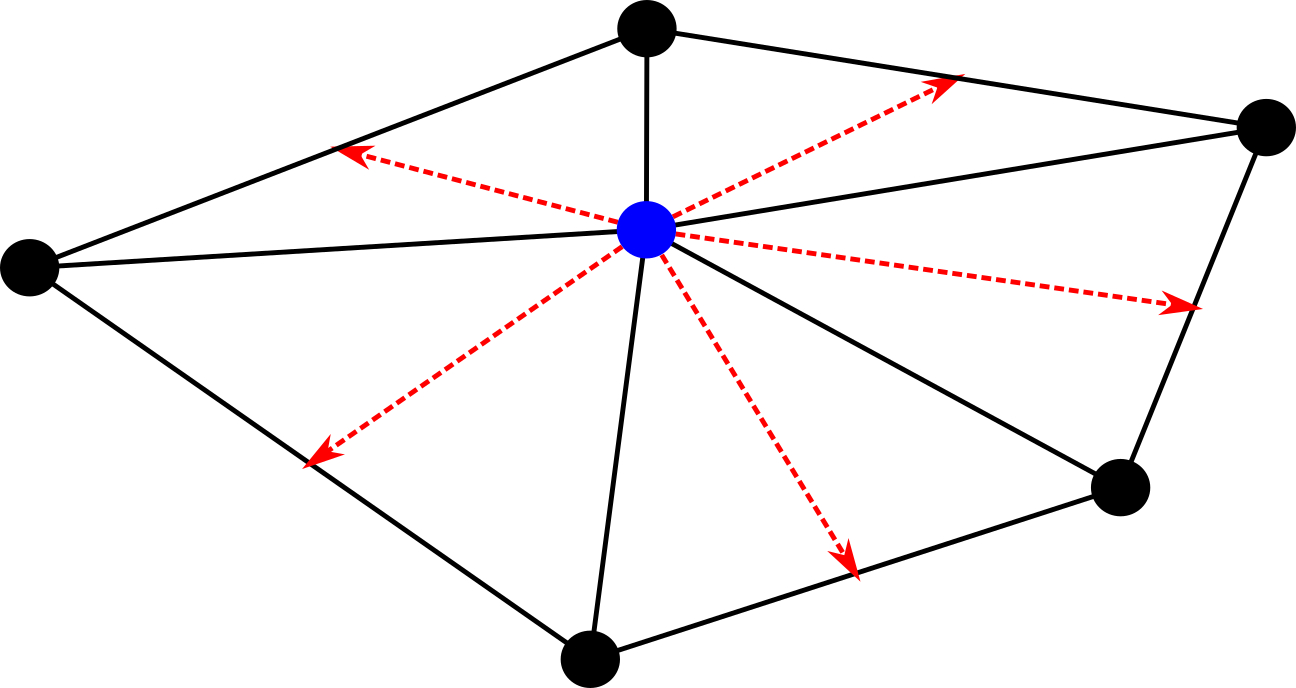}
	\caption{Search Space for Smoothing Operation.
}
	\label{fig:smoothing}
\end{figure}

During the Quality Improvement step of the isotropic mesh generation pipeline
of \cdt the user can use a combination of Laplacian~\cite{buell_mesh_1973}
and optimal point placement smoothing~\cite{drakopoulos_finite_2017}.
Utilizing these methods
for metric-based adaptation did not yield a substantial improvement in mesh
quality while, on the other hand, adding a significant overhead in the running
time.
As an alternative, a different vertex relocation strategy was employed. First,
only vertices with at least one attached element that has a mean ratio shape
measure (equation \eqref{eq:quality})  below $0.1$ are considered.
This value is based upon common practices present in the literature~\cite{ibanez_first_benchmark_UGAWG_2017,tsolakis_parallel_2021}.
Moreover, a non-smooth
optimization method similar to the one presented in~\cite{freitag_tetrahedral_1997} was employed by optimizing the minimum value of
the mean ratio measure  among all the elements attached to the vertex. For
simplicity, the current implementation does not utilize the integer
programming-based solution presented in~\cite{freitag_tetrahedral_1997} or
the computation of the active set of the gradients used in~\cite{klingner_improving_2008} in order to determine the optimal search
direction. Instead, it uses  a reduced search space comprised by the segments
that connect the vertex to be relocated with the centroids of the faces of its
cavity (see Figure~\ref{fig:smoothing} left). This search space was found to
be sufficient for the cases of this study.
Once the search space is determined,
the vertex will be moved incrementally along all the search directions
and the position that optimizes the quality will be selected.

For vertices lying  on the surface, the search space is constrained to the
segments that connect the moving vertex to the midpoints of the edges of the
corresponding surface cavity (see Figure~\ref{fig:smoothing} right). If the
vertex lies on a ridge the search directions are only two; toward either end
of the ridge. Along with the optimization criterion, the method always ensures
that no elements are inverted and thus no subsequent untangling step is needed.
When it comes to speculative execution, vertex smoothing follows a similar approach to edge collapse.

\subsection{Effect of the complete set of operations}

\begin{figure}[!htpb]
	\centering
	\includegraphics[width=0.5\linewidth]{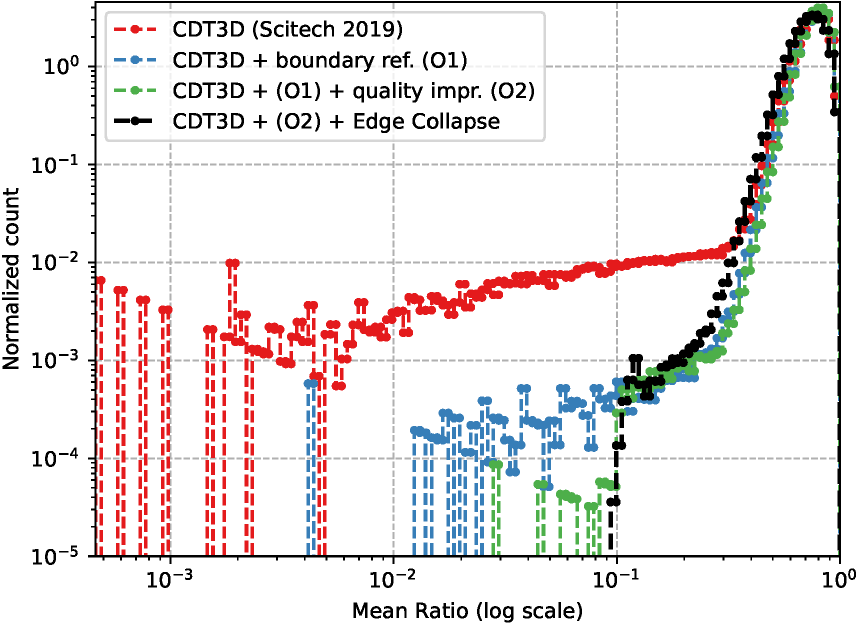}%
	\includegraphics[width=0.5\linewidth]{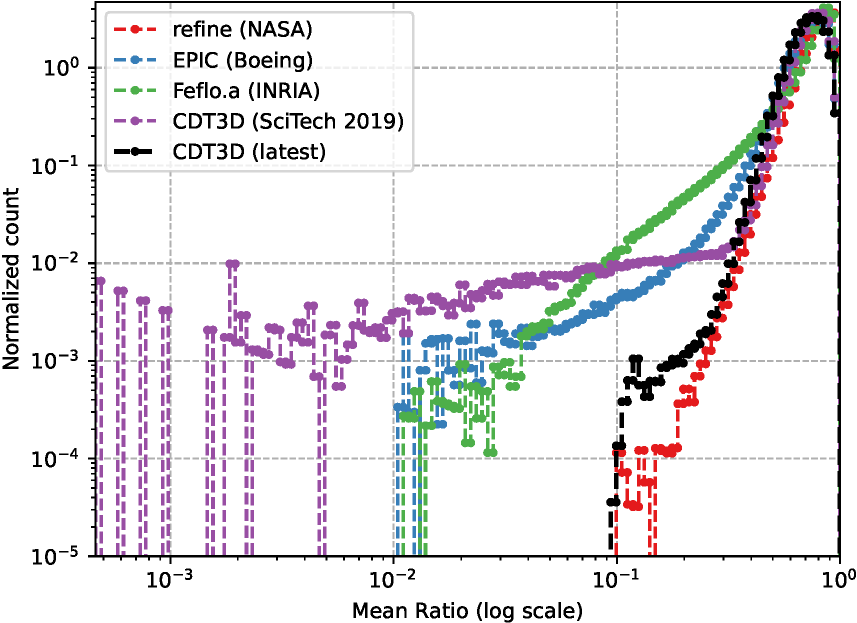}%
	\caption{Left: The effect of adding more metric-aware operations.
    Right: The improved approach versus our previous results presented in~\cite{tsolakis_parallel_2019}.}
	\label{fig:third_improvement}
\end{figure}

During the process of developing metric-based adaptation in \cdt we often
wondered if  the full suite of operations was indeed required in order to
achieve reasonable results.  \autoref{fig:third_improvement} compares an early
implementation of the method  presented in \cite{tsolakis_parallel_2019} along
with the current version compared with  other state-of-the-art mesh adaptation
methods \cite{tsolakis_parallel_2021}.  The earlier attempt utilized an
external tool for surface adaptation and created an initial mesh by means of
boundary recovery. The mesh was then adapted using metric-based local
reconnection only for the inner volume elements with no other metric-aware
operation. During the procedure, the surface mesh was kept fixed.  The ability
to adapt the boundary of the mesh at the same time as the volume mesh, along
with the vertex smoothing operation and the addition of a metric-based edge
collapse enhanced the quality of the generated mesh by $3$ orders of magnitude(as measured by the mean ratio shape measure \eqref{eq:quality})
with respect to our initial attempt.

\section{Handling Geometry through metric spaces}
\label{sec:introducing-geometry}

Industrial applications as well as several high-quality research-focused
workshops such as the
High-Lift prediction workshop\footnote{\url{https://hiliftpw.larc.nasa.gov} (Accessed 2023-04-25).},
the International Workshop on High-Order CFD Methods\footnote{\url{https://how5.cenaero.be} (Accessed 2023-04-25).},
the Sonic Boom Prediction Workshop\footnote{\url{https://lbpw.larc.nasa.gov} (Accessed 2023-04-25).},
and  the Geometry and Mesh Generation workshop\footnote{\url{http://www.gmgworkshop.com} (Accessed 2023-04-25),Internet Archive \url{https://web.archive.org/web/20230407233207/https://www.gmgworkshop.com/} (Accessed 2024-04-06).}
make extensive use of analytical/ geometrical descriptions of the domain. These descriptions are of particular importance since many flow quantities of interest depend
on the exact shape characteristics of the components which could be lost in a discrete model. Thus building a mesh based on the geometrical description of the
domain is essential for these studies. Moreover, accessing geometrical
information while adapting the mesh leads to better domain discretization and
thus more accurate solution.

There are many methods that can be used to build this representation but traditionally
CFD simulations, and the engineering community in general,
leaned towards the use of the Boundary Representation method (BREP or B-rep)~\cite{taylor_geometry_2018}.
%
 The B-rep method uses a combination of topological entities
(Faces, Edges, and Vertices) along with their geometrical description as (analytic)
surfaces, curves, and points. B-rep data can also hold boolean operations
such as intersection and union between entities as well as higher-level
operations such as extrusion and sweeping~\cite{haimes_engineering_2013}.
B-rep data are usually handled by a Computer-Aided Design (CAD) kernel which is
responsible both for generating B-rep data and performing queries to them.

The authors of ~\cite{park_geometry_2019} describe in detail
two approaches that can be used to incorporate geometrical
information to a meshing procedure.
The first option is to build a surrogate geometry by constructing a
discrete, often high-order, surface mesh that captures all the features of the
input geometry at a desired resolution.
 This approach has the advantage of
controlling the fidelity of the constructed representation. Also, it allows
fixing inconsistencies of the continuous representation that can occur due to
different tolerances of each continuous patch. One can construct a surrogate
geometry even when a geometrical description is not available based on the input
surface mesh. This approach is currently utilized by the \fefloa and
\epic mesh adaptation software~\cite{park_geometry_2019} (among others).

The second approach, which is used in this work, is to maintain an association
between each boundary vertex of the mesh and its adjacent geometric entities.
This approach allows querying the appropriate  Geometrical entity through the CAD kernel.
It has the disadvantage of inheriting the issues present in the B-rep model
but, it provides access to the CAD kernel in a simple manner.
Moreover, it aligns better with our goal which is to introduce preliminary support
for B-rep data to our mesh adaptation method. Currently, this approach is also favored by the \refine
mesh mechanics suite~\cite{refine_github} and \avro~\cite{avro_gitlab}.
In practice, we introduced at each mesh vertex a pointer to the lowest dimension geometric entity
that is adjacent. This information together with topological and
geometrical queries to the geometry kernel allows to evaluate $uv$ (or $t$)
parametric coordinates for any mesh vertex.  The current implementation makes use of
the EGADS geometry kernel~\cite{egads_haimes_2012} through a generic API which
could be adapted for another CAD kernel in the future.

Geometry information is used throughout the mesh adaptation procedure
in several ways.
First, newly introduced boundary points are projected to the surface
using a dedicated module (see Figure~\ref{fig:cad_pipeline}) right
after vertex insertion.
The projection of a mesh vertex $p$ is evaluated by first querying the kernel for
the closest point $p'$ of the B-rep.
The mesh vertex $p$ is then relocated to $p'$ only if this operation
does not create any inverted elements attached to vertex $p$.
If it does, we try to move $p$ to $(p+p')/2$, i.e., the midpoint
between the two points. This procedure is applied iteratively until
we find a valid  position or reach a recursion limit which we set to $5$.
Mesh vertices that were placed in an intermediate position are recorded
and they are included for projection in the next iteration of the algorithm.
The local reconnection that will be applied in the vicinity of point $p$
may enable moving the point closer to its projection  in a subsequent iteration.
Projecting a mesh vertex to the B-rep involves a Newton-Raphson
root finding method and therefore its speed and final solution depend heavily on the initial
guess of the projected point. To speed up the procedure, we approximate the
$uv$ (or $t$) parameters of a newly created point during the point creation module
based on the average of the $uv$ (or $t$) values of the vertices belonging to the triangle or edge being split.
These approximate $uv$ (or $t$) parameters  are then cached for this point and it is
used as an initial guess during the next projection stage.

\begin{figure}[!htpb]
	\centering
	\includegraphics[width=0.6\linewidth]{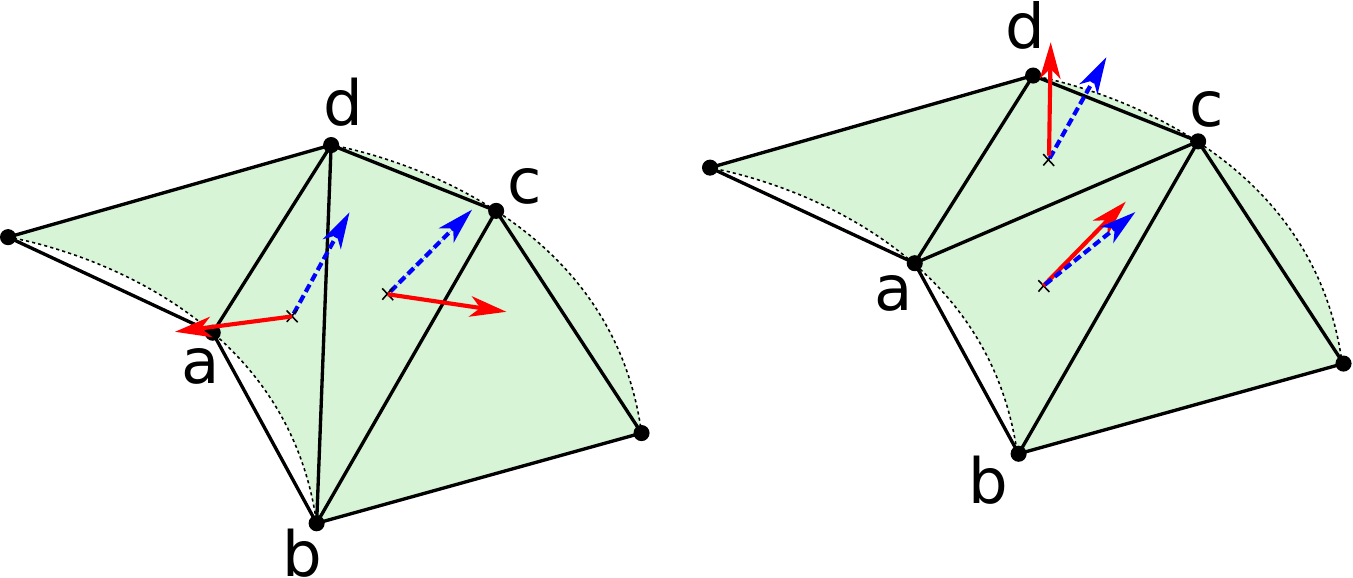}
	\caption{Deviation Improvement. A flip of the edge $bd$ for $ac$ reduces the deviation between the
		discrete normals (red solid vectors) and the analytic normals computed at
		the centroids of the triangles (blue dashed vectors).}
	\label{fig:deviation}
\end{figure}

Information on the  analytic expression of the underlying surface is also used
to minimize the
deviation of the discrete surface mesh from its analytic description. In
practice, the deviation is evaluated as the dot product between the normal of a
discrete triangle and the normal of the geometry surface evaluated at the
centroid of discrete triangle~\cite{park_geometry_2019},
(see also Figure~\ref{fig:deviation}).
The deviation is minimized as part of the local reconnection pass (see Figure~\ref{fig:cad_pipeline}) using 2-2 Flips (see Figure~\ref{fig:flips:flip22}).
The Edge Collapse operation can also use deviation of the surface cavity from the analytic surface as an extra quality criterion
when deciding whether a surface edge should be collapsed.
Controlling the deviation not only produces a mesh that approximates the
surface better, but makes the operations more robust avoiding cases that will
lead to a tangled mesh.

When utilizing  CAD projection, we also found it beneficial to perform vertex
smoothing at the end of the adaptive iteration (see~\autoref{fig:cad_pipeline})
allowing to improve the quality of the vicinity of the projected vertices.

\section{Evaluation}\label{sec:evaluation}
\cdt has been already compared in the literature
to state-of-the-art methods in terms of both mesh quality
and scalability. In particular, the analysis presented in
\cite{tsolakis_parallel_2021} indicates that \cdt offers competitive mesh
quality and scales well within a single shared-memory node.

In contrast to our previous work that
focused mainly on element quality measures, we compiled a suite 
of cases that help re-focusing on the  main goal of mesh adaptation:
capturing features of the underlying simulation. At the same time, we test
how well \cdt operates as a part of an adaptive pipeline composed of
open-source components that include a CFD solver and publicly available test cases
enabling thus future comparisons with other methods.

Section \ref{sec:experimental-setup} describes the setup of our adaptive pipeline in terms of software as well as data flow between the various components.
The test cases are of increasing difficulty and focus on different aspects of
\cdt. First, in section \ref{sec:results-analytic} the adaptation procedure is
verified by adapting meshes based on analytic scalar fields.
Then, section \ref{sec:delta-wing-pipeline} presents results of \cdt with a CFD
solver and a laminar flow case that involves a piece-wise linear
geometry. Section \ref{sec:oneram6_inviscid} increases the difficulty by introducing curved geometry and solving an inviscid flow.
As a stress test in section \ref{sec:high_lift_jaxa} we use \cdt to perform an analysis of a turbulent flow over a complex geometry.
To provide a complete picture of \cdt characteristics we perform a
detailed speedup
and efficiency  analysis of each operation in
section \ref{sec:parallel_performance_evaluation}.

The adaptation is performed using the multiscale metric
\cite{loseille_achievement_2007} which is a feature-based approach that
controls the $L^p$ norm of the interpolation error of a user-defined
scalar variable. The metric field is evaluated using the $L^2$ projection
scheme  provided by \refine \cite{park_refine_2020} software suite.
The final step in constructing a metric field is the application of gradation.
The gradation aims at putting constraints over the largest size deduced from a
metric  and smoothing the metric transition between edge-connected
vertices. In this work we adopt the  \emph{mixed-space-gradation} method presented
in~\cite{alauzet_size_2010} and utilize its implementation in \refine~\cite{refine_github}.

The cases that utilize a CFD solver use SU2 7.0.6 \cite{economon_su2_2016},
version 1.18 of EGADS is used while for \refine we used version 1.9.4-d3ffb79d28.
ParaView was used to visualize results that involve meshes, FreeCAD was used to visualize
geometry, and matplotlib was used for the 2D plots.

\subsection{Experimental Setup}\label{sec:experimental-setup}
\begin{figure}[!htb]
	\centering
	\includegraphics[width=0.7\textwidth]{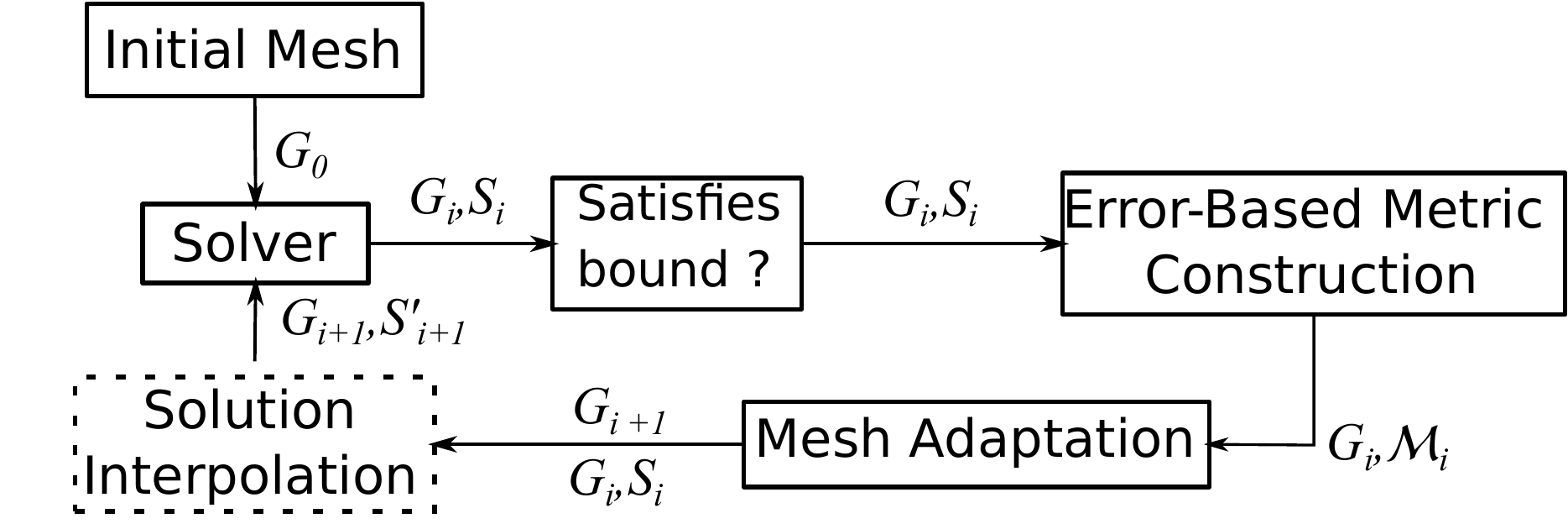}
	\caption{Mesh Adaptation pipeline. $G_i$ denotes the mesh at the $i$-th iteration.
		$S_i,S'_i$ is the solver solution and the interpolated solution at the vertices of $G_i$, respectively.
		$\metric_i$ corresponds to the metric field associated with the vertices of $G_i$ and derived from $S_i$.}
	\label{fig:adaptive_pipeline}
\end{figure}

In order to meet the ever-evolving and growing demands of the CFD community, a
simulation pipeline should be able to integrate a plethora of different tools.
The T-infinity project~\cite{oconnell_tinfinity_2018} demonstrates a series of
different use cases where a high-level Python interface can be used to build
sophisticated pipelines. In this work, we focus on a single pipeline depicted
in~\autoref{fig:adaptive_pipeline} which is pertinent  to mesh adaptation.

In~\autoref{fig:adaptive_pipeline},  the process is  initialized with a (usually
coarse) mesh $G_0$ that captures all the geometrical features of the input
model at some user-defined accuracy. The solver then evaluates a discrete
solution of the problem of interest
and stores it in each mesh element. For
simplicity, we assume that the solver in this case is vertex-based and the
solution is stored at each vertex of the mesh $S_i$. The next block captures a
user-defined condition that controls the exit of the iterative process. It can
be based on some target simulation quantity or on the total number of iterations
of the adaptive loop. The Metric Construction step creates a metric field
$\metric_i$ at each vertex of $G_i$ using $S_i$ that drives the adaptation process.
Mesh Adaptation modifies the mesh based on the provided metric field and generates a new mesh $G_{i+1}$.
Optionally, one can interpolate the solution of the previous iteration to the new mesh thus producing
$S'_{i+1}$.
This step allows the solver to restart the calculation
from a state closer to the converged solution instead of the freestream
conditions which is the default.
Finally, the new mesh (and optionally the interpolated solution $S'_{i+1}$) are
passed to the solver for the next iteration of the loop.

\begin{figure}[!htbp]
	\centering
	\includegraphics [width = 0.7\textwidth]{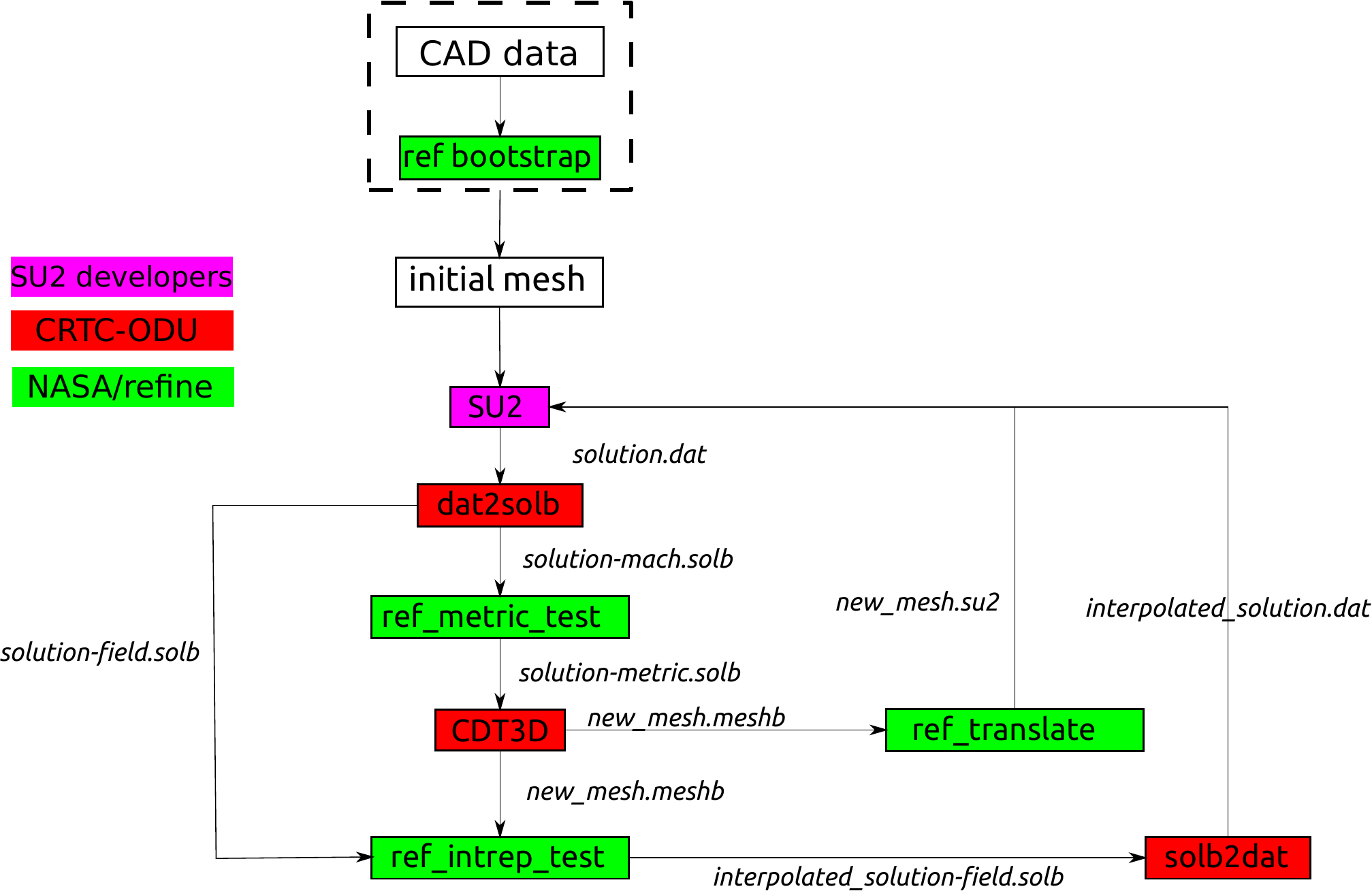}
	\caption{Software pipeline utilized in the adaptive pipelines of this study.}
	\label{fig:su2_pipeline}
\end{figure}

The corresponding  software pipeline can be seen in
Figure~\ref{fig:su2_pipeline}. For the cases of this study, the input volume mesh is
either given or created out of a CAD file using \texttt{ref bootstrap} which is
part of the \refine mesh mechanics suite\cite{refine_github}.
\texttt{ref bootstrap} uses the EGADS~\cite{egads_haimes_2012} kernel in order to generate an initial
surface triangulation of the input CAD file. The surface mesh is then adapted
based on the curvature and other geometrical features. Adapting the surface
in the absence of a volume mesh gives greater flexibility since
the software is not constrained by the requirement of conformity to a volume
after each operation. A volume mesh is then
generated using an external tool such as TetGen~\cite{si_tetgen_2015} or
AFLR~\cite{marcum_unstructured_1995} and finally the volume mesh is adapted based on
a metric field derived by the geometrical features of the CAD input.
\texttt{SU2} will then produce a solution file that holds values of the discrete
solution at each vertex of the input mesh. \texttt{dat2solb} is used to convert the solution
to a libMeshb-compatible file~\cite{libmeshb_github}.
The extracted Mach field
(\texttt{solution-mach.solb}) is then passed to \texttt{ref\_metric\_test} that
creates a multiscale metric field based on it
(\texttt{solution-metric.solb}). The multiscale metric field can be optionally
intersected with a curvature- and feature-based metric built based on the
geometrical features of the input model.
\cdt will then use the metric field along
with the mesh used by \texttt{SU2} to generate an adapted mesh
(\texttt{new\_mesh.meshb}). If solution interpolation is utilized, we pass the
new mesh along with a \texttt{.solb}  version of the \texttt{SU2} solution to
\texttt{ref\_intrep\_test} which we then convert using  \texttt{solb2dat} to an
SU2-compatible file (\texttt{interpolated\_solution.dat}).
The values of the previous solution are interpolated using linear
interpolation.
Finally, the adapted
mesh is passed to \texttt{SU2} after being converted to a \texttt{.su2} mesh file along
with the interpolated solution if this was generated.

It should be noted that the metric field can be built using any solution variable besides the
local Mach number\footnote{The local Mach number is defined as the ratio of the local flow speed over the local speed of sound.}.
However, the use of the local Mach number is favored in the literature,
since it provides a ``compound'' scalar variable that varies in most flow regions,
thus allowing us to capture most of the flow
features~\cite{bourgault_use_2009,habashi_anisotropic_2000}.

Since the solver, a major part of the pipeline is an external and sophisticated project,
fine-tuning its parameters and detailed convergence and error-analysis is outside the
scope of this work. Instead, the goal
of this section is to show the capability of our method to function as part of
an adaptive pipeline.

\subsection{Analytic scalar fields}\label{sec:results-analytic}

The adaptation pipeline  of Figure~\ref{fig:adaptive_pipeline} has
many components and the errors in each one can have accumulative and
unpredictable effects in the final calculation.
In an effort to mitigate these issues, we first test \cdt by replacing
the CFD solver with analytic metric fields. In particular, instead of solving
a flow problem at each iteration, we evaluate an analytic function
at the vertices of the mesh. The adaptive iterations will create a mesh that
is expected
to drive the interpolation error down. For this test, we will be using the
three analytic cases of the benchmark described in~\cite{galbraith_verification_2020} and
implemented in the \refine suite. The multiscale metric implementation
of \refine has been combined with several mesh adaptation tools and
verified
separately in~\cite{galbraith_verification_2020} and thus, we will only focus on
the verification of the adaptation procedure in \cdt.

For each of the three analytic scalar fields
(\eqref{eq:sinfun3},\eqref{eq:tanh3},\eqref{eq:sinatan3}) the adaptation
pipeline starts with a uniform tetrahedral mesh of the unit cube domain $[0,1]
\times [0,1] \times [0,1] $ with $64$ vertices.
In each subsequent iteration, a multiscale
metric  field is computed using $F(x,y,z)$ as a scalar field.
The metric is computed in the $2$-norm and the gradation value is set to $3$.
The metric is then passed to our method along with the mesh of the previous iteration with the goal of adapting the previous mesh to the new metric field.

\begin{align}
	& \texttt{sinfun3} &:= F(x,y,z) &=
	\begin{cases*}
		0.1\sin(50xyz) \text{ if } xyz \leq \frac{-1}{50}\pi \\
		\sin(50xyz) \text{ if } xyz \leq \frac{2}{50}\pi \\
		0.1\sin(50xyz) \text{ else }
	\end{cases*},\nonumber \\
	& & & \text{ where } 	 xyz=(x-0.4)(y-0.4)(z-0.4)  \label{eq:sinfun3}  \\
	& \texttt{tanh3}  &:= F(x,y,z) &= \tanh\left( (x+1.3)^{20} (y-0.3)^9z) \right) \label{eq:tanh3} \\
	& \texttt{sinatan3} &:= F(x,y,z) &= 0.1\sin(50xz) + \tan^{-1}\left(0.1/(\sin(5y) -2xz)  \right)
	\label{eq:sinatan3}
\end{align}

\begin{figure}[h]
	\centering
	\includegraphics[width=0.33\linewidth]{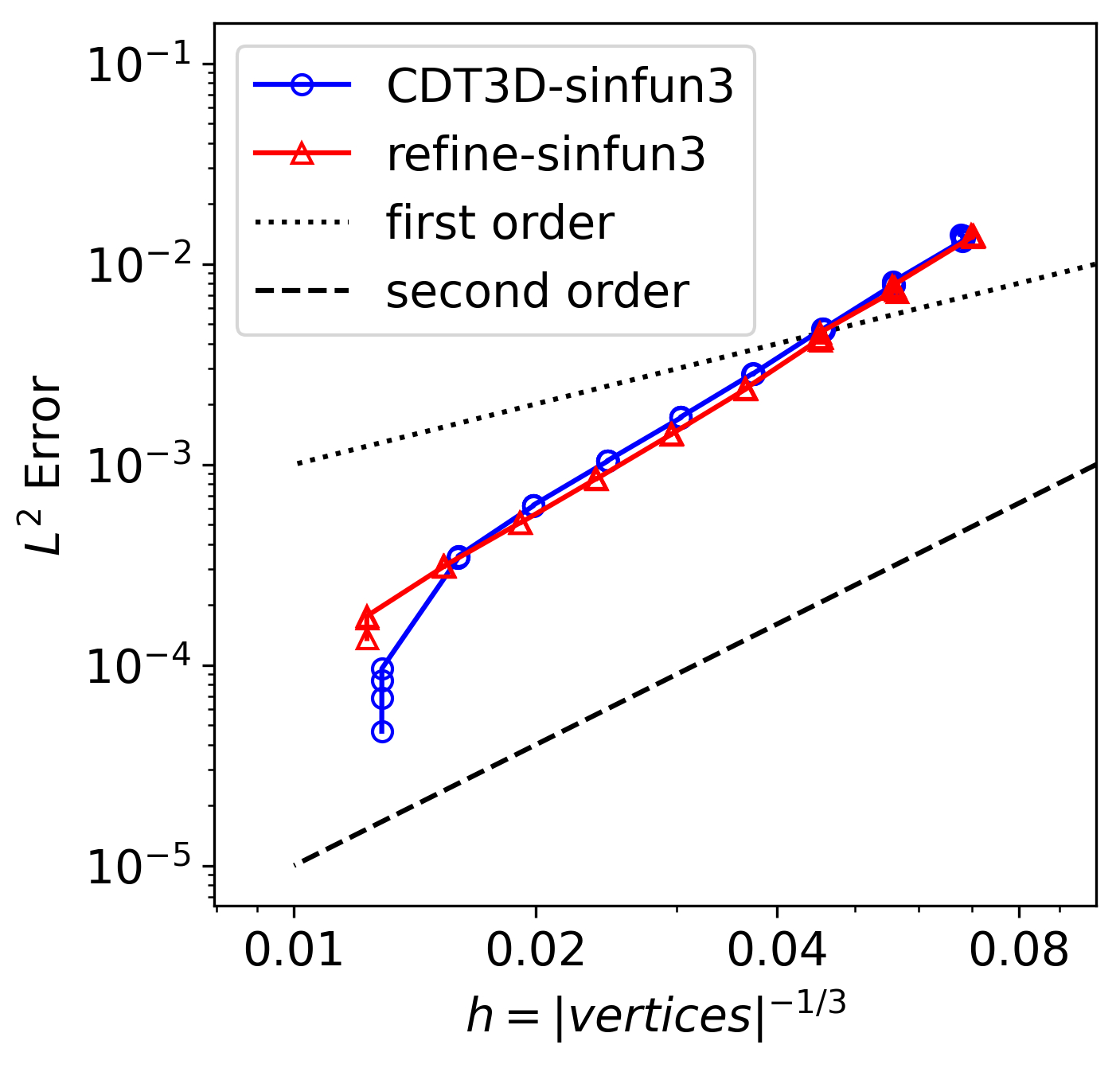}%
	\hfill
	\includegraphics[width=0.33\linewidth]{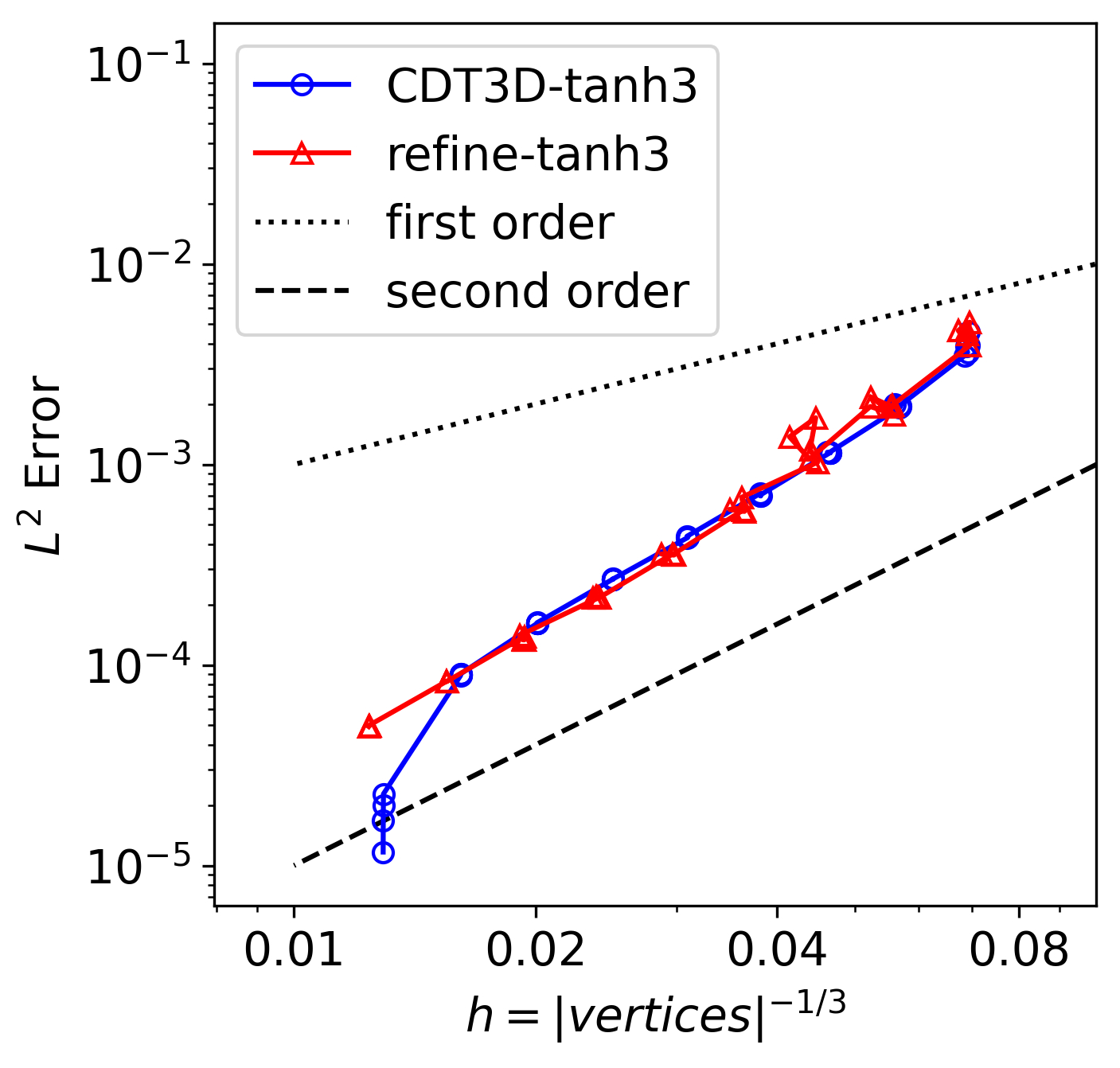}%
	\hfill
	\includegraphics[width=0.33\linewidth]{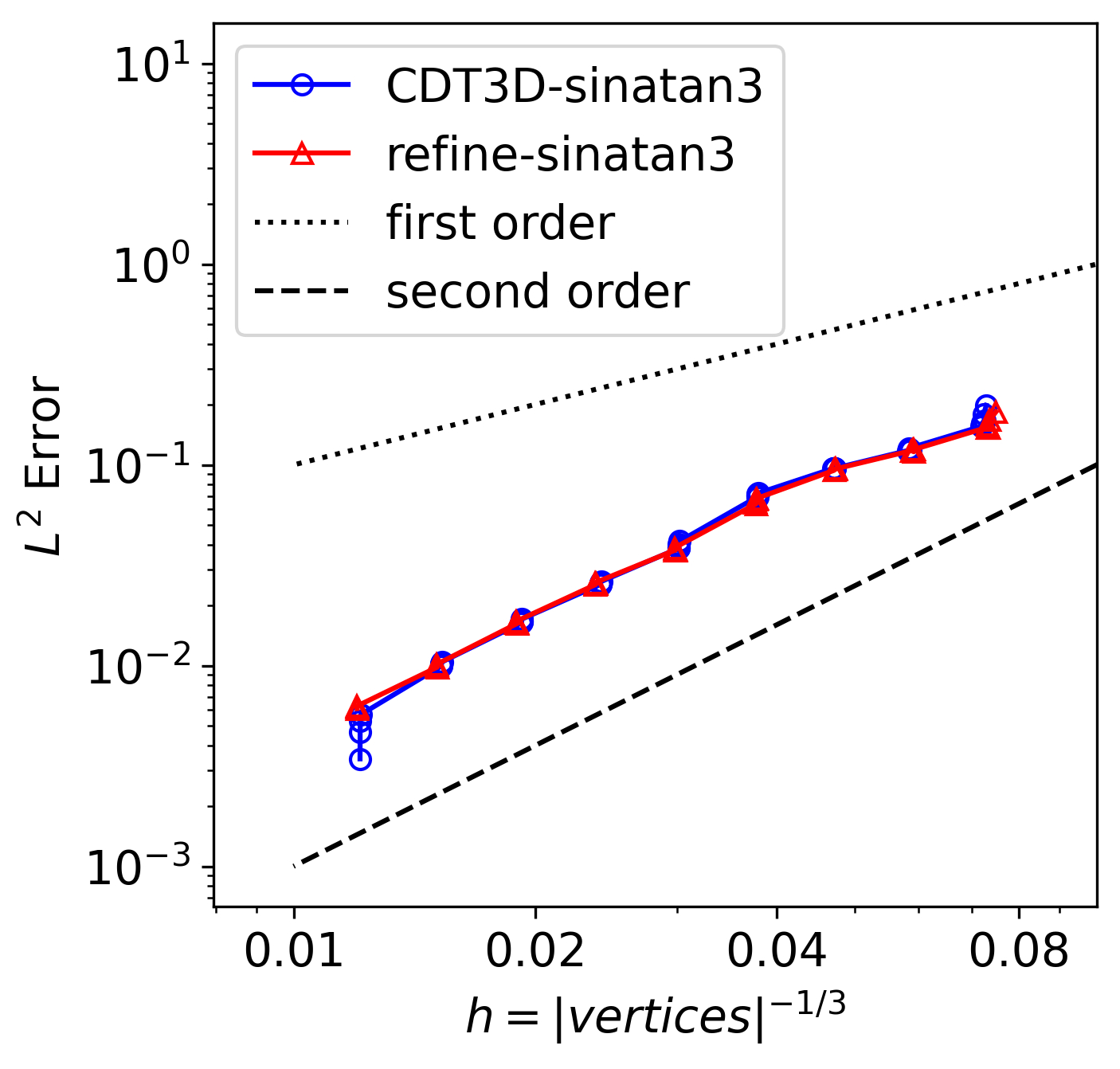}
	\caption{Convergence rates for \cdt and \refine for the three scalar fields.}
	\label{fig:scalar-fields-error}
\end{figure}

For each field, $90$ adaptive iterations are performed with the metric complexity (see equation \eqref{eq:complexity})
increased at every $10$ iterations.  The convergence plots in Figure~\ref{fig:scalar-fields-error}  show the interpolation error of the last $5$
iterations at each complexity with respect to the finest generated mesh. Since,
the multiscale metric approximates linear interpolation error via a Hessian
reconstruction, all results are expected to exhibit second-order convergence
rate. For comparison, the same adaptation procedure was performed using
\refine.

\begin{figure}[h]
	\centering
	\includegraphics[width=\linewidth]{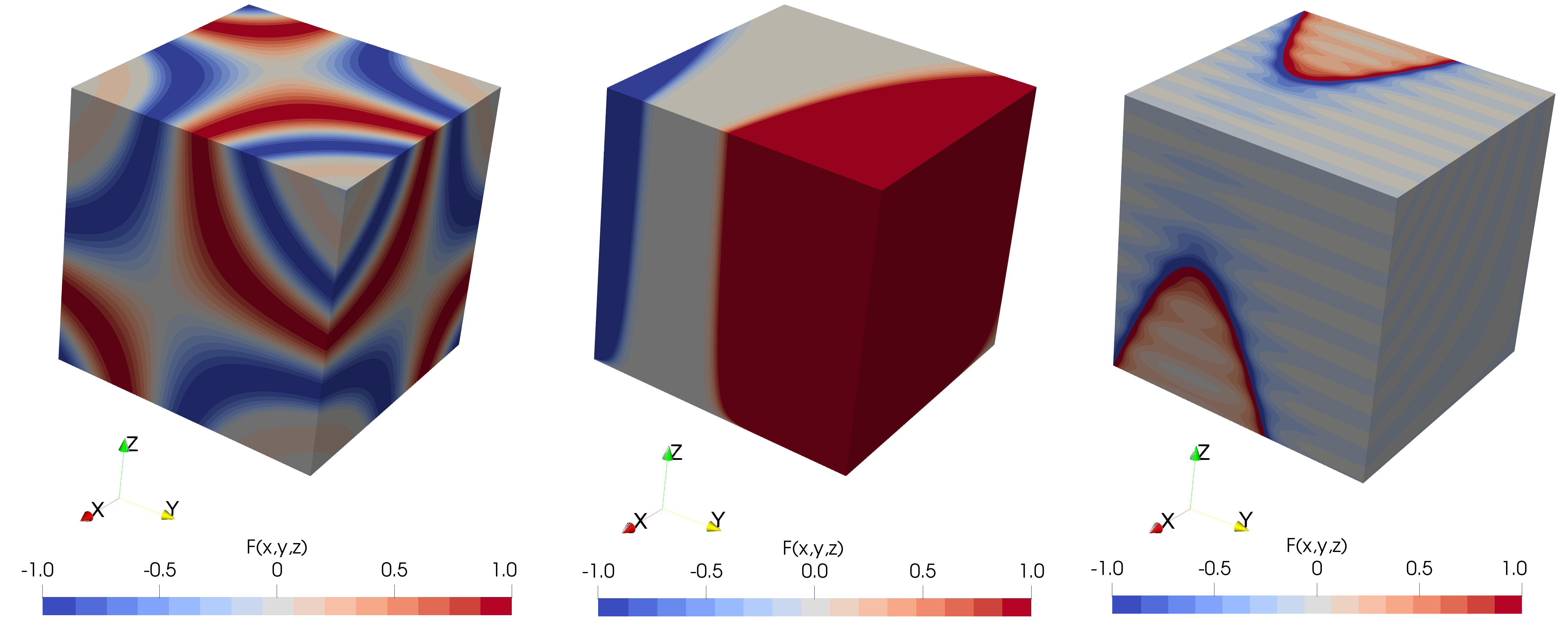}
	\includegraphics[width=\linewidth]{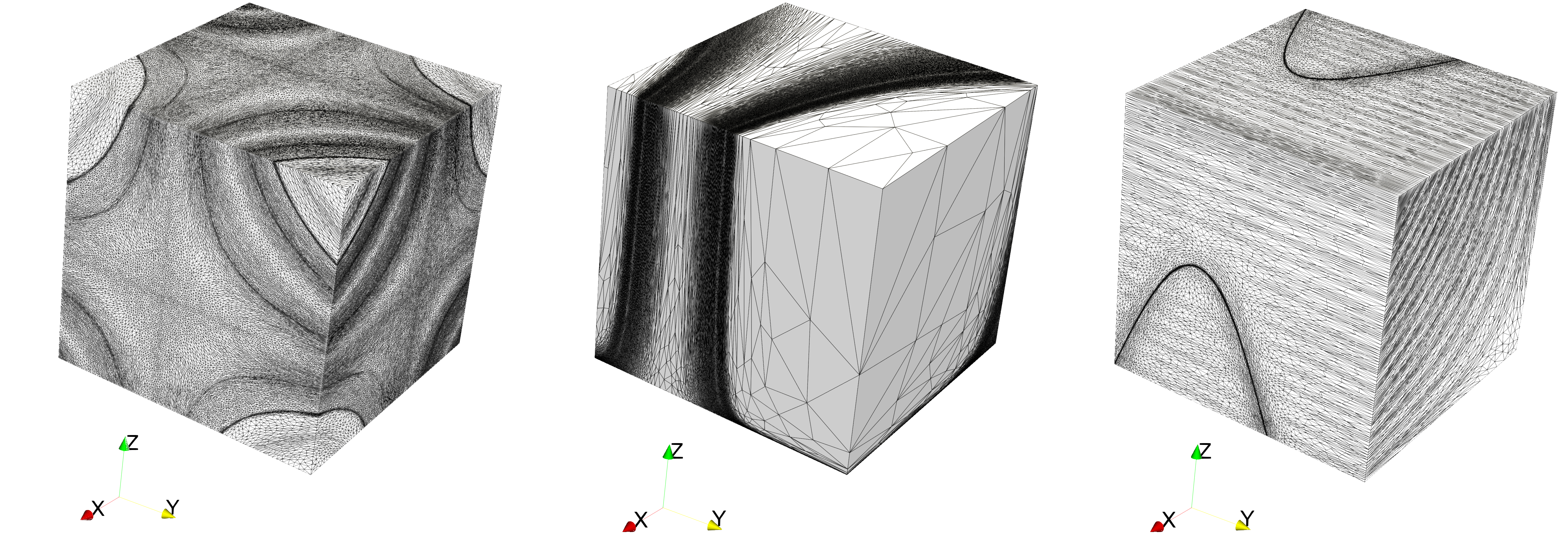}
	\caption
	{Adapted meshes for the three fields. Top: \texttt{sinfun3}, \texttt{tanh3} and  \texttt{sinatan3}
		fields. Bottom: Corresponding \cdt adapted meshes at 256,000 target
		complexity.}
	\label{fig:scalar-fields}
\end{figure}

\autoref{fig:scalar-fields-error} indicates that the convergence rate of \cdt
matches closely the rate of \refine and they both exhibit 2nd-order
convergence.
\autoref{fig:scalar-fields} presents the adapted meshes. \cdt is able to
recover the features of the scalar fields at both small and large scales.


\subsection{Laminar Subsonic Flow over a Delta Wing}
\label{sec:delta-wing-pipeline}

For the next case, \cdt is coupled with an adaptive pipeline that includes a CFD solver.
The input geometry is a delta wing with planar faces.
The 3D delta wing simulation conditions  have been
set so that they match the case used in the first three High-Order Workshops~\cite{wang_high-order_2013}.
This case is well studied in the literature and it is preferred due to its simple
geometry and yet non-trivial flow features.
Adaptive results in terms of multiblock meshes
appear in~\cite{leicht_error_2010}, verification results for the multiscale,
MOESS and output-based metrics appear in~\cite{galbraith_verification_2020}
and~\cite{balan_verification_2020}.

The freestream conditions are $0.3$ Mach, $4000$ Reynolds number based on a
unit root chord length and  $12.5^\circ$ angle of attack. The wing surface is
modeled as an isothermal no-slip boundary with the
freestream temperature equal to $273.15$K. The Prandtl number is
$0.72$ and the viscosity is assumed constant.
SU2 is configured with an initial CFL number of $1$ and a final value of
$5$ with a ramping of $1.001$. As a linear solver, FGMRES is used with the ILU
preconditioner. The error for the linear solver is set to $10^{-10}$ and the
number of the linear iterations to $10$. The Roe convective scheme is used with MUSCL
reconstruction and the Van Albada edge limiter.

For each iteration except the
first, we also supplied an interpolated solution on the new mesh based on
the solution of the previous iteration.
The metric is constructed based on the  local Mach field
of the solution and the metric gradation value is set to $2.0$. The
complexity of  the metric (as defined by \eqref{eq:complexity}) is doubled every  $5$ iterations.
The solution-based metric is intersected
also with a curvature- and feature-based metric built based on the
geometrical features of the wing. Although the geometry, in this case, is planar, the
CAD kernel is utilized to validate its implementation and coupling with
\cdt.
We considered $7$ metric complexity values
for this study: $[50\,000,100\,000,200\,000,400\,000,800\,000,1\,600\,000,3\,200\,000]$.

\begin{figure}[!htpb]
	\centering
	\includegraphics[width=0.27\linewidth]{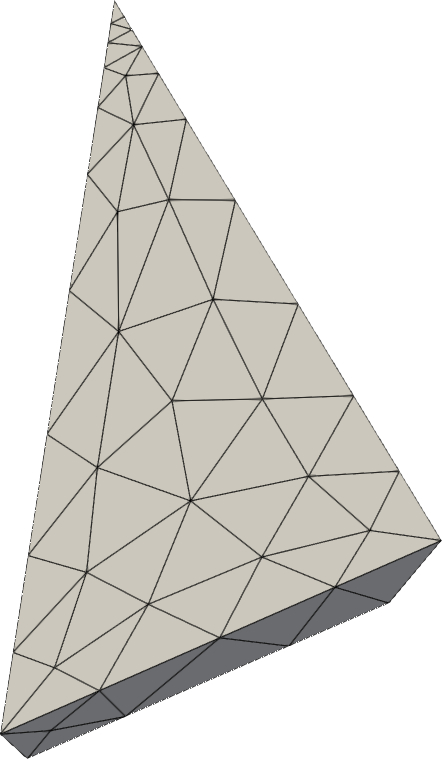}%
	\includegraphics[width=0.27\linewidth]{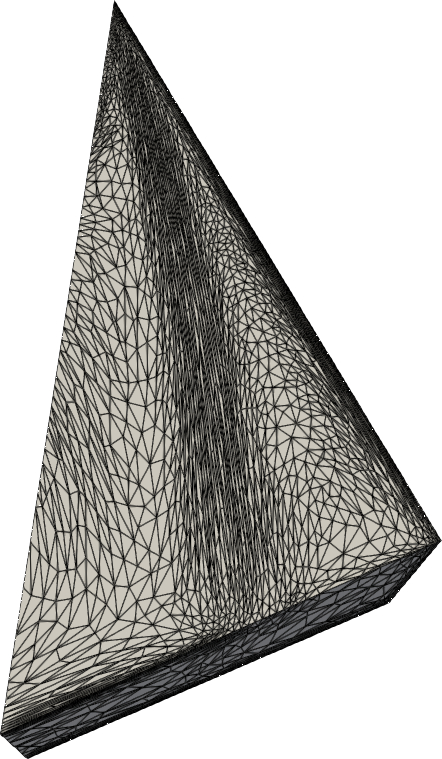}%
	\includegraphics[width=0.27\linewidth]{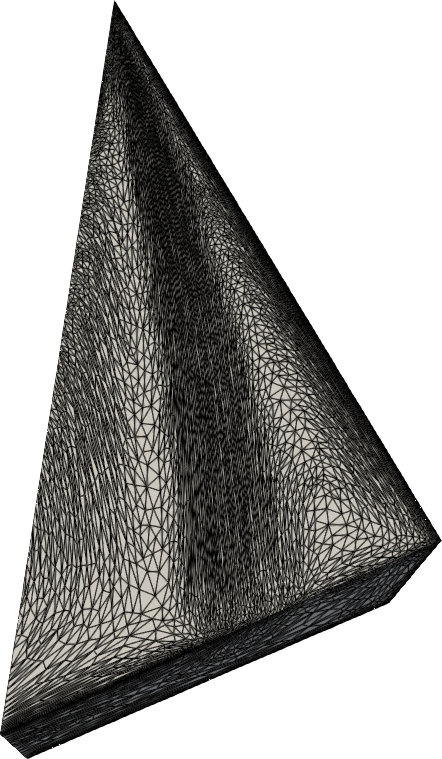}%
	\caption{
		Adapted mesh at three different complexities.
		Left: Initial mesh,
		Middle: mesh at $100,000$ complexity,
		Right:  mesh at $800,000$ complexity.}
	\label{fig:deltawing_meshes}
\end{figure}

\begin{figure}[!htpb]
	\centering
	\includegraphics[width=0.9\linewidth]{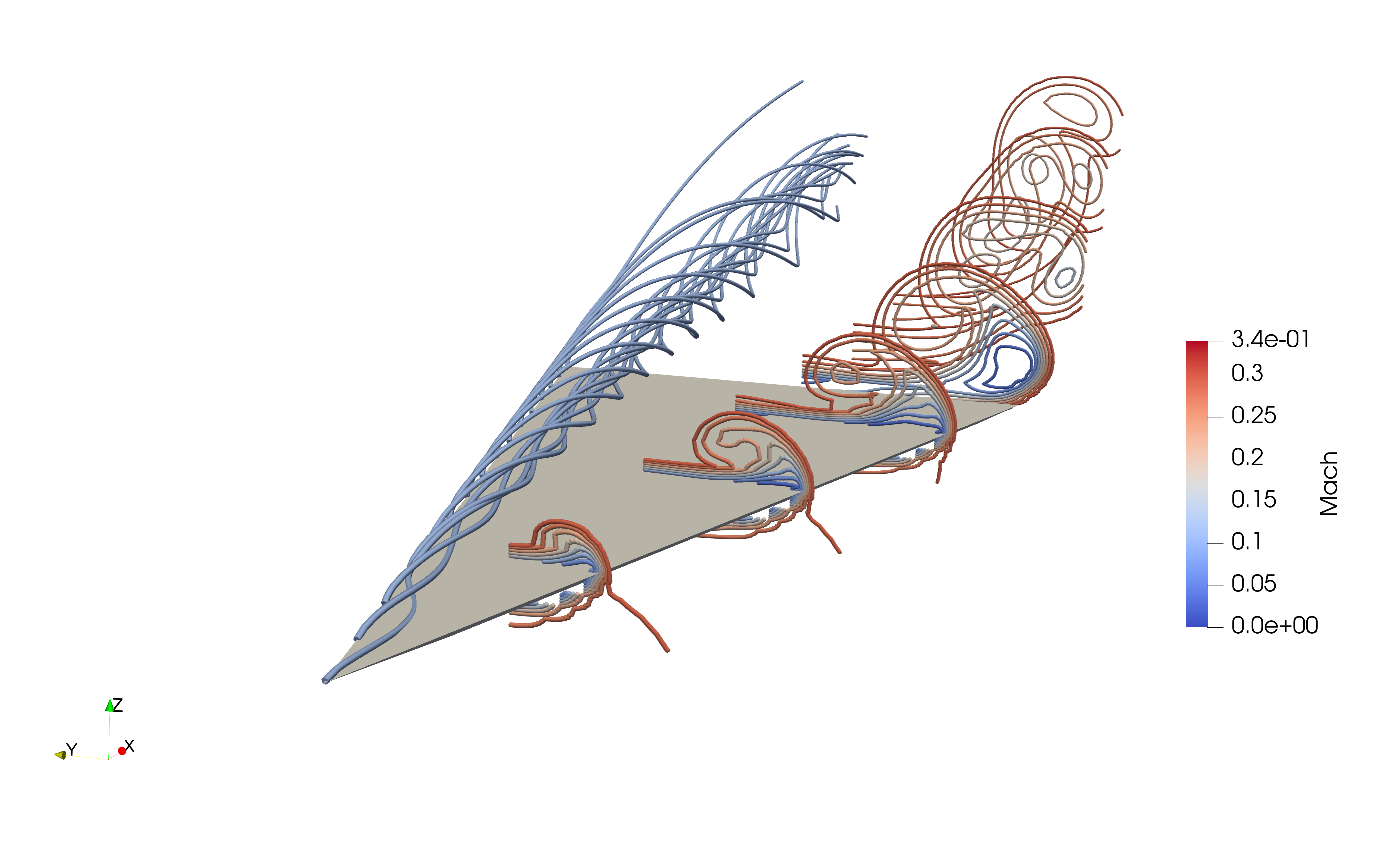}
	\caption{Streamlines and Contour slices of the Mach number of
		the solution. (Simulation performed on the half model).}
	\label{fig:deltacontour}
\end{figure}

Figure~\ref{fig:deltawing_meshes} depicts the initial surface mesh of the
delta wing as well as adapted meshes at $100,000$ and $800,000$ complexity
respectively. The initial mesh has $901$ vertices, the middle corresponds
to the $15$th iteration with $377,569$ vertices and the last corresponds to the
$25$th iteration and has $1,467,922$ vertices. Figure~\ref{fig:deltacontour}
depicts streamlines and contour slices of the final solution.

\begin{figure}[!htpb]
	\centering
	\includegraphics[width=0.8\linewidth]{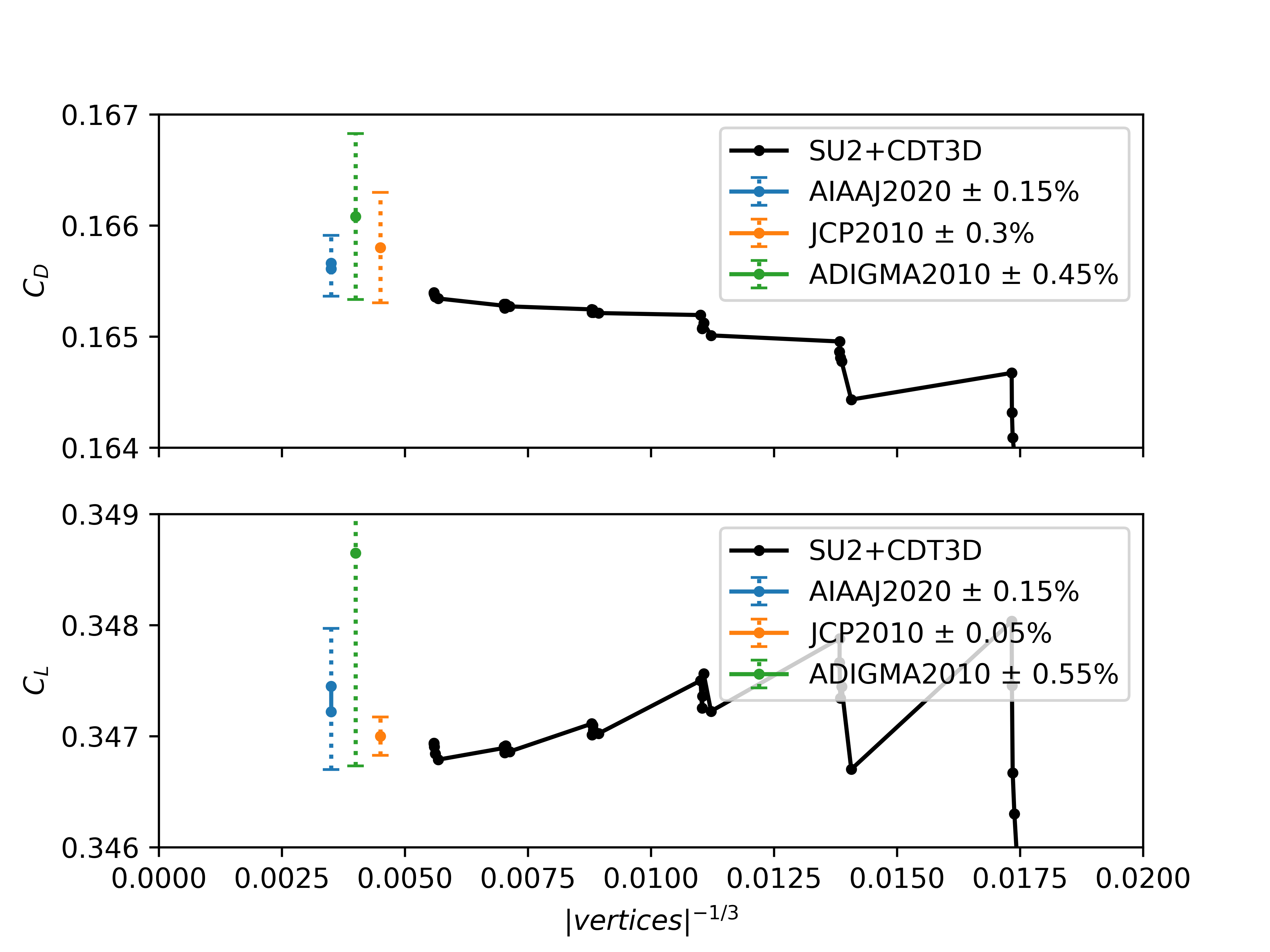}
	\caption
	{Lift and drag coefficients as evaluated by SU2 compared against
		results presented in~\cite{galbraith_verification_2020}(AIAA2020)
		\cite{leicht_error_2010}(JCP2010) and~\cite{hartmann_error_2010}(ADIGMA2010).}
	\label{fig:delta_wing_coeffs}
\end{figure}

To access the quality of the results of the adaptation procedure
and its  coupling with \cdt, drag and lift
coefficients are compared against the results presented
in~\cite{leicht_error_2010,hartmann_error_2010}, and~\cite{galbraith_verification_2020}.
Figure~\ref{fig:delta_wing_coeffs} presents the results.
Both the drag and the lift coefficients are within less than $0.55\%$
of all the reference values.
The final values as evaluated by SU2 on the $35$th iteration are
$C_D=0.165396$ and $C_L=0.346937$.

When it comes to execution time, we collected the time  required for the solver and \cdt which
are the dominant components of the adaptation pipeline.
Table~\ref{tab:deltawing} presents performance data for every 5 iterations of the adaptive pipeline.
SU2 is deployed on the ODU's \texttt{turing} cluster\footnote{\url{https://wiki.hpc.odu.edu/en/Cluster/Turing} (Accessed 2023-05-06).}
that houses nodes with a variety of different specifications.
The number of cores used by the solver was set so that it corresponds to about 10,000 vertices per core
and it was constrained to 300 to reduce the waiting time in the job scheduler queue of the cluster.
\cdt is using one of \texttt{turing}'s nodes with two sockets each one with a Intel\R Xeon\R CPU E5-2698 v3 @ 2.30GHz
(16 cores) for a total of 32 cores.

To ease the comparison
that involves different core counts and hardware specifications, we include a
\emph{core-hours}\footnote{core-hours = the number of cores used by application * hours required for the execution.}
column. Using core-hours allows the evaluation of the performance of the application with respect to the cost of running it on a shared cluster where charging is common to take place in terms of
core-hours.  The running time of \cdt occupies only a small fraction of the
adaptive pipeline. As mentioned in section \ref{sec:experimental-setup} we
used regular files to exchange data between \cdt and the solver for simplicity
but also to allow the solver and \cdt to run on different jobs on the cluster
which minimizes the total time of the experiment. The overhead of using files
is small, for example in iteration $5$ SU2 and
\cdt spent $0.18$ and $0.71$ seconds respectively for writing out the results,
while for the last iteration the respective times are $11.16$ and $86.13$ seconds.

\begin{table}[h]
	\caption{Performance data of adaptive iterations.}
	{\small
		\begin{tabular}{rrrrrrrrr}
            \toprule
			iter. & vertices  & tetrahedra & solver  & solver     & \cdt     & \cdt \\
                  &           &            &   (s)  & core-hours  &	(s)    &  core-hours      \\
           \midrule
			0              & 901       & 3,444      & 57.55           &     0.16   & -        &     -  \\
			5              & 97,896    & 563,930    & 2,833.51        &     7.87   & 63.53    &  0.56  \\
			10             & 192,098   & 1,114,412  & 3,301.11        &    18.34   &  49.26   &  0.44  \\
   			15             & 377,569   & 2,203,660  & 2,897.73        &    32.20   & 111.95   &  1.0   \\
			20             & 749,290   & 4,391,974  & 3,865.75        &    85.91   & 207.08   &  1.84  \\
			25             & 1,467,922 & 8,641,694  & 3,476.85        &   154.53   & 392.55   &  3.49  \\
			30             & 2,897,903 & 17,108,219 & 2,777.01        &   232.96   & 808.27   &  7.18  \\
			35             & 5,726,724 & 33,883,975 & 3,281.82        &   273.49   & 1622.15  & 14.32  \\
            \botrule
		\end{tabular}
	}
	\label{tab:deltawing}
\end{table}

\subsection{Inviscid Onera M6 case}\label{sec:oneram6_inviscid}

The next case introduces CAD data to the adaptation pipeline.
We use an inviscid flow based on the description of a turbulent case included in
 NASA's Turbulence Modeling Resource
(TMR)\footnote{\url{https://turbmodels.larc.nasa.gov/onerawingnumerics_val.html} (Accessed 2023-05-06).}.
As mentioned in NASA's website\footnote{\url{https://www.grc.nasa.gov/www/wind/valid/m6wing/m6wing.html} (Accessed 2023-05-06).}
\emph{``The ONERA M6 wing is a classic CFD validation
case for external flows because of its simple geometry combined with complexities of transonic flow [..]
It has almost become a standard for CFD codes because of its inclusion as a validation case in numerous CFD papers over the years.''}
The flow conditions for this study are $3.06^\circ$ angle of attack, $0.84$ Mach
number, and freestream temperature equal to $300$K.
This case utilizes the ONERA M6 wing geometry 
\autoref{fig:oneram6_verification_input_mesh} depicts
the initial mesh generated by \texttt{ref bootstrap}.

SU2 is configured similarly to the previous case but using the JST as the convective scheme
which we found to converge faster for this case.
For each iteration except the
first, we also supplied an interpolated solution on the new mesh based on
the solution of the previous iteration.
The metric is constructed based on the local Mach field
of the solution and the metric gradation value is set to $10$. The
complexity of  the metric (see equation \eqref{eq:complexity}) is doubled every  $5$ iterations.
The solution-based metric is intersected
also with a curvature- and the feature-based metric is built based on the
geometrical features of the wing. We considered $3$ metric complexity values
for this study: $[50\,000, 100\,000, 200\,000]$. 

These flow conditions produce the typical ``lambda'' shock along the upper surface wing.
\autoref{fig:oneram6_mesh_solution} depicts the mesh as well as
the corresponding contour plot of the local Mach number
for the final iteration of the adaptive loop.

To verify our results, we compare the pressure coefficient against two different datasets from the literature.
First, the experimental values of Case 2308 of~\cite{schmitt_pressure_1979}
acquired from the TMR website\footnote{\url{https://turbmodels.larc.nasa.gov/onerawingnumerics_val.html}
(Accessed 2023-05-06).} that corresponds to our configuration. Also, we executed
SU2 with the same configuration on a structured grid generated using a custom publicly available  mesh generation
code~\cite{nishikawa_customized_2018} using the input parameters suggested by the TMR
website\footnote{\url{https://turbmodels.larc.nasa.gov/onerawingnumerics_grids.html} (Accessed 2023-05-06).}.
In particular, we used the level 2 mesh (L2) that has $36,865$ points across the surface of the wing.
For comparison, the final mesh of our pipeline has $6,898$ points across the surface of the wing.
The top left subfigure in \autoref{fig:oneram6_cp_sections} depicts the $7$ sections along which the experimental and numerical
results are compared. The rest of the plots in \autoref{fig:oneram6_cp_sections} compares the results generated
using \cdt and the pipeline of~\autoref{fig:su2_pipeline},
the structured mesh, and the experimental values.
The $x$ axis denotes the $x$-coordinate of the cross-section normalized by the local cord length of the wing.
The $y$ axis represents the local pressure coefficient which measures the pressure at a point relative to the freestream conditions.
The combination of \cdt with SU2 generates results very close
to the experiment and the numerical solution obtained on the structured mesh.
The differences with the experimental values are in part due to
the  inviscid method used in this simulation. We attempted to perform a viscous simulation
using the same configuration but we did not succeed in obtaining converged results.
Still, these results indicate that the meshes produced by \cdt in the presence of simple curved geometries
supplied as CAD data are suitable for inviscid calculations and the results are close to the reference values.

\renewcommand{\w}{0.41}
\begin{figure}[!htpb]
	\centering
 	\begin{subfigure}[b]{\w\textwidth}
  	\includegraphics[width=\linewidth]{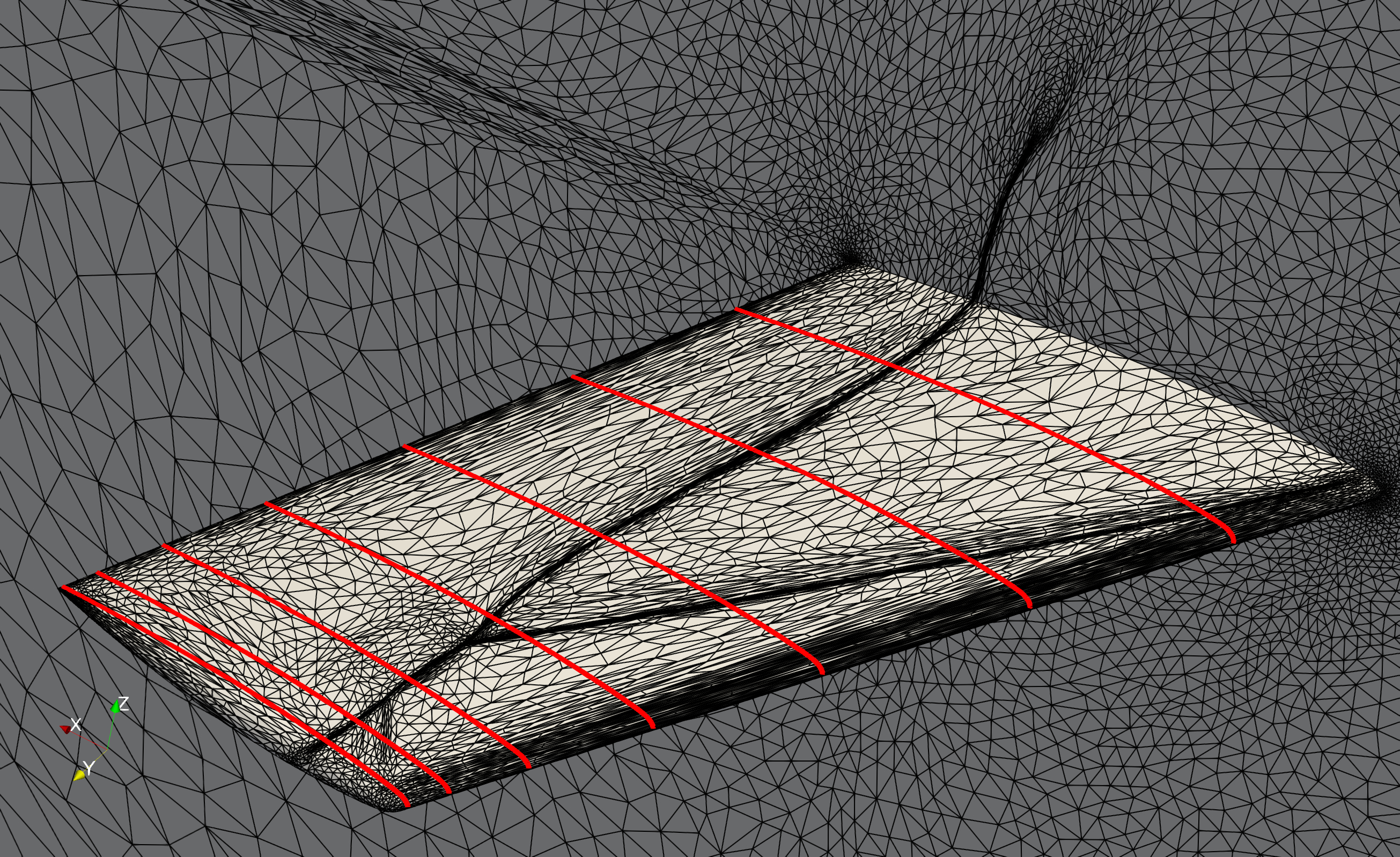}
   	\label{fig:oneram6_cp_lines}
	\end{subfigure}
	\begin{subfigure}[b]{\w\textwidth}
	\includegraphics[width=\linewidth]{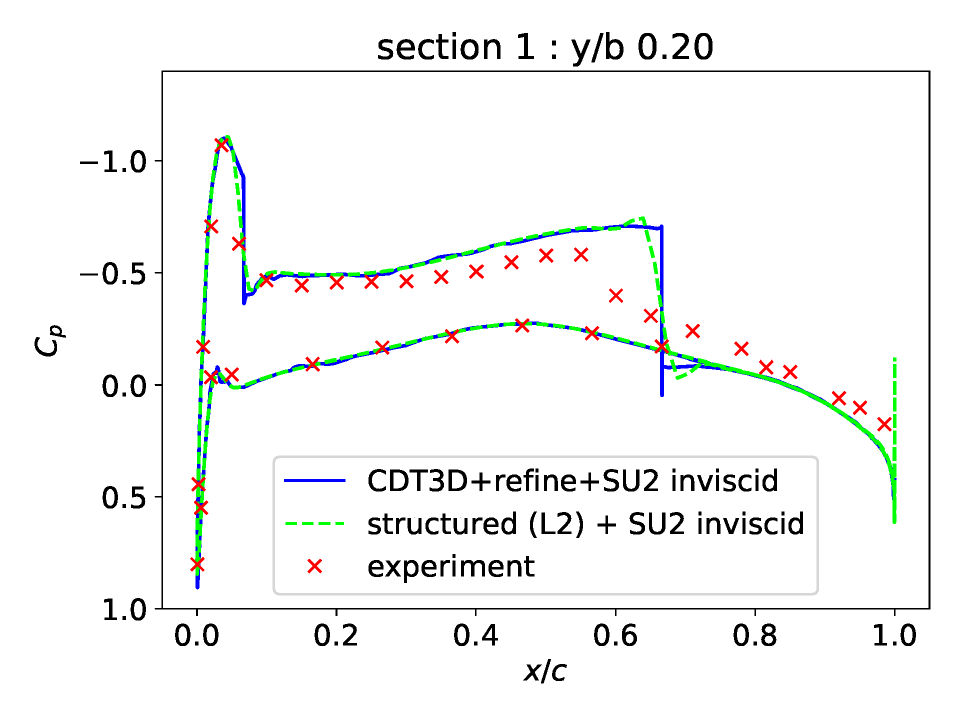}%
	\end{subfigure}
	\begin{subfigure}[b]{\w\textwidth}
	\includegraphics[width=\linewidth]{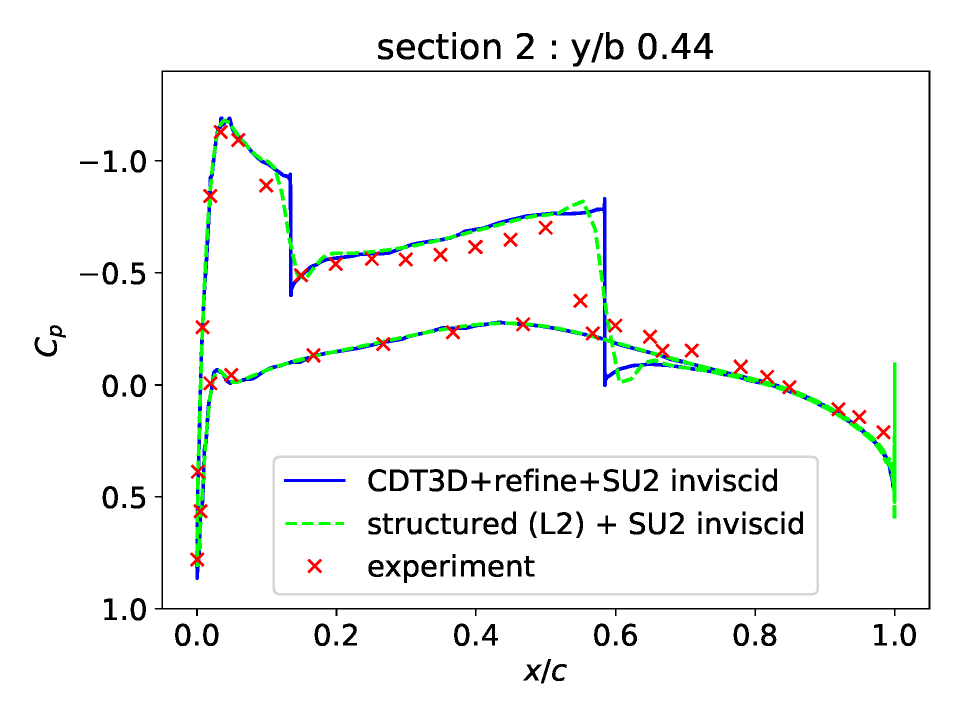}
	\end{subfigure}%
	\begin{subfigure}[b]{\w\textwidth}
	\includegraphics[width=\linewidth]{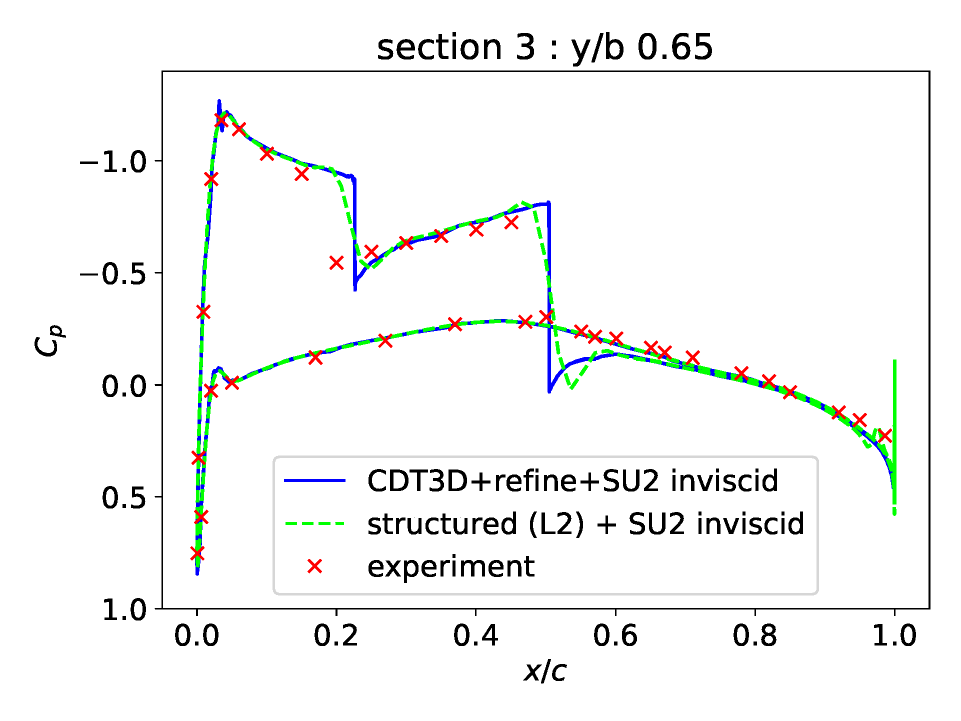}%
	\end{subfigure}
	\begin{subfigure}[b]{\w\textwidth}
	\includegraphics[width=\linewidth]{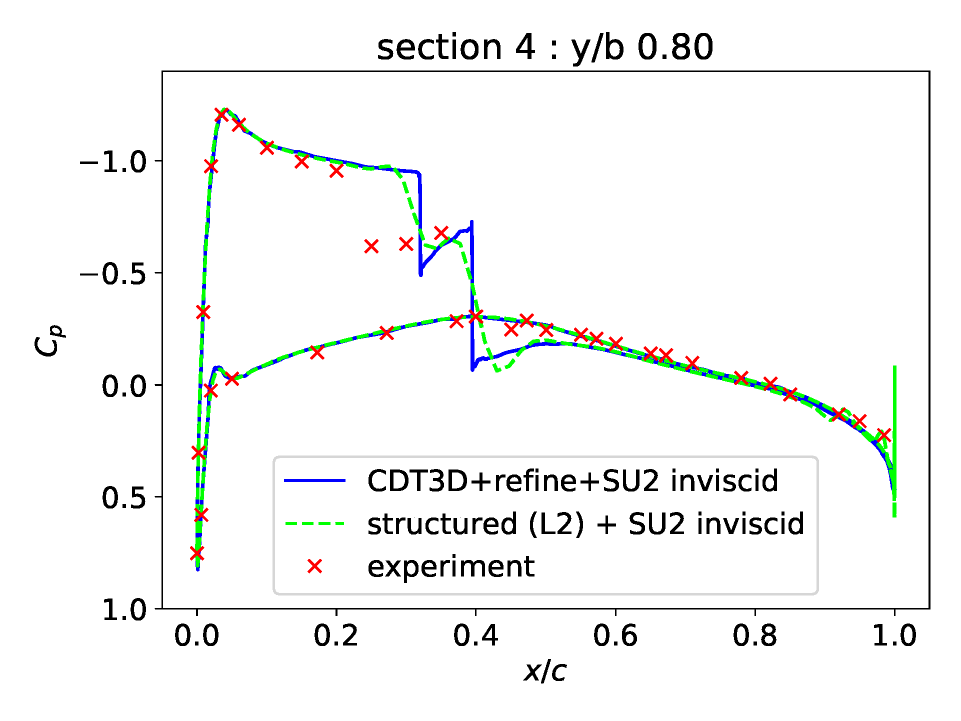}
	\end{subfigure}%
	\begin{subfigure}[b]{\w\textwidth}
	\includegraphics[width=\linewidth]{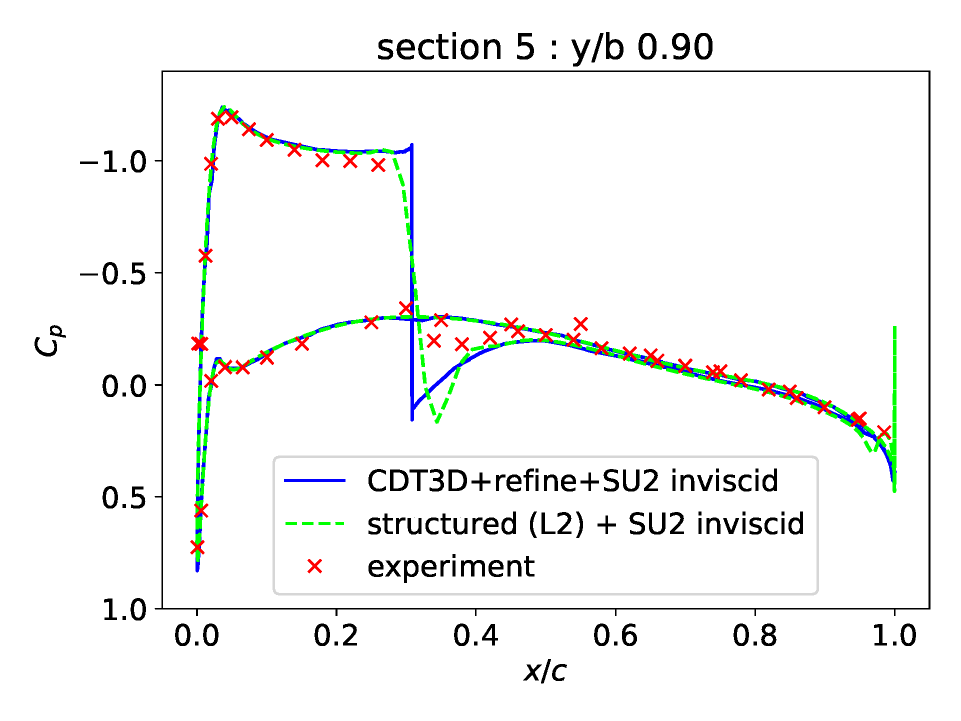}%
	\end{subfigure}
	\begin{subfigure}[b]{\w\textwidth}
	\includegraphics[width=\linewidth]{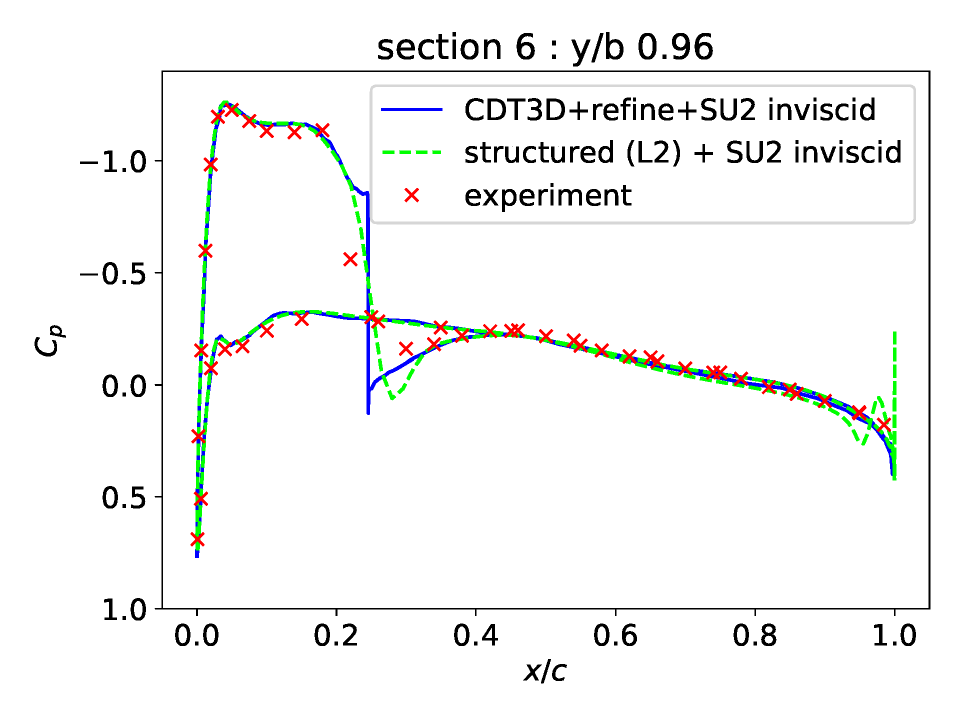}
	\end{subfigure}%
	\begin{subfigure}[b]{\w\textwidth}
	\includegraphics[width=\linewidth]{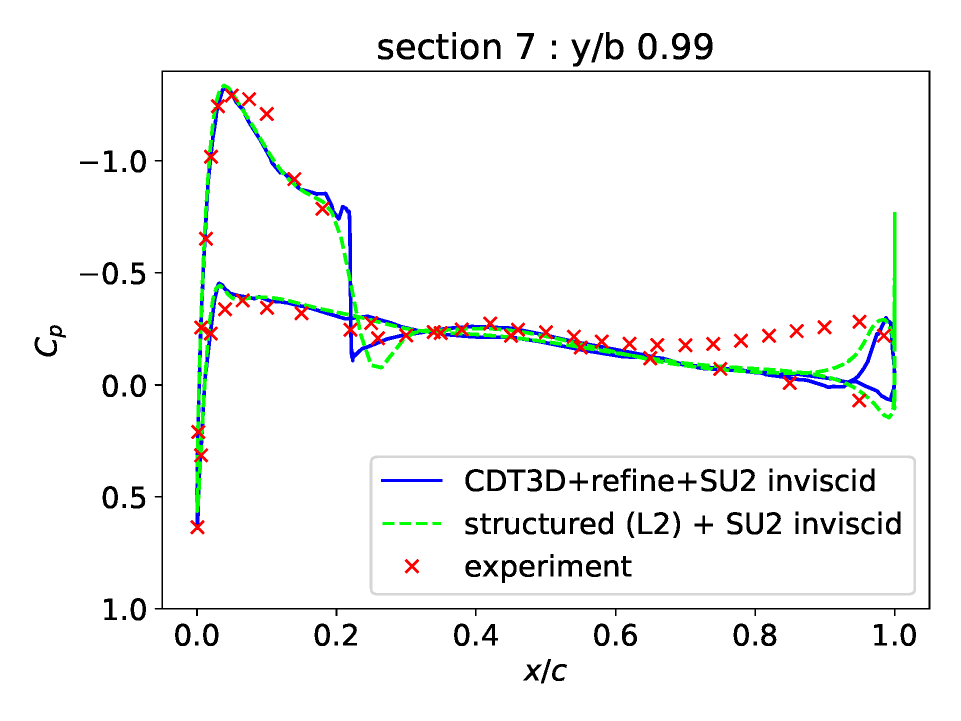}%
	\end{subfigure}

	\caption{Values of the pressure coefficient as evaluated by the solver versus the experiment
		across the 7 sections of the ONERAM6 wing. The location of the sections are depicted in the first figure.}
	\label{fig:oneram6_cp_sections}
\end{figure}

\subsection{Inviscid flow over the JAXA Standard Model}
\label{sec:high_lift_jaxa}

As a final stress test, we use the
Japan Aerospace Exploration Agency (JAXA) Standard
Model (JSM). JSM was built as an attempt to study flow effects
over a fairly complete configuration instead of isolated aircraft components
that were commonly used. There are several experimental  data available,
see for example~\cite{ito_high-lift_2006,	yokokawa_investigation_2008,yokokawa_aerodynamic_2010}
but, we will focus on the use of the JSM
in the context of the 3rd AIAA CFD High-Lift Prediction Workshop\footnote{\url{https://hiliftpw.larc.nasa.gov/index-workshop3.html}
	(Accessed 2023-05-06).}.
A summary of the workshop's results appears in~\cite{rumsey_overview_2018}.
In particular, we will study the case 2b which excludes the pylon and the nacelle
of the original model and uses an angle-of-attack equal to $4.36^\circ$
and a Mach number of $0.172$.
The JSM geometry is combined out of 200+ surfaces, modeling details of the aircraft including
brackets, flaps and slats (see~\autoref{fig:jsm_geometry}).

SU2 is configured similarly to the inviscid ONERA M6 case of the previous section.
For each iteration except the
first, we also supplied an interpolated solution on the new mesh based on
the solution of the previous iteration.
For the first iteration, we used the coarse mesh of~\autoref{fig:jsm_initial_mesh} of appendix \ref{sec:supplementary_figures}
created by \texttt{ref bootstrap} of the \refine mesh mechanics suite~\cite{refine_github}.
The metric is constructed based on the Mach field
of the solution and the metric gradation value is set to $1.5$. The
complexity of  the metric (see equation \eqref{eq:complexity}) is doubled every  $5$ iterations.
The solution-based metric is intersected
also with a curvature- and feature-based metric built based on the
geometrical features of the model. We considered $8$ metric complexity values
for this study: $[50\,000, 100\,000, 200\,000,
400\,000, 800\,000, 1\,600\,000, 3\,200\,000, 6\,400\,000]$.
The final mesh contains  $13,227,952$
vertices, $478,518$ triangles
and $78,479,450$
tetrahedra.

Figure~\ref{fig:jsm_wing} in appendix \ref{sec:supplementary_figures}
 depicts the upper surface of the wing of the final iteration
along with the distribution of the local Mach number around it. Notice
that the method inserts more points around the regions of higher variability of
the local Mach number as expected.
In particular, the wakes of the slat brackets are resolved on the upper surface.
These wakes are initiated at the sharp edges of the brackets.
Figure~\ref{fig:jsm_vortex} depicts the final mesh along with the final solution colored my the local Mach number.
Zoom-in views of one of the generated vortices are also provided.

To verify our results we compare the pressure coefficient values as evaluated by
the solver  against experimental results acquired from High-lift workshop
website\footnote{\url{https://hiliftpw.larc.nasa.gov/Workshop3/pressures.html} (Accessed 2021-06-17).}.
\autoref{fig:jsm_cp_lines} depicts the locations of $C_p$ extraction along the wing of JSM.
The rest subfigures of~\autoref{fig:jsm_cp_sections} present results generated using our
approach and the pipeline of~\autoref{fig:su2_pipeline}. In general, the obtained results
are close to the experimental values. Notice, however, that our results overpredict the $C_p$
values on the upper surface of the wing which corresponds to the upper section of the blue datapoints.
This is in part attributed to the fact that we used an inviscid simulation instead of a viscous one.
Viscous simulations were attempted starting from a coarse mesh but we didn't succeed in
obtaining a converged solution. Even though we didn't investigate the reasons in depth and we are not familiar with the internal workings of SU2,
we believe that these failed attempts were due to a combination of factors.
First, the numerical schemes used in SU2 in combination with the complex geometry of
this case might require a mesh better aligned to the geometrical features of the problem.
Also, a boundary layer mesh might be required to achieve convergence for this
configuration. A different solver configuration might
yield better results but exploring the configuration space of the solver is out
of the scope of this work.
Still, this case indicates that the new functionality of
\cdt allows the method to handle fairly complicated CAD data in combination with solution-based
metric derived from inviscid calculations.

\renewcommand{\w}{0.45}
\begin{figure}[!htpb]
	\centering
	\begin{subfigure}[b]{\w\textwidth}
		\includegraphics[width=0.85\linewidth]{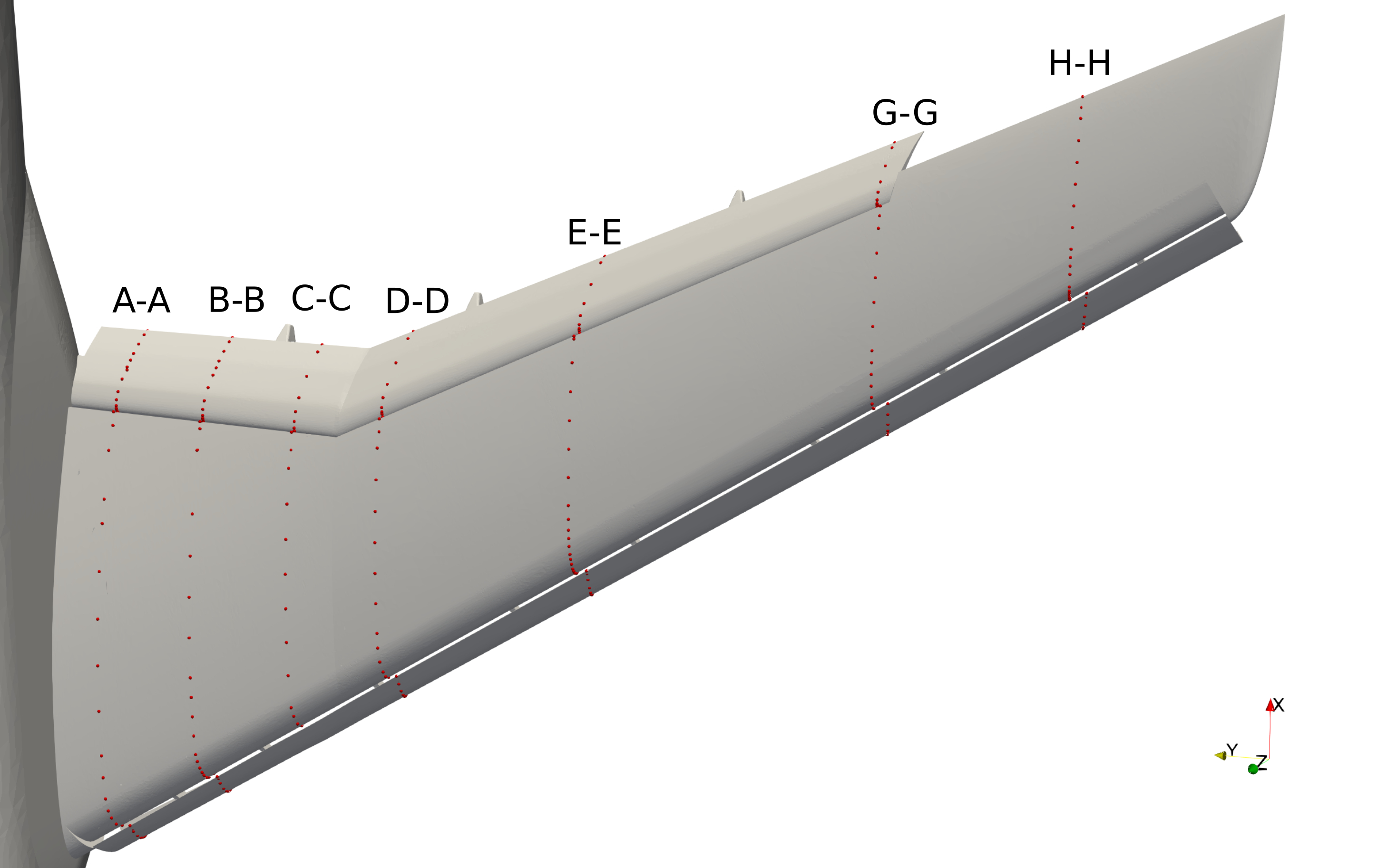}%
		\caption{Locations of experimental data measurements.}
		\label{fig:jsm_cp_lines}
	\end{subfigure}
	\begin{subfigure}[b]{\w\textwidth}
		\includegraphics[width=\linewidth]{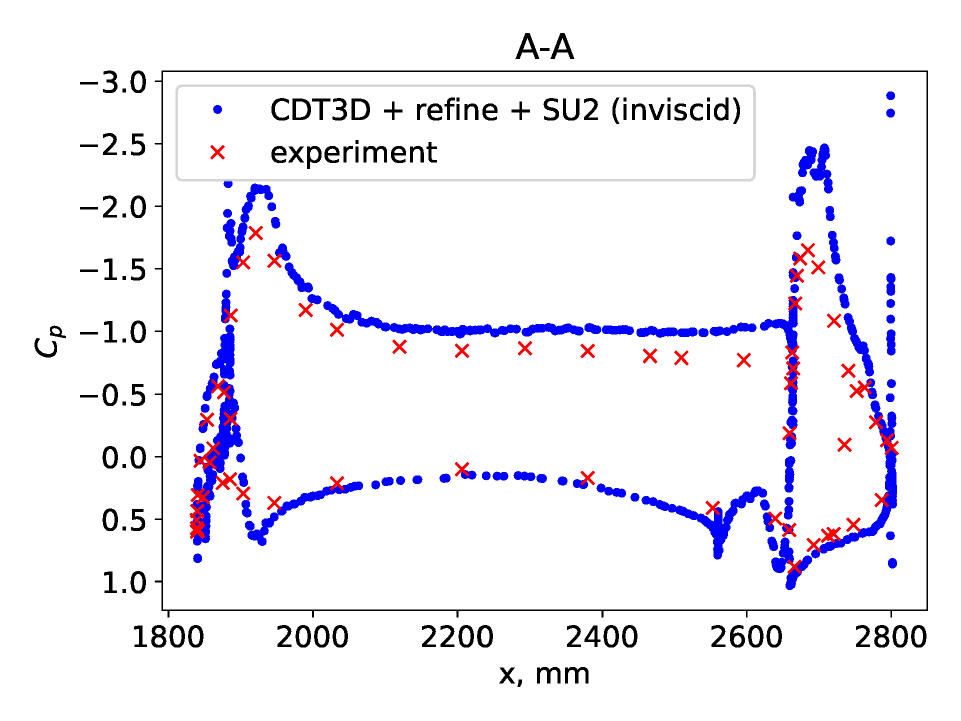}%
	\end{subfigure}
	\begin{subfigure}[b]{\w\textwidth}
		\includegraphics[width=\linewidth]{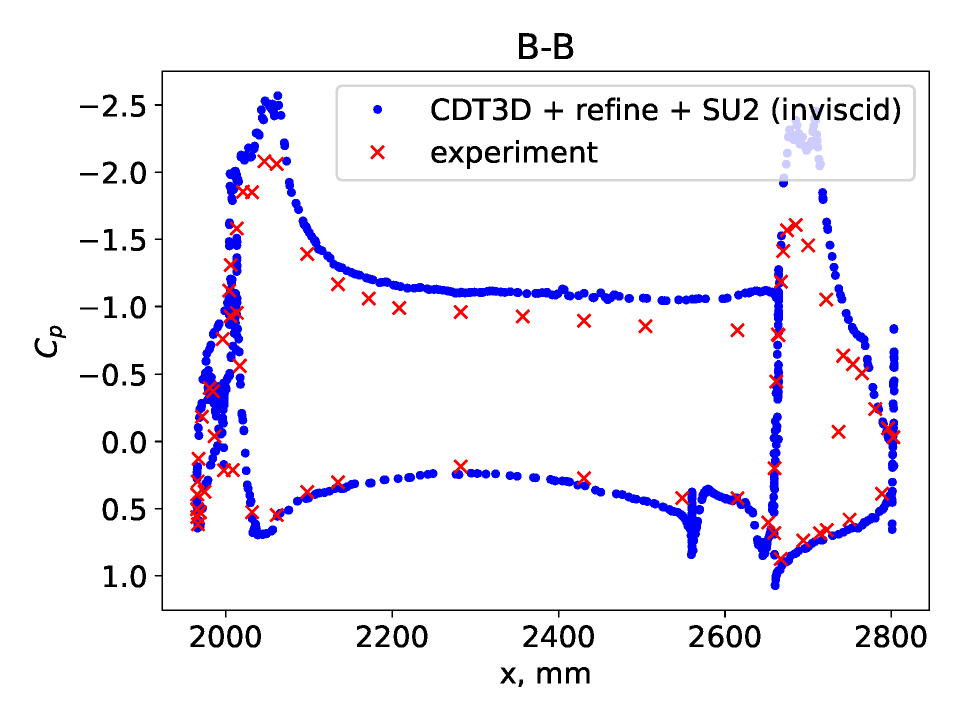}
	\end{subfigure}%
	\begin{subfigure}[b]{\w\textwidth}
		\includegraphics[width=\linewidth]{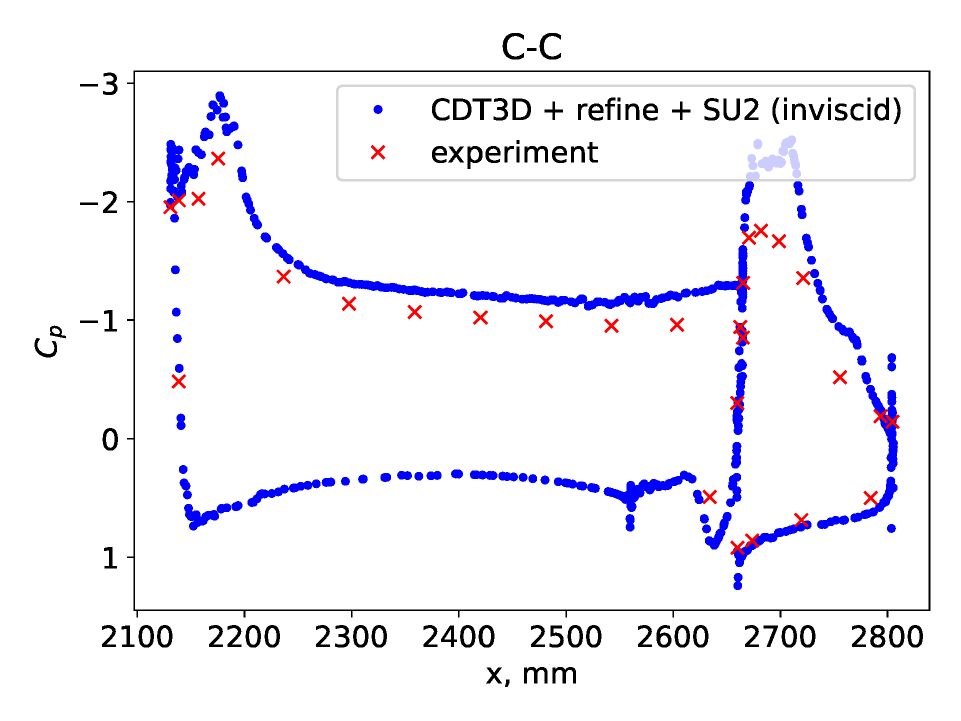}%
	\end{subfigure}
	\begin{subfigure}[b]{\w\textwidth}
		\includegraphics[width=\linewidth]{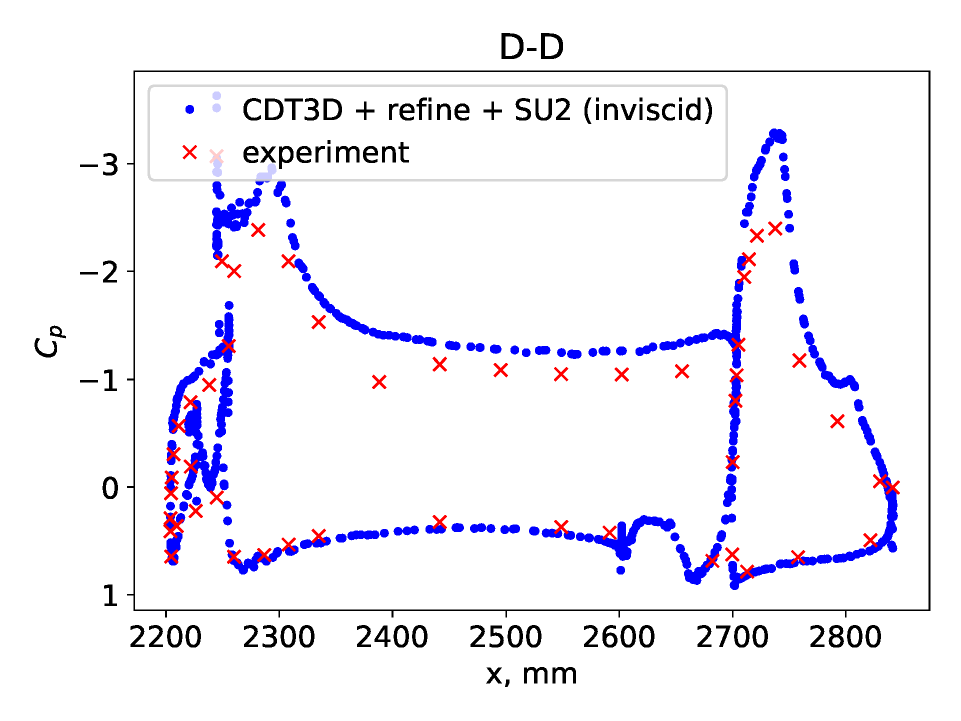}
	\end{subfigure}%
	\begin{subfigure}[b]{\w\textwidth}
		\includegraphics[width=\linewidth]{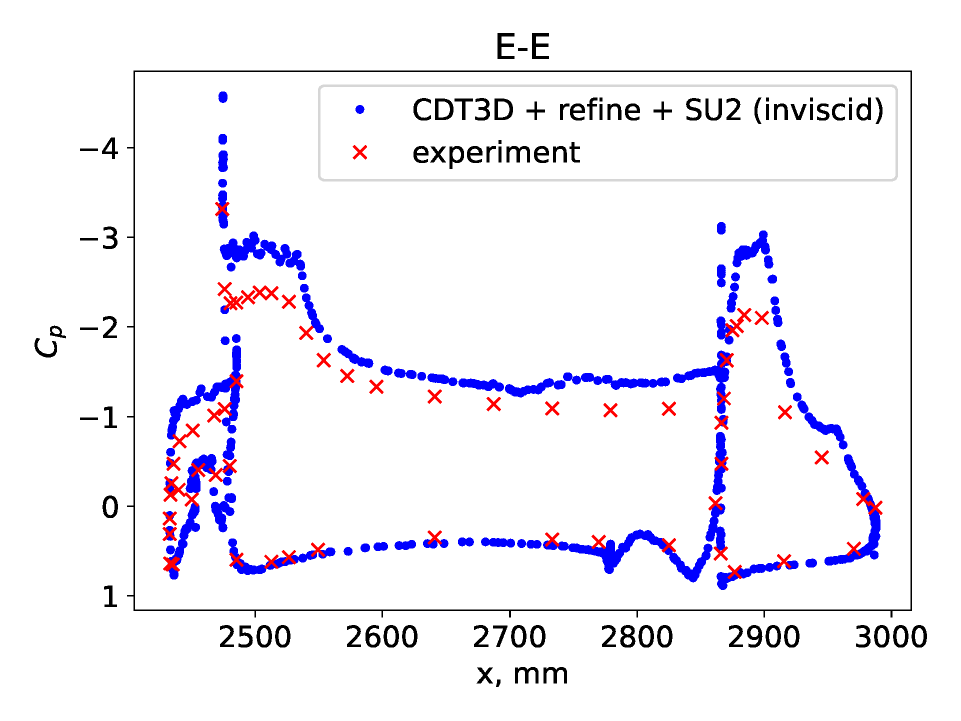}%
	\end{subfigure}
	\begin{subfigure}[b]{\w\textwidth}
		\includegraphics[width=\linewidth]{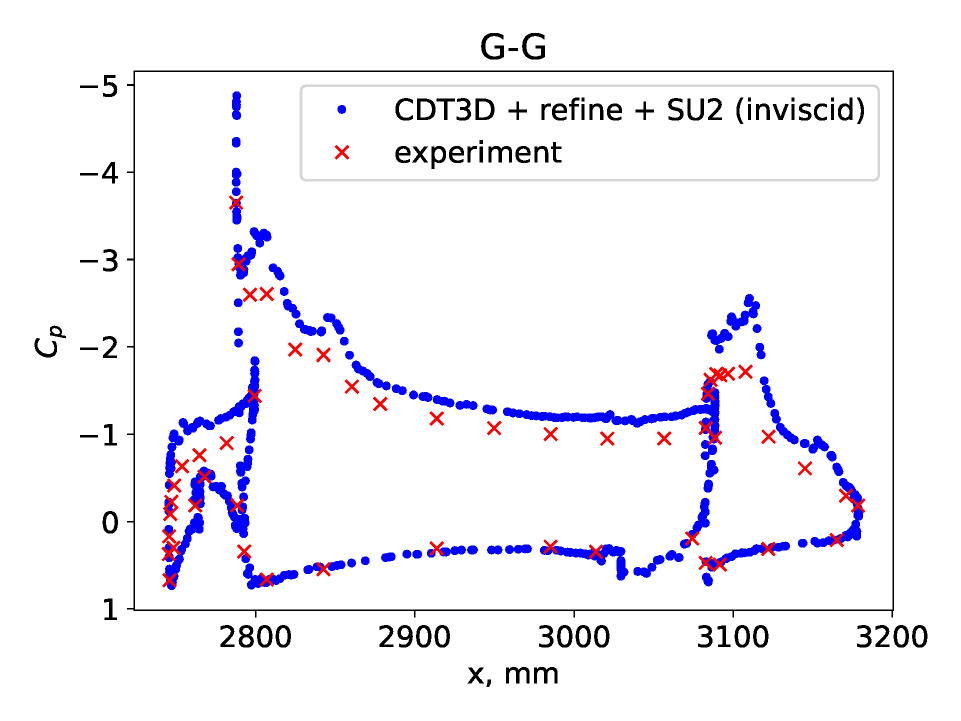}
	\end{subfigure}%
	\begin{subfigure}[b]{\w\textwidth}
		\includegraphics[width=\linewidth]{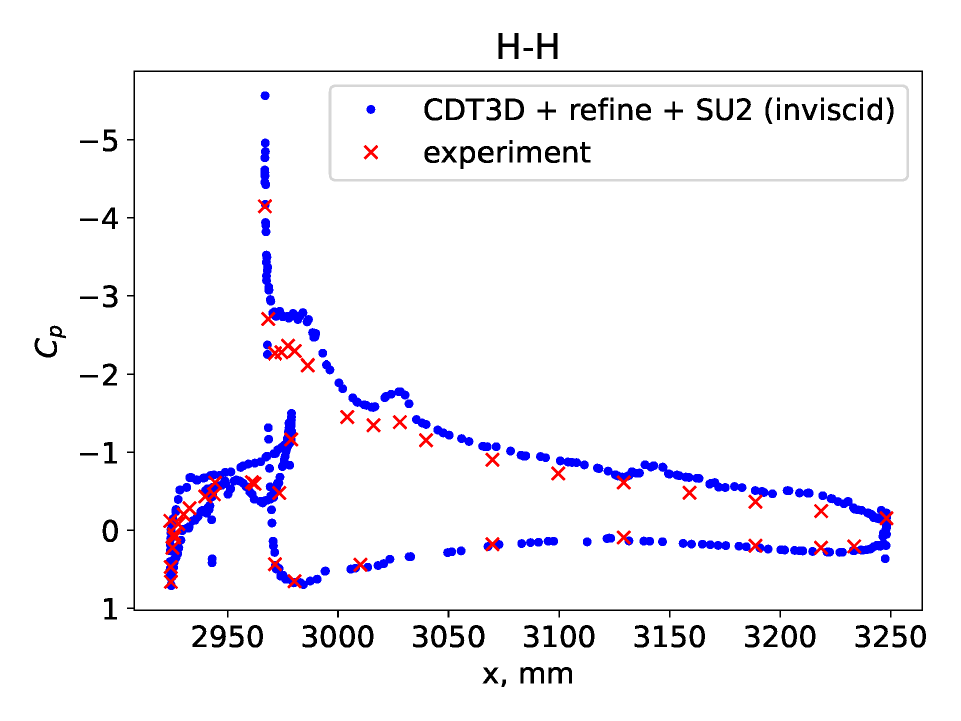}%
	\end{subfigure}

	\caption
	{Values of the pressure coefficient as evaluated by the solver versus the experiment
		across the 7 sections of~\autoref{fig:jsm_cp_lines}.}
	\label{fig:jsm_cp_sections}
\end{figure}

\subsection{Parallel Evaluation}
\label{sec:parallel_performance_evaluation}

In this section, we discuss the efficiency of our parallel implementation
for the metric-based operations presented in the previous sections
as well as for the end-to-end mesh adaptation process.

As input, we use a mesh of a	 delta wing with planar faces and $800,000$ metric
complexity.
The input mesh has $1,439,310$ vertices and $8,470,523$ tetrahedra while the
target metric has a complexity (as defined by \eqref{eq:complexity}) of $1,600,000$. The difference in complexity causes the
mesh size to double during adaptation.
The experiments were performed on the \texttt{wahab} cluster of Old Dominion
University using  dual-socket nodes equipped with
two Intel\R Xeon\R
Gold 6148 CPU @ 2.40GHz (20 slots) and  368 GB of memory.
The compiler is \texttt{gcc 7.5.0} and the compiler flags \texttt{-O3 -DNDEBUG -march=native}.
Each run was repeated $5$ times and the results were averaged using the geometrical mean~\cite{fleming_how_1986}.
For the base case, we ran the parallel code using one core.

\begin{figure}[!h]
	\centering
	\includegraphics[width=0.485\linewidth]{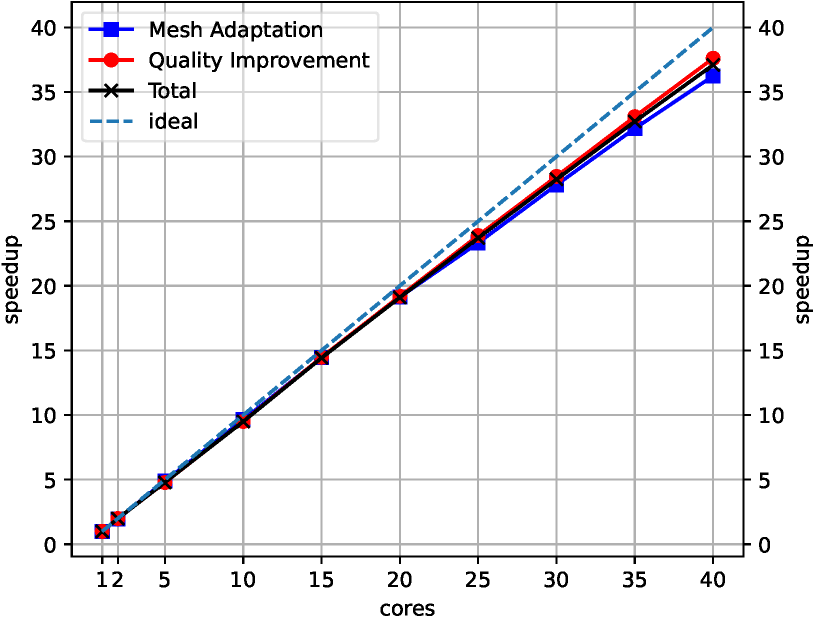}%
	\hfill%
	\includegraphics[width=0.485\linewidth]{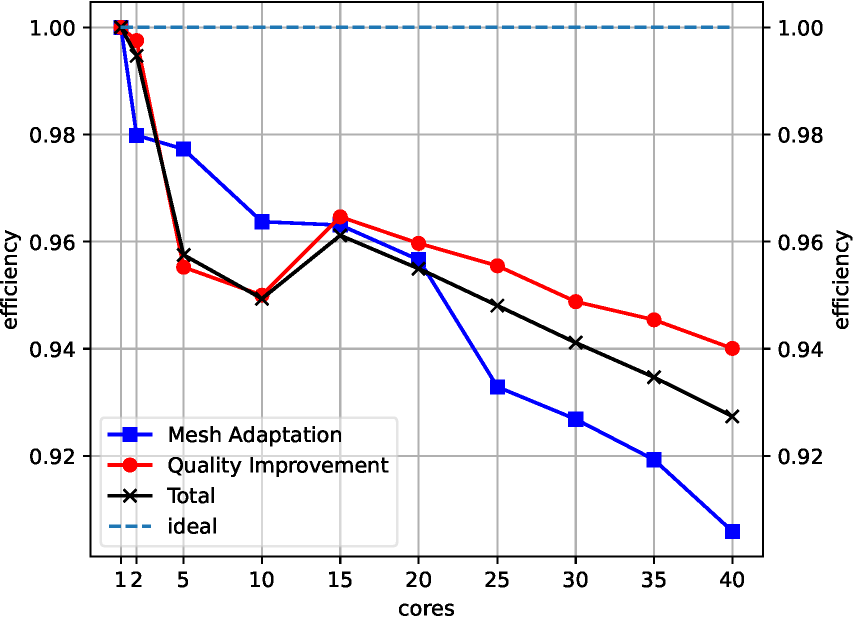}
	\caption {Speedup and efficiency of the two main modules of \cdt (see also
		Figure~\ref{fig:cad_pipeline}).}
	\label{fig:performace-total}
\end{figure}

\begin{figure}[!thpb]
	\centering
	\begin{subfigure}{0.485\textwidth}
		\includegraphics[width=\linewidth]{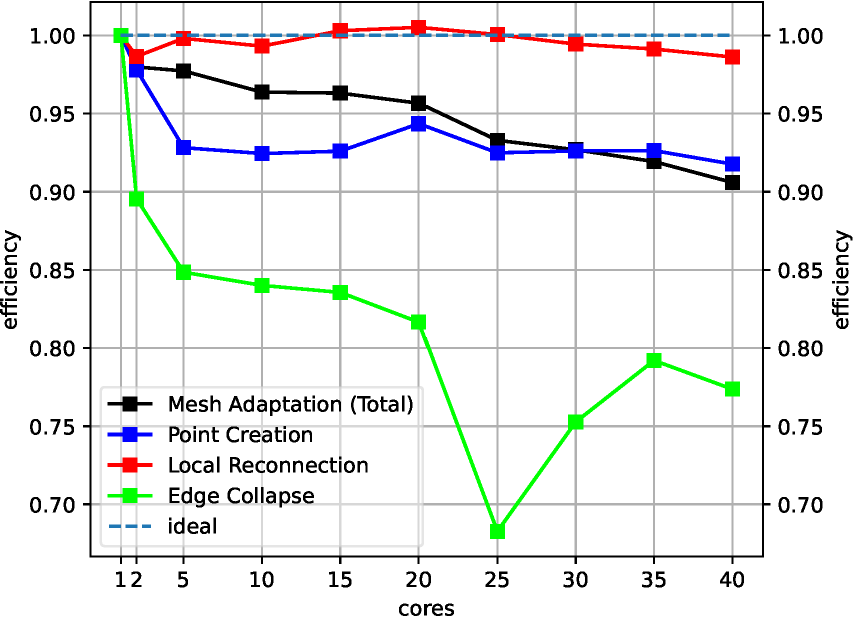}%
		\caption{}
		\label{fig:efficiency-mr}
	\end{subfigure}%
	\hfill%
	\begin{subfigure}{0.485\textwidth}
		\includegraphics[width=\linewidth]{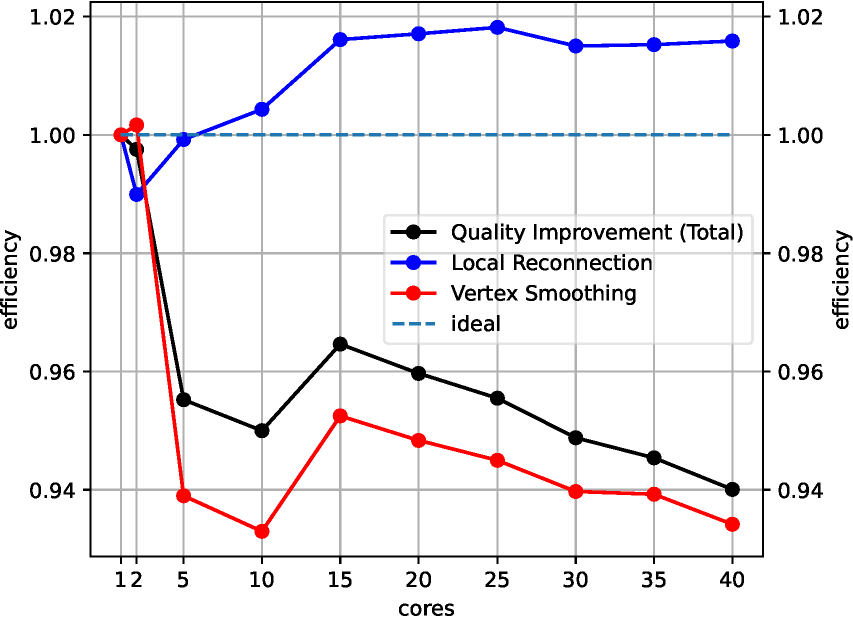}%
		\caption{}
		\label{fig:efficiency-qi}
	\end{subfigure}
	\caption
	{Efficiency breakdown of the mesh adaptation and  quality improvement modules of
		\cdt (see also Figure~\ref{fig:cad_pipeline}).}
	\label{fig:efficiency-breakdown}
\end{figure}

Figure~\ref{fig:performace-total} depicts the total efficiency of the method as
well as its breakdown with respect to the two main modules of
\cdt (see also Figure~\ref{fig:cad_pipeline}). The end-to-end efficiency is
$92.3\%$ at $40$ cores. The efficiency of the \emph{Mesh Adaptation} and
\emph{Mesh Quality Improvement} modules are $90.5\%$ and $94\%$ respectively.
Figure~\ref{fig:efficiency-mr} presents a breakdown of the efficiency of the
\emph{Mesh Adaptation} module. The \emph{Local Reconnection} operation performs the best with
more than $98\%$ efficiency. The super-linear speedup is caused by the ``buckets''.
 Splitting the list of active
elements into buckets and repeatedly performing reconnection over the same
bucket improves the cache locality. \emph{Point Creation} benefits
from the same construct but its spin-lock implementation for updating
the internal list of the contained element
(see subsection~\ref{sec:point_creation}) results in a lower
efficiency.
\emph{Edge Collapse} exhibits a lower speedup in comparison to the other two operations
due to the generic OpenMP implementation that was used to exploit parallelism.
Still, this implementation of \emph{Edge Collapse} delivers $80-85\%$ efficiency for up to $20$ cores
and $75-80\%$ efficiency for more cores. At $25$ cores  the edge collapse
efficiency drops significantly. This is in part attributed to the dual-socket nature of the
host machine. At $25$ cores the code uses one and a half sockets and the
OpenMP back-end of the operation does not have any special treatment for accessing memory
from a different socket. 
For \emph{Quality improvement} (see Figure~\ref{fig:efficiency-qi}) the super-linear
performance of \emph{Local Reconnection} is more prominent due to the (approximately)
constant size of the mesh during the \emph{Quality Improvement} phase
(no vertices are introduced). The Vertex
\emph{Smoothing} operation exhibits $93.4\%$ efficiency on $40$ threads.
Notice also that the efficiency curve of the \emph{Quality Improvement} stage (black)
follows the trends of the smoothing operation. This is due to the fact that
smoothing is the dominant operation in terms of time and also because
the efficiency of \emph{Local Reconnection} is approximately constant.
\autoref{fig:total-time-breakdown-mr} presents a breakdown of the mesh adaptation module
of \cdt. \emph{Local Reconnection} accounts for more than
$75\%$ of the total mesh adaptation time. The other two major parallel mesh operations  \emph{Point Creation} and \emph{Edge Collapse},
are responsible for less than $10\%$ of the mesh adaptation time.
The effect of the sequential point insertion is becoming increasingly higher as expected by Amdahl's law~\cite{amdahl_law}
but still, it is less than $4\%$ of the total time.

Figure~\ref{fig:total-time-breakdown} depicts the percentage of the total
time that corresponds to each operation.  The time to smooth the vertices
corresponds to about $60\%$ of the total running time while, \emph{Mesh Adaptation} takes about $12\%$
of the total time.

\section{Conclusion}

In this work, we introduced a new parallel metric-based mesh adaptation method
that can serve as the parallel optimistic mesh adaptation module of the
\emph{Telescopic Approach} in the context of CFD simulations. In particular, we
extended the \cdt library  by adding new parallel mesh operations,
incorporating metric adaptivity (Section~\ref{sec:metric-aware}) and the ability
to interface with a CAD kernel (Section~\ref{sec:introducing-geometry}).
The combination of these methods along with the simultaneous adaptation of
surface and volume improve the quality of the mesh (as measured by
element-based measures) by 3 orders of magnitude (see Figure
\ref{fig:third_improvement}) over our initial attempt
\cite{tsolakis_parallel_2019} that used external software for surface
adaptation.

These improvements are not implemented in the void by just adding new
capabilities in the mesh generation toolkit.  Instead, they allow us to put
the adapted meshes into use and verify that are really suitable for a class of
simulations. To verify our implementation and its capability to function as a
component of an adaptive pipeline -  a crucial part of many real-world
applications - in Sections~\ref{sec:results-analytic}-\ref{sec:high_lift_jaxa}
we tested \cdt through a series of cases of increasing difficulty.

Section~\ref{sec:results-analytic} served as a verification test leaving the
solver outside the setup.  The results match the theoretical expected values
and other state-of-the-art mesh adaptation software.  The solver is then
introduced back in the adaptive pipeline in
section~\ref{sec:delta-wing-pipeline} where the evaluated quantities are within
less than $0.55\%$ of the reference values in the cited literature.  At the
same time, Table \ref{tab:deltawing} indicates that the mesh adaptation stage
of our adaptive pipeline takes only a small fraction of the total
core-hours required by the simulation.  In section~\ref{sec:oneram6_inviscid}
we increased the complexity of the problem by adding a curved geometry model
based on one of the most cited CFD cases.  The results closely matched
the experimental data as well as the ones obtained by using the solver with a
carefully crafted structured mesh which is often treated as the best-case
scenario in terms of mesh quality.  Our last test case in the section
~\ref{sec:high_lift_jaxa} increased the difficulty once more by utilizing a
well-known complex simulation configuration.  Although our numerical results
were not as close as the other cases (mainly due to the lower fidelity inviscid
configuration we used), \cdt was able to handle the fairly complicated CAD data
in combination with the solution-based metric.

Finally, the data of section~\ref{sec:parallel_performance_evaluation} indicate
that \cdt still follows a \emph{scalability-first}
\cite{tsolakis_parallel_2021} approach offering a well-optimized implementation
attaining more than $92\%$ end-to-end efficiency on a single node.

\section{Future work}
\begin{figure}[htb]
	\centering
	\includegraphics[width=0.9\linewidth]{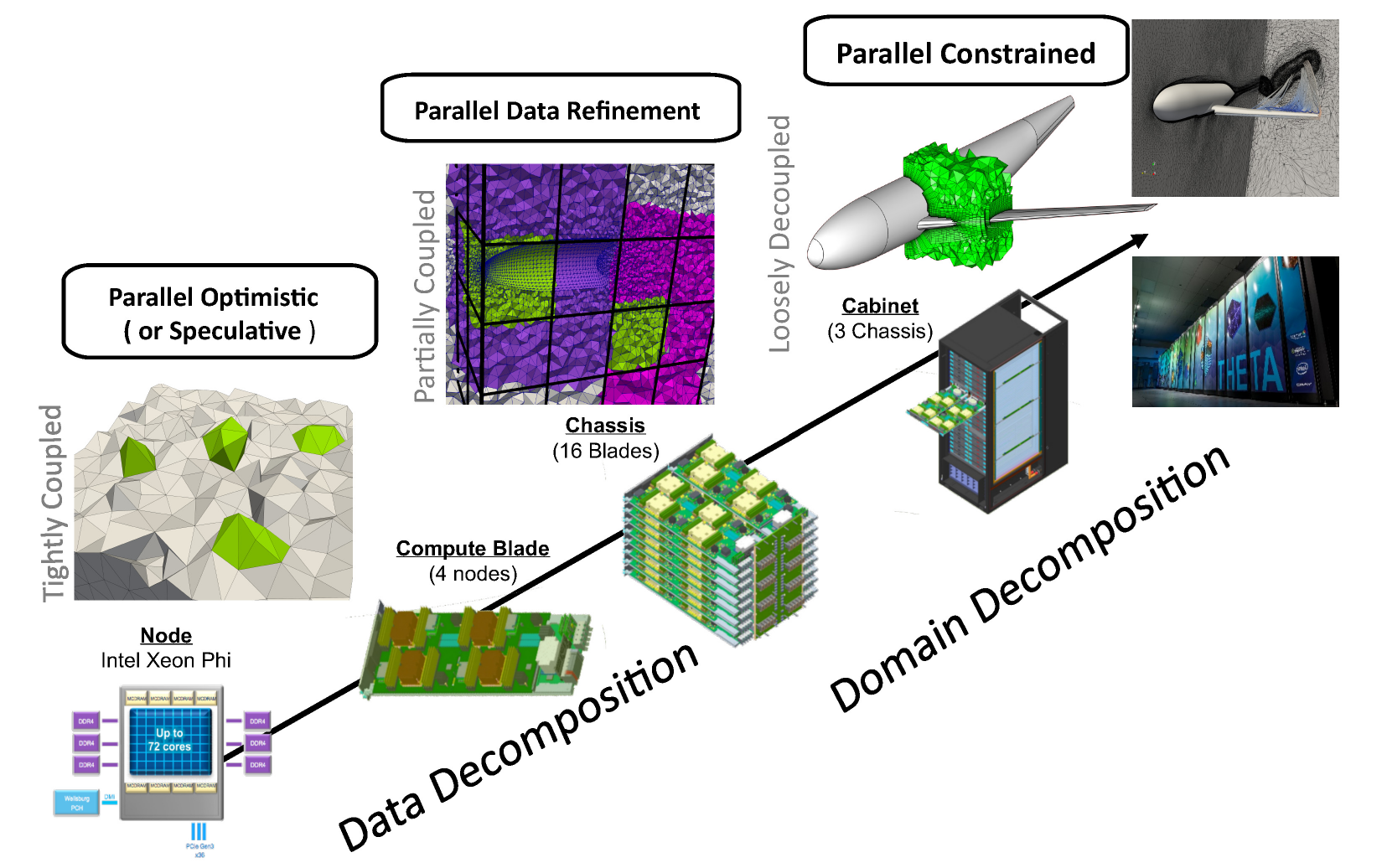}
	\caption{The Telescopic Approach
	\cite{chrisochoides_telescopic_2016}.}
	\label{fig:telescopic}
\end{figure}

In this work we presented a software that uses shared memory as a basis to create a building block for scalable parallel mesh generation and adaptivity.
A promising path toward extracting concurrency of mesh adaptation methods on HPC systems, the \emph{Telescopic Approach} (see Figure \ref{fig:telescopic}) was
proposed in \cite{chrisochoides_telescopic_2016}.
The \emph{Telescopic Approach} handles the software and hardware complexity
by defining different work decomposition methods at each level of the memory hierarchy.
Each method is designed to exploit the concurrency
at multiple levels in parallel and adaptive simulations.
However, there are some more challenges to overcome ~\cite{garner_towards_distributed_speculative_2024} and outside the scope of this paper.  
The next layer is the Parallel Data Refinement layer which has been
already implemented with both Delaunay-based~\cite{feng_scalable_2017,thomadakis_experience_2023} and
Advancing-front based methods~\cite{garner_pseudoConstrained_2019}. 
The potential scalability of these
approaches were hindered by challenges related to minimizing the amount of data
movement as identified by~\cite{feng_hybrid_2018,thomadakis_experience_2023} and reducing the effect of the constrained faces on the final mesh
quality as discussed in~\cite{kevin_Thesis}. An alternative approach would be to
revisit the distributed speculative approach presented in~\cite{nave_guaranteed-quality_2004} and adapt it for \cdt.
This approach would
benefit from the asynchronous message passing, automatic work-load balancing,
and migration (based on data dependencies) capabilities provided by the runtime
system, PREMA 2.0\cite{thomadakis_multithreaded_2022,thomadakis_toward_2023}  which is designed to support the Telescopic
Approach.

The vertex smoothing operation can be further improved by incorporating a more complete
search space for the optimal node position such as the methods presented
in~\cite{freitag_tetrahedral_1997,klingner_improving_2008}. Moreover, it could
be extended to all vertices and not just the ones attached to low-quality
elements providing an overall smoothed result which may provide better
convergence rates for the solver. Also, CAD information such as local curvature
and local feature size could be incorporated in order to optimize the quality of curved surfaces.

GPUs (Graphic Processing Units) are common in today's supercomputers however,
currently, \cdt make no use of them. Extending the presented meshing operations
so that they can take advantage of the accelerators is expected to improve the
running speed of certain operations significantly.
Figure~\ref{fig:efficiency-breakdown} reveals that almost $95\%$ of the total
time is spent on just two operations: the local reconnection and the vertex
smoothing operation. Although they both exhibit more than $90\%$ efficiency,
they can still be improved by the use of accelerators. In particular, porting
the inner floating-point-heavy kernels such as the predicates of the Delaunay
criterion and the  min-max edge-weight measure to GPUs could potentially 
reduce the running time significantly. We believe that such an implementation will benefit significantly 
from heterogeneous runtime systems like the one presented in \cite{thomadakis_heterogeneous_2023}
since they allow for separating the concerns of mesh adaptation and efficient execution on accelerators
reducing thus the problem complexity significantly.

As mentioned in subsections \ref{sec:oneram6_inviscid} and \ref{sec:high_lift_jaxa}
we didn't succeed in obtaining viscous results with our pipeline.
This is in part attributed to the absence of boundary layer mesh that many numerical methods expect.
Although  fully unstructured results have been reported with other solvers (see for example~\cite{park_ugawg_solver_tech_2018}),
to the best of our knowledge, SU2 has not been tested thoroughly within this context.
The boundary layer can be provided as an external procedure and integrated similarly to our
work in~\cite{zhou_hybrid_2019}. A similar approach that combines
state-of-the-art boundary layer generation with an adaptive anisotropic method appears in~\cite{marcum_unstructured_2013}.
Another path to explore is the generation of metric-aligned meshes such as
the ones presented in~\cite{loseille_metric_orthogonal_2014,marcum_aligned_2014}.
Moreover, the use of specialized metrics may also be suitable.
For example, in~\cite{sukas_hemlab_2021} the authors
report success by combining wall-distance to their metric creation
while in \cite{fidkowski_review_2011}
the authors review various output-based metric construction schemes
that could be evaluated.

\section*{Acknowledgments}
This research was sponsored in part by the NASA Transformational Tools and
Technologies Project (NNX15AU39A) of the Transformative Aeronautics Concepts
Program under the Aeronautics Research Mission Directorate, NSF grant no.
CCF-1439079, the Richard T.  Cheng Endowment, the Modeling and Simulation
Fellowship of Old Dominion University, and the Dominion scholar fellowship of
Old Dominion University. The authors would like to thank Fotis Drakopoulos for
the initial  implementation of the parallel vertex creation operation and the
insightful discussions. Mike Park and Charles Hyde for feedback on an earlier
draft of this document, Jayant Mukhopadhaya for his directions on
how to configure SU2 efficiently for the delta-wing case and Kevin Garner for his insights regarding the upcoming challenges regarding the future use of \cdt in a distributed memory environment.

\begin{appendices}

\section{Metric Spaces in the context of Mesh Adaptation}
\label{sec:background}

Mesh adaptation modifies an existing mesh so that it can
accurately and efficiently (i.e. fewer  number of elements) capture features of
both the solution and the underlying domain.
One of the earliest methods of mesh adaptation appears
in \cite{berger_adaptive_1984} where the authors create overset meshes of
different resolutions that are driven by estimates of the
local truncation error.
Other approaches refine or regenerate a mesh based on the magnitude and
the direction of the error, utilizing an advancing front method
\cite{Peraire_AdaptiveRemeshingCompressible_1988}
and in
\cite{Mavriplis_AdaptiveMeshGeneration_1990} via a Delaunay-based method.
The common goal of all these approaches is to extract information about
the error of the numerical solution and communicate it to the mesh adaptation
method.

In this work, we adopt the \emph{unit mesh} approach \cite{alauzet_size_2010}
where the error of the numerical solution is communicated 
via metric tensors.
This method transforms the problem of creating the \emph{best mesh}
to creating a \emph{unit mesh} (i.e. all edge lengths equal to $1$) in a
transformed space. The transformation is built such that a unit element in the
transformed space reduces the error of the numerical solution back in the
physical space.
In practice, the method is applied by incorporating metric tensors in various geometric quantities that are used to drive decisions during mesh adaptation
while staying in the physical space.

A metric tensor in three dimensions corresponds to a $3 \times 3$ symmetric positive
definite matrix $\metric$\footnote{
	A real symmetric $n \times n$ matrix is called \emph{positive definite} if
	$v^TMv > 0, \forall v \neq 0$
}. 
Metric tensors in 3 dimensions can be visualized using ellipsoids.
For example, the identity matrix corresponds to the unit sphere,
while the matrix
$M=\begin{bsmallmatrix} 2 & 0 & 0 \\ 0 & 1 & 0 \\ 0 & 0 & 1\end{bsmallmatrix}$
can be represented by an ellipsoid aligned to the coordinate axes 
with axes lengths $1/2,1$ and $1$. For a detailed derivation of 
this correspondence see \cite{tsolakis_unified_2021}.

Using as a starting point the fact that a positive definite matrix induces an
inner product through $ \inner{u,v}_\metric := u^T\metric v $,
one can use $\inner{u,v}_\metric$ to redefine how common geometric
quantities are measured. For example:
\begin{align}
	&\text{ length of a segment } &\ell_\metric(x,y) = \sqrt{\inner{x-y,x-y}_\metric} \label{eq:euclidean_length}\\
	&\text{ angle between non-zero vectors} &\cos( \theta_\metric) = \frac{ \inner{u,v}_\metric }{ \norm{u}_\metric \norm{v}_\metric},\quad \theta_\metric \in [0,\pi]
	\label{eq:euclidean_angle}
\end{align}
Reference
\cite{loseille_continuous_framework_I_2011} presents a thorough introduction to
the properties of metric tensor fields and their relation to the interpolation error of the discretized solution.

In the context of mesh adaptation, the error metric $\metric$ will vary from
point to point and so there is a need for a way to calculate the above
quantities at any point of the domain.  Following the widely used approach, the
Log-Euclidean framework introduced in~\cite{Arsigny_LogEuclidean_2006} is used
as interpolation scheme which can be formulated as follows:
Given  $x_i\,,\,i=1 \ldots k$ be a
set of vertices and $\metric_i = (\metric(x_i))$ their corresponding metrics, then for a point $x$ of the domain with barycentric coordinates $a_i$ , $	x = \sum_{i=1}^k a_i \cdot x_i ,\quad \text{ with } \quad \sum_{i=1}^{k} a_i = 1 $
the interpolated metric is defined by:
\begin{equation}\label{eq:metric_interpolation}
	\metric(x) = \metric_{mean}(x) = \exp\left( \sum_{i=1}^{k} a_i\ln \metric_i
	\right)
\end{equation}
Note that since $\metric_i$ is positive definite, it has positive eigenvalues and therefore the exponential and logarithm of the metric are well
defined and given by $\ln(\metric) := P \ln(D) P^T, \quad \exp(\metric) :=
P\exp(D) P^T$.
Where $\metric = P D P^T $ is the spectral decomposition of $\metric$.
The complexity of a discrete metric $M$ over a mesh with $n$ vertices is given by 
\begin{equation}\label{eq:complexity}
    C(M) = \sum_{i=1}^{n}\sqrt{\det{M_i}}V_i,
\end{equation}
where $V_i$ is the volume of the Voronoi dual element surrounding each node.

Finally, the quality measures used to report mesh quality also need to be
adapted. In this work, we adopt the mesh quality measures used by the
Unstructured Grid Adaptation Working Group\footnote{\url{https://ugawg.github.io/} Retrieved 2023-03-30}.
The edge lengths are evaluated
using:
\begin{equation}\label{eq:length}
	\begin{split}
		L_e & =
		\begin{cases}
			\frac{L_a -L_b}{\log(L_a/L_b)} & \|L_a -L_b\| > 0.001 \\
			\frac{L_a +L_b}{2}             &  otherwise
		\end{cases} \\
		L_a & = (v_e^T M_av_e)^{\frac{1}{2}},\,L_b = (v_e^T M_bv_e)^{\frac{1}{2}}
	\end{split}
	,
\end{equation}
where $v_e$ is the vector along edge $ab$.
Since the goal is to create a unit
mesh, edges of length below or above $1$ are considered sub-optimal.
The Mean Ratio shape measure is also approximated in the discrete metric,
\begin{equation}\label{eq:quality}
	\begin{split}
		Q_K & = \frac{36}{3^{1/3}}
		\frac{ \left( \abs{K} \sqrt{\det(M_{\text{mean}})} \right)^{\frac{2}{3}} }
		{\sum_{e \in L} v_e^TM_{\text{mean}}v_e} \\
	\end{split}
	,
\end{equation}
where $v_e$ is the vector along the edge $e$ of element $K$, $\abs{K}$ the
isotropic volume of the element and $M_{\text{mean}}$ is the interpolated
metric tensor evaluated at the centroid of element $K$. The measure is
normalized by the volume of an equilateral element and as such its range is
$[0,1]$ with $1$ being the optimal quality for an element.
As described in detail in \cite{loseille_continuous_framework_I_2011},
the first criterion is directly related to the goal of the \emph{unit mesh}
approach, which is to produce edges of length one in the transformed space.
Using this criterion alone however is not enough, since it can lead to elements
of volume equal or near to zero. The existence of these elements can create
numerical issues in the mesh generation process which operates in finite
precision and it can also cause numerical stability issues to the solver 
that processes the generated mesh. Combining the edge length measure 
with the mean ratio shape measure which achieves $1$ for equilateral elements and approaches zero as the volume of the element (in the transformed space) approaches zero
allows to avoid nearly flat elements.

Having established a way to communicate between the numerical solution  and the
mesh adaptation software it remains to specify how the error metric $\metric$
is evaluated. There are a number of different methods including but not limited
to the multiscale metric \cite{alauzet_high-order_2010}, output-based metrics
\cite{fidkowski_review_2011}, and  optimization schemes
\cite{yano_optimization-based_2012}. For this work, the multiscale error metric
is selected due to its wide use, simplicity, and availability in open-source
projects like the \refine adaptation mechanics \cite{park_refine_2020}.
This approach has been
confirmed both  theoretically \cite{loseille_continuous_framework_I_2011} and
experimentally \cite{loseille_continuous_framework_II_2011} in a number of
applications.

\section{Supplementary figures}\label{sec:supplementary_figures}

\begin{figure}[!htbp]
	\centering
	\includegraphics[width=\linewidth]{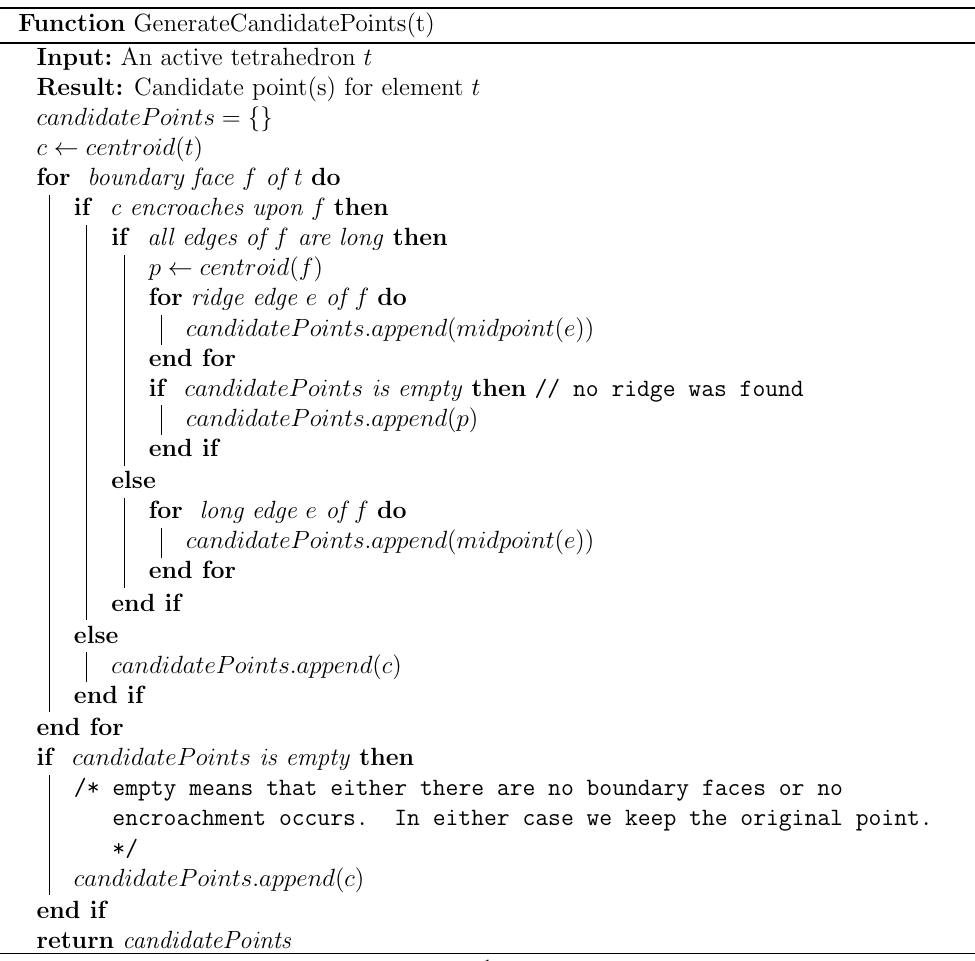}
	\caption{Encroachment rules of the centroid-based point-creation method.}
	\label{alg:centroidalgo}
\end{figure}

\begin{figure}[!hptb]
	\centering

\begin{subfigure}{\linewidth}
	\includegraphics[width=0.3\linewidth]{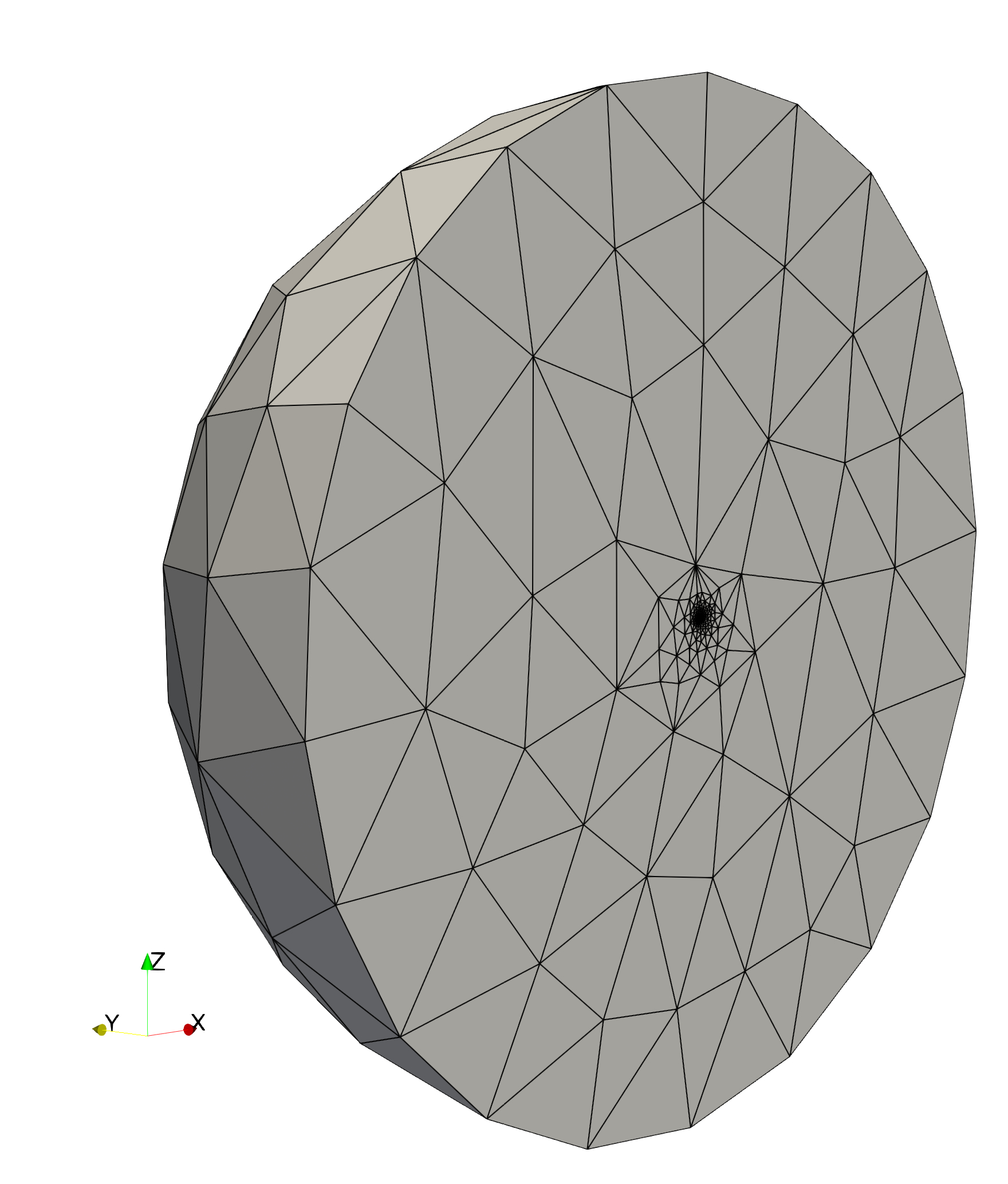}%
	\hfill%
	\includegraphics[width=0.5\linewidth]{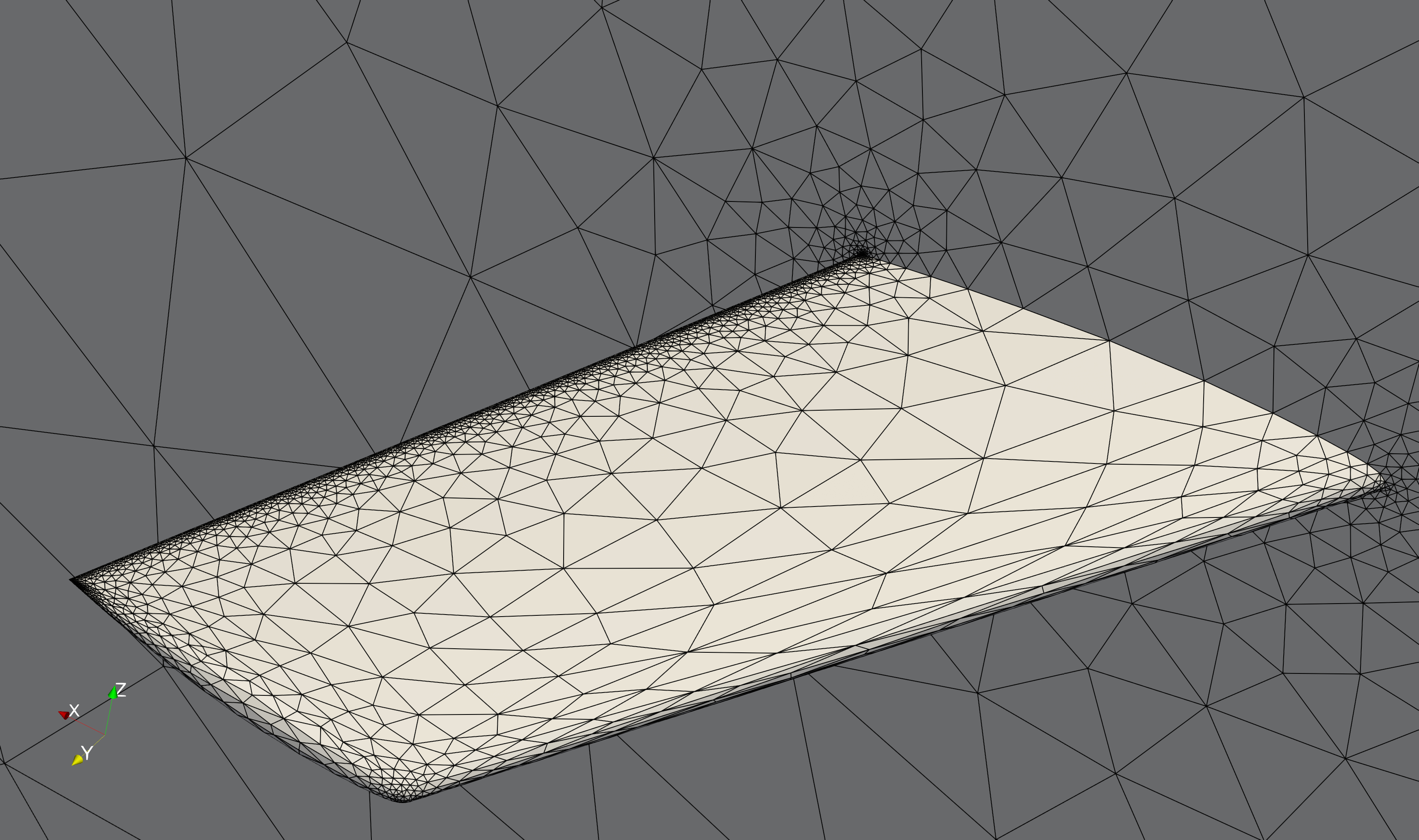}
	\caption[Initial mesh generated by \texttt{ref bootstrap}.]{Initial mesh generated by \texttt{ref bootstrap}. The mesh conforms to the geometrical features
	of the wing.}
	\label{fig:oneram6_verification_input_mesh}
\end{subfigure}
\begin{subfigure}{\linewidth}
	\includegraphics[width=0.5\linewidth]{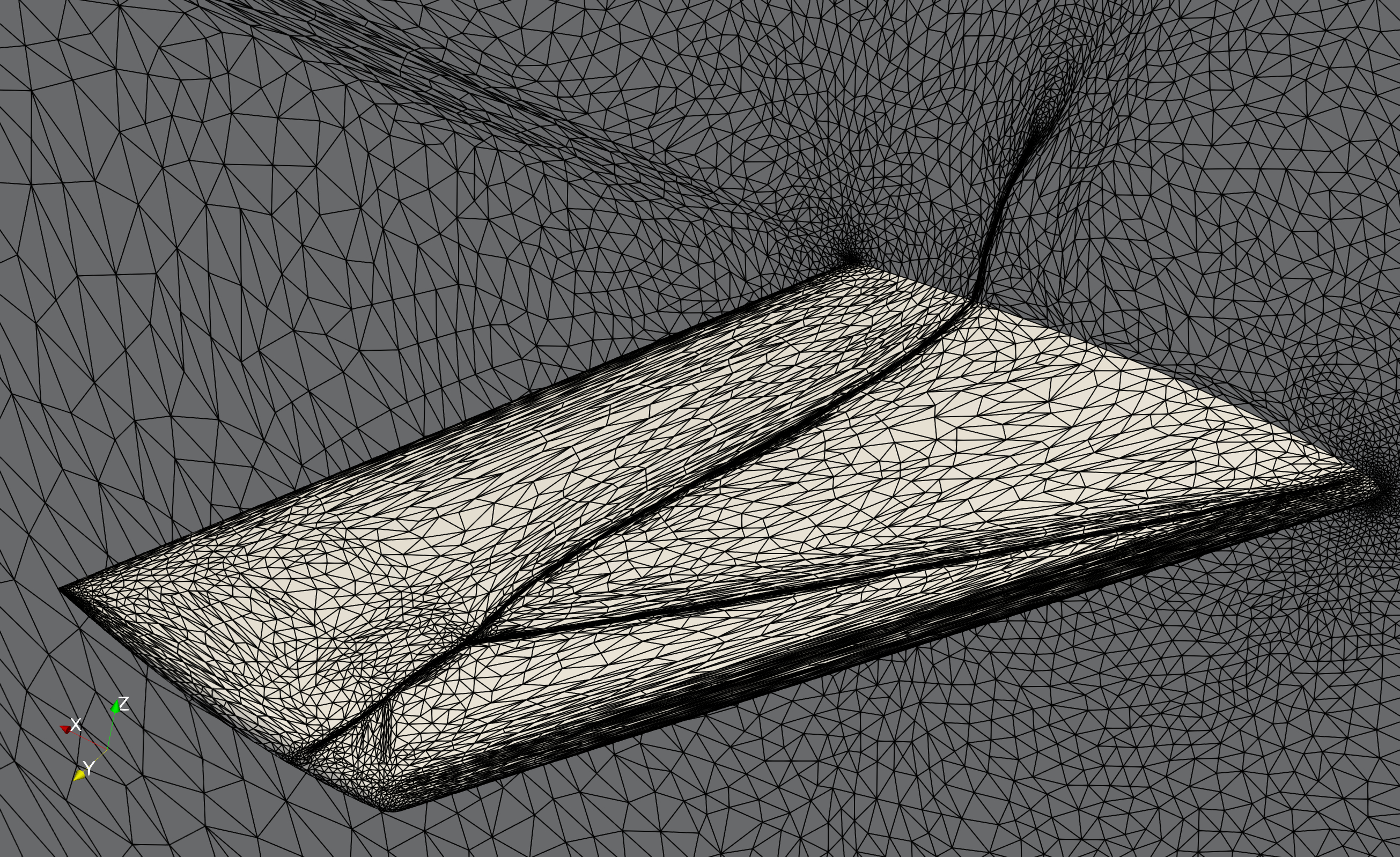}%
	\hfill
	\includegraphics[width=0.5\linewidth]{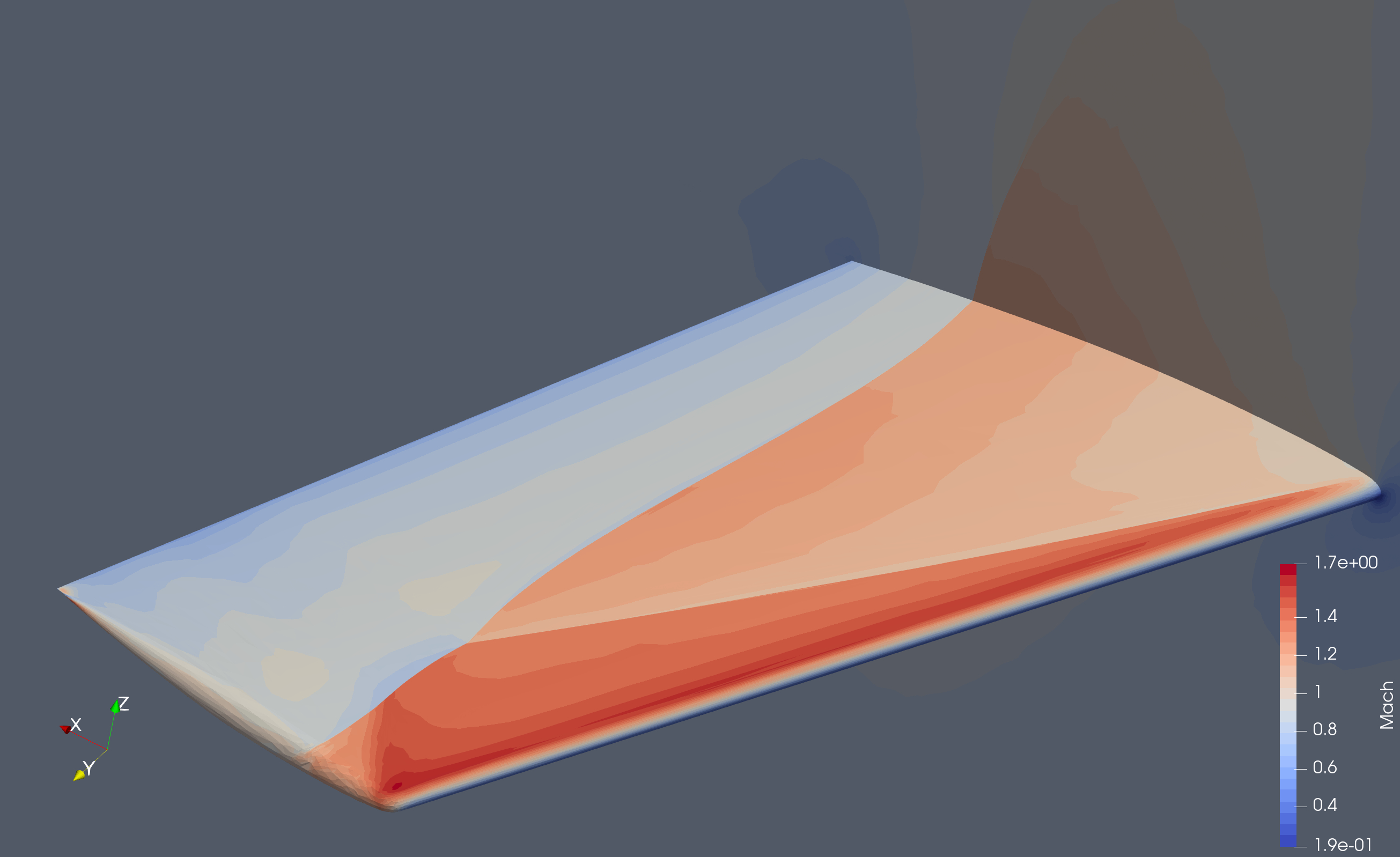}%
	\caption{Final iteration of the adaptive loop.}
	\label{fig:oneram6_mesh_solution}
\end{subfigure}
\caption{First and last  mesh of the adaptation pipeline.}
\end{figure}%
\begin{figure}[!thpb]
	\centering
	\includegraphics[width=\linewidth]{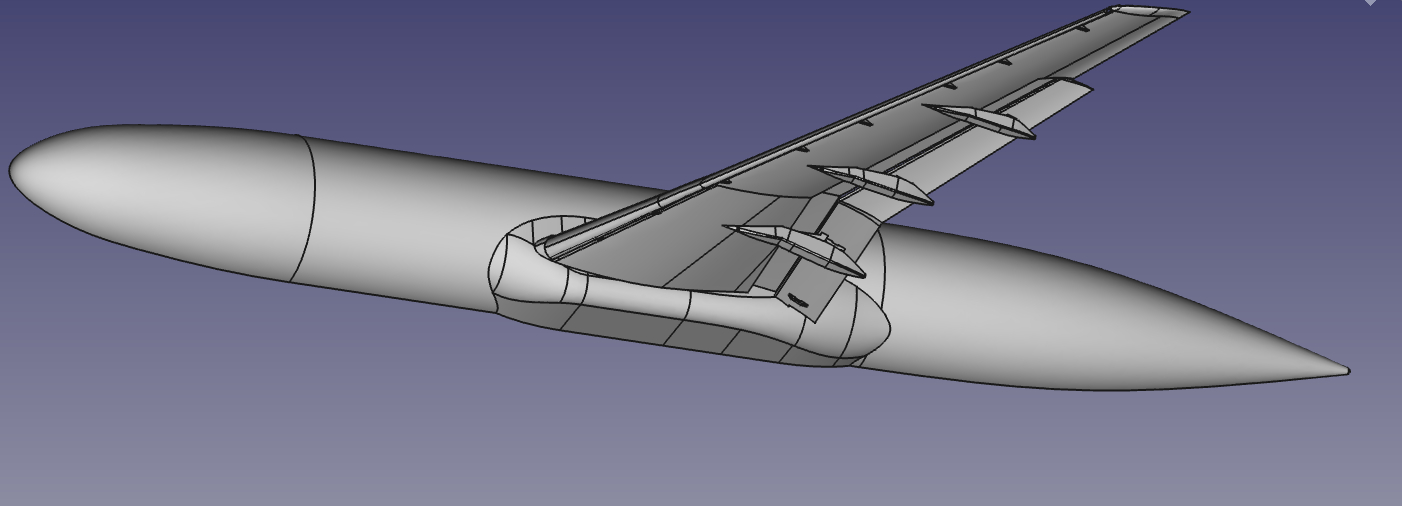}
	\caption{The JSM geometry.}
	\label{fig:jsm_geometry}
\end{figure}
\begin{figure}[!thpb]
  	\includegraphics[width=\linewidth]{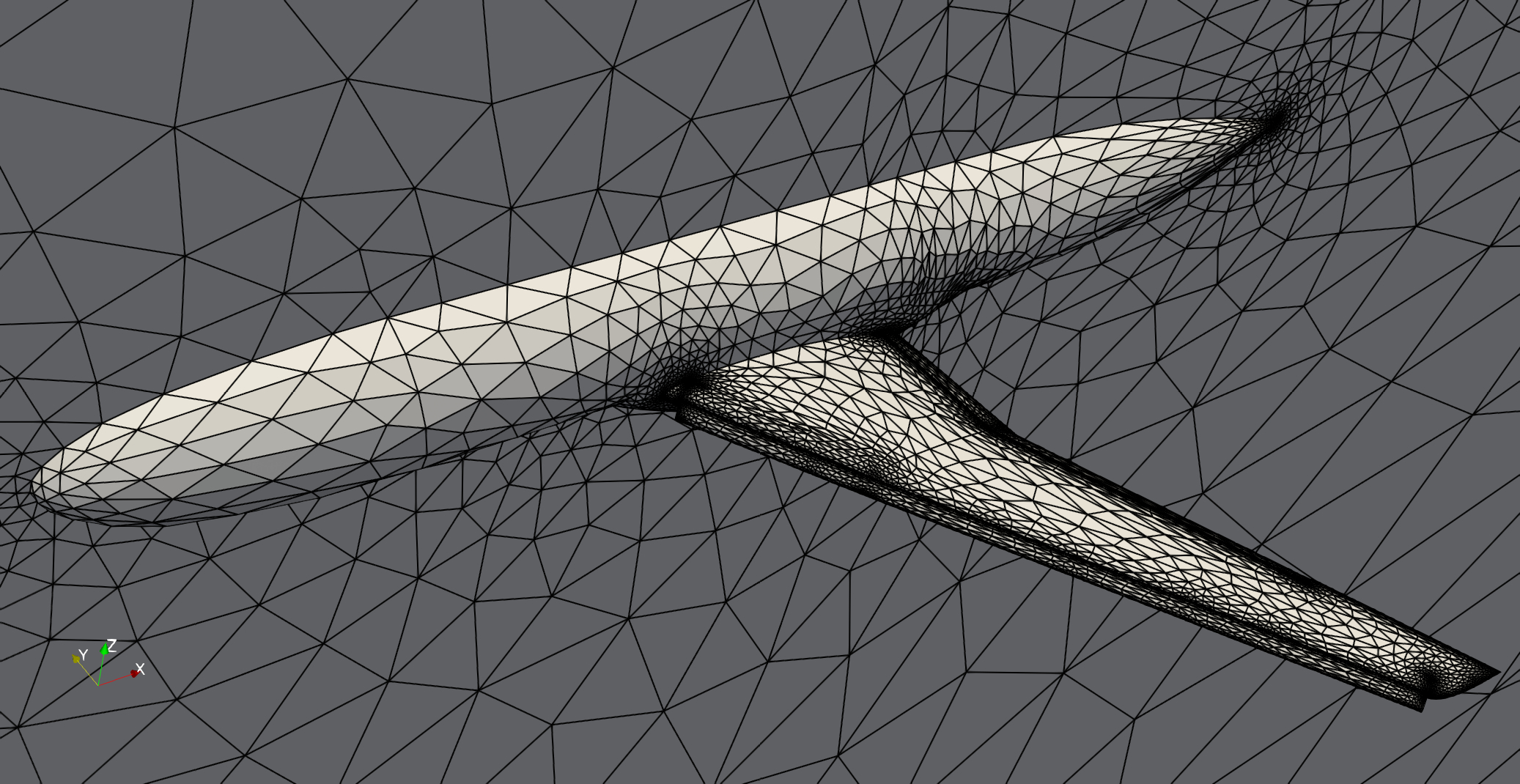}
	\caption{Initial coarse mesh created by \texttt{ref bootstrap}.
		\# vertices 52,265, \# triangles  57,240, \# tetrahedra : 219,230.}
	\label{fig:jsm_initial_mesh}
   \caption{Geometry and Initial mesh of the JSM case.}
\end{figure}

\begin{figure}[!htbp]
	\centering
	\includegraphics[width=\linewidth]{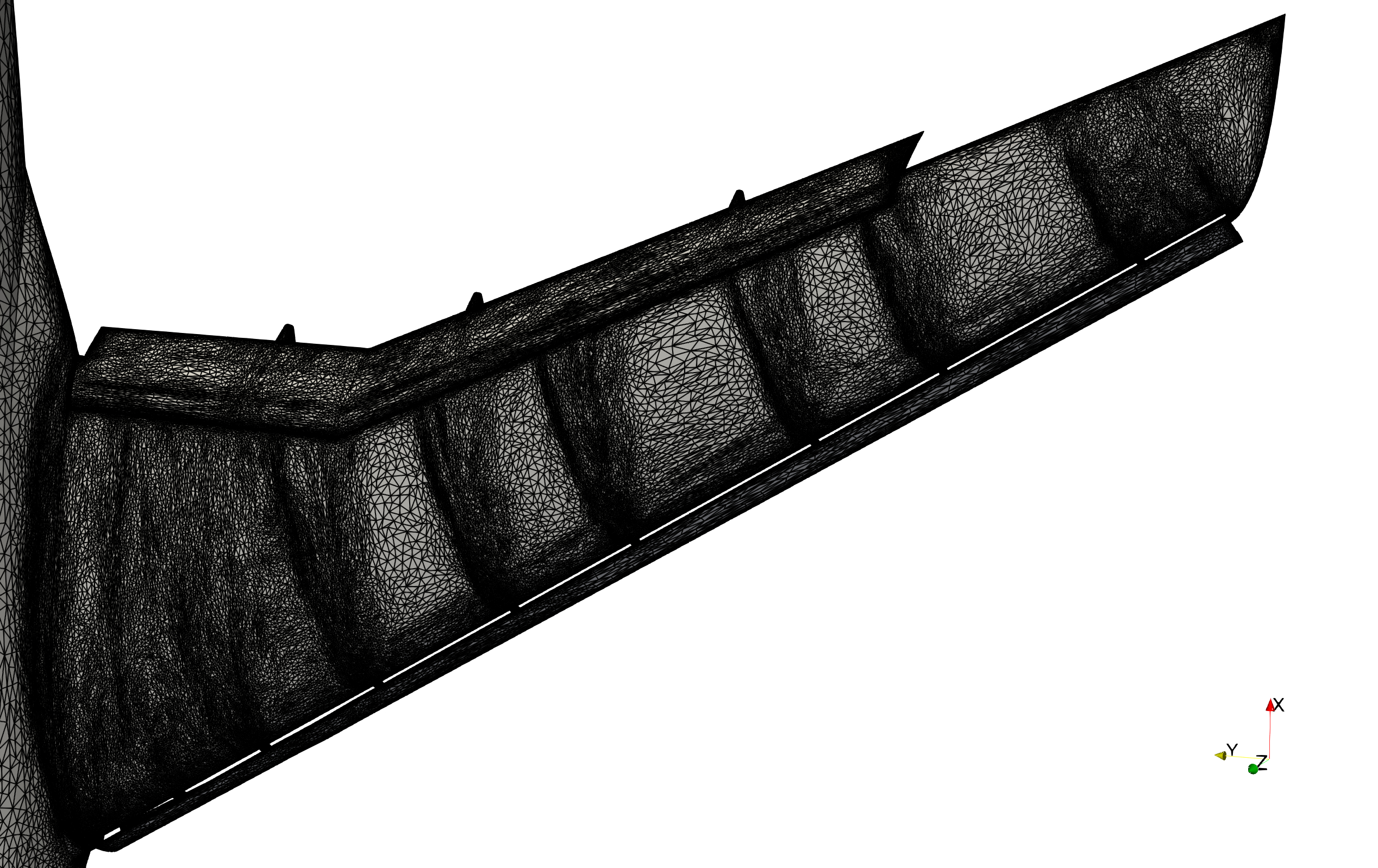}
	\includegraphics[width=\linewidth]{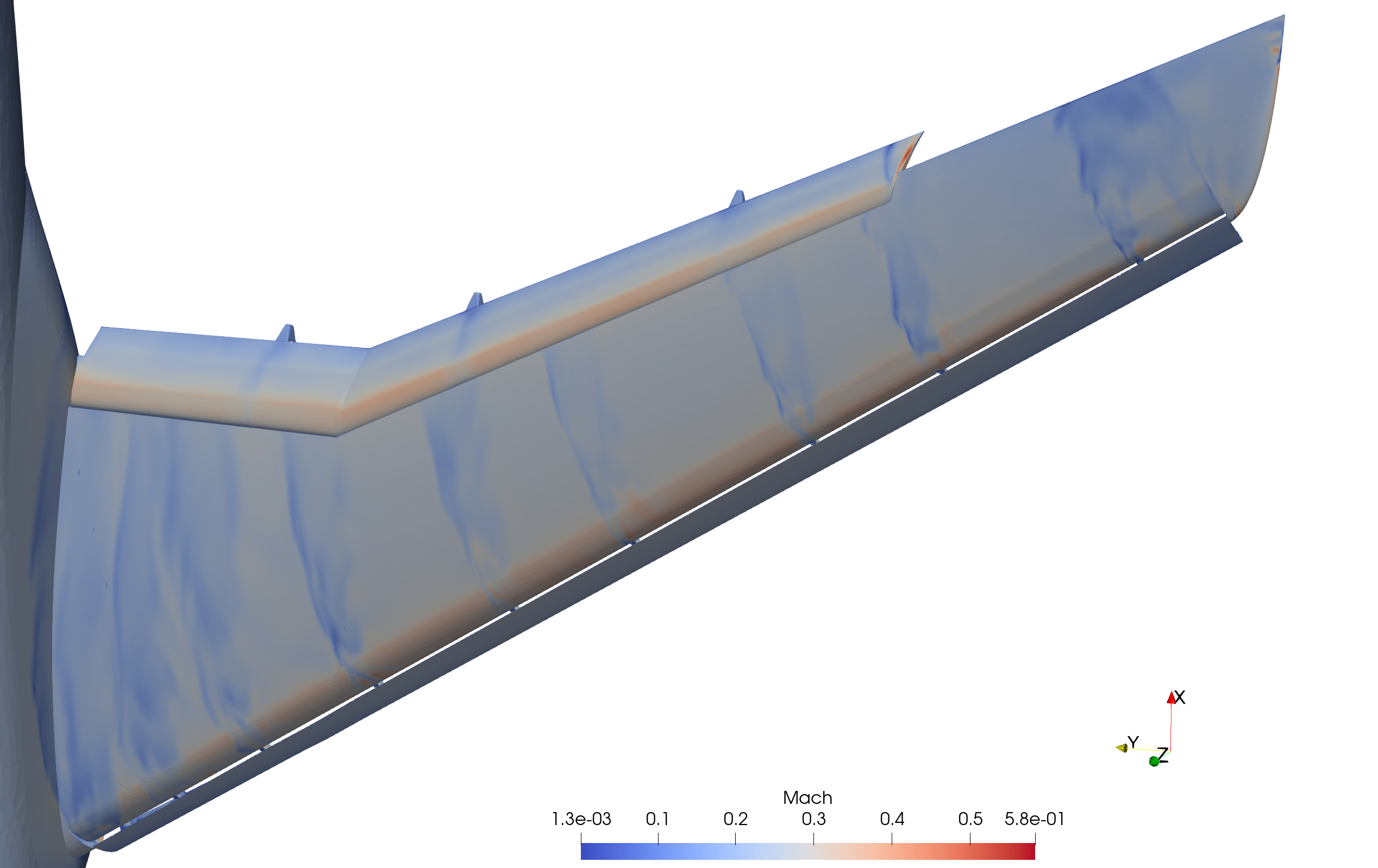}
	\caption{Final mesh and coloring of the wing by the local Mach number for the JSM case.}
	\label{fig:jsm_wing}
\end{figure}

\begin{figure}[!htbp]
	\centering
	\includegraphics[width=\linewidth]{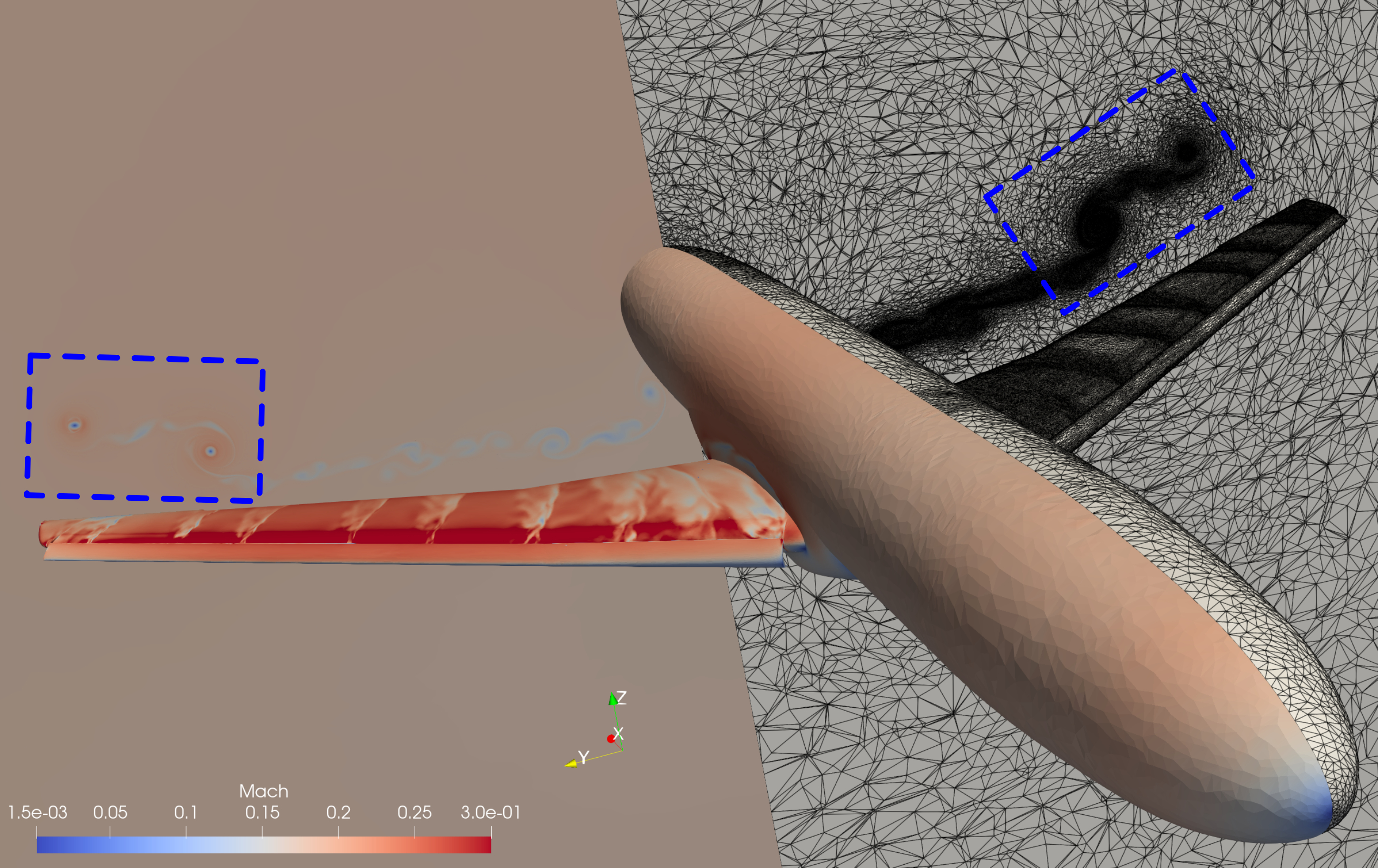}%

	\hspace{0.5em}

	\includegraphics[width=0.49\linewidth]{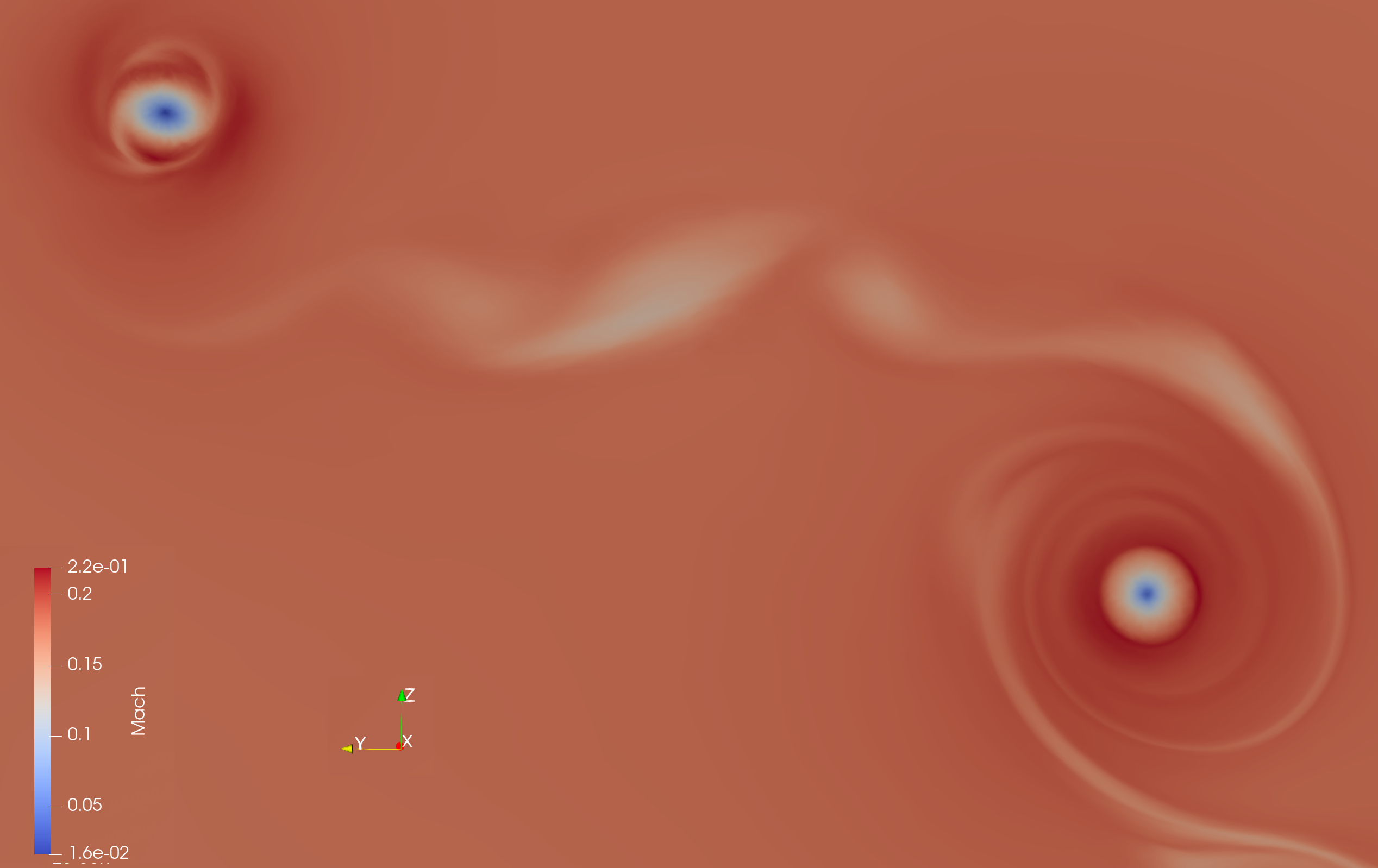}
 	\vspace{0.5em}
	\includegraphics[width=0.49\linewidth]{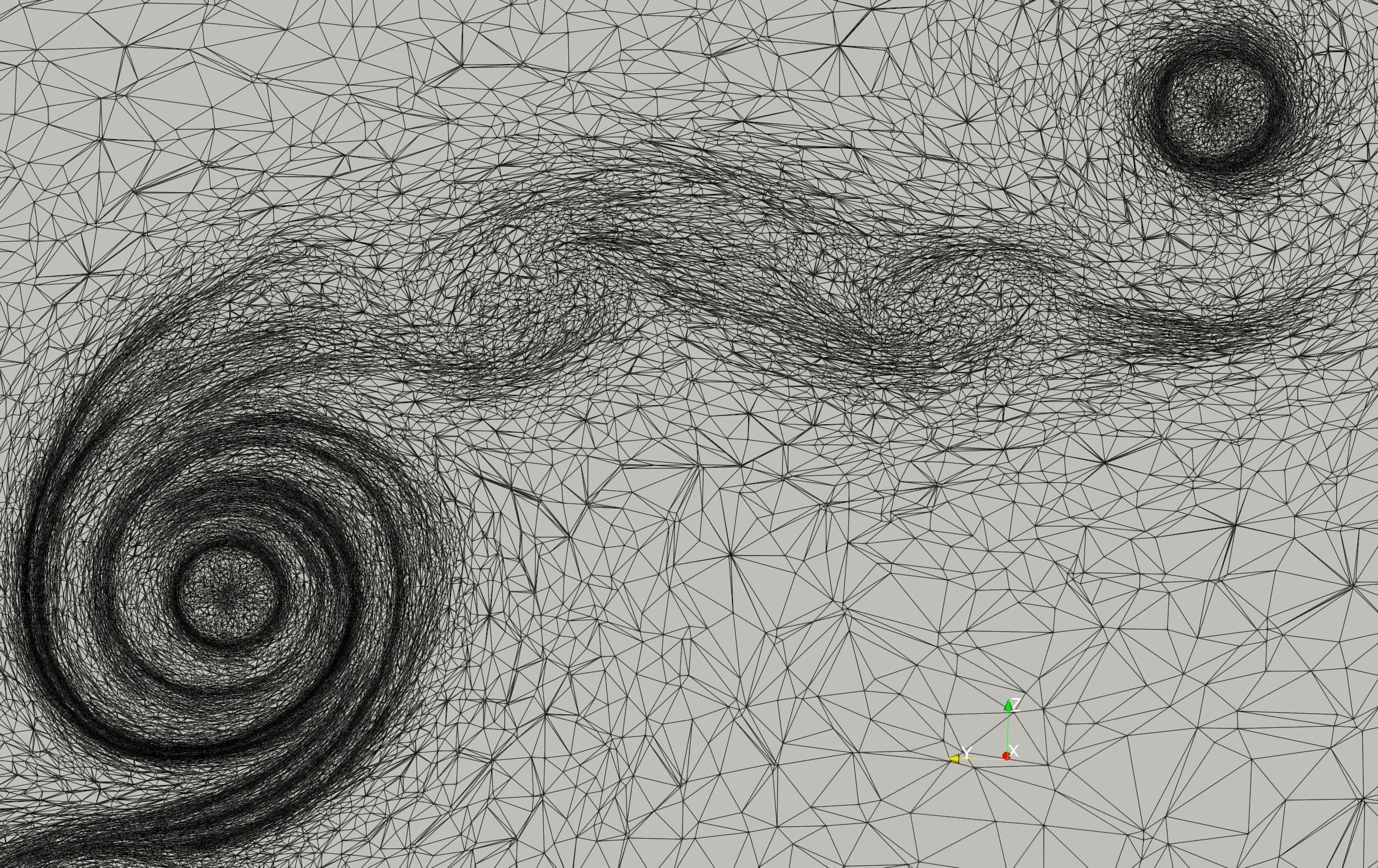}
	\caption{Simulation results. Top: Final mesh alongside the corresponding solution.
		Bottom: Zoom-in on the blue regions of the top figure.}
	\label{fig:jsm_vortex}
\end{figure}

\begin{figure}[h]
	\centering
	\includegraphics[width=\linewidth]{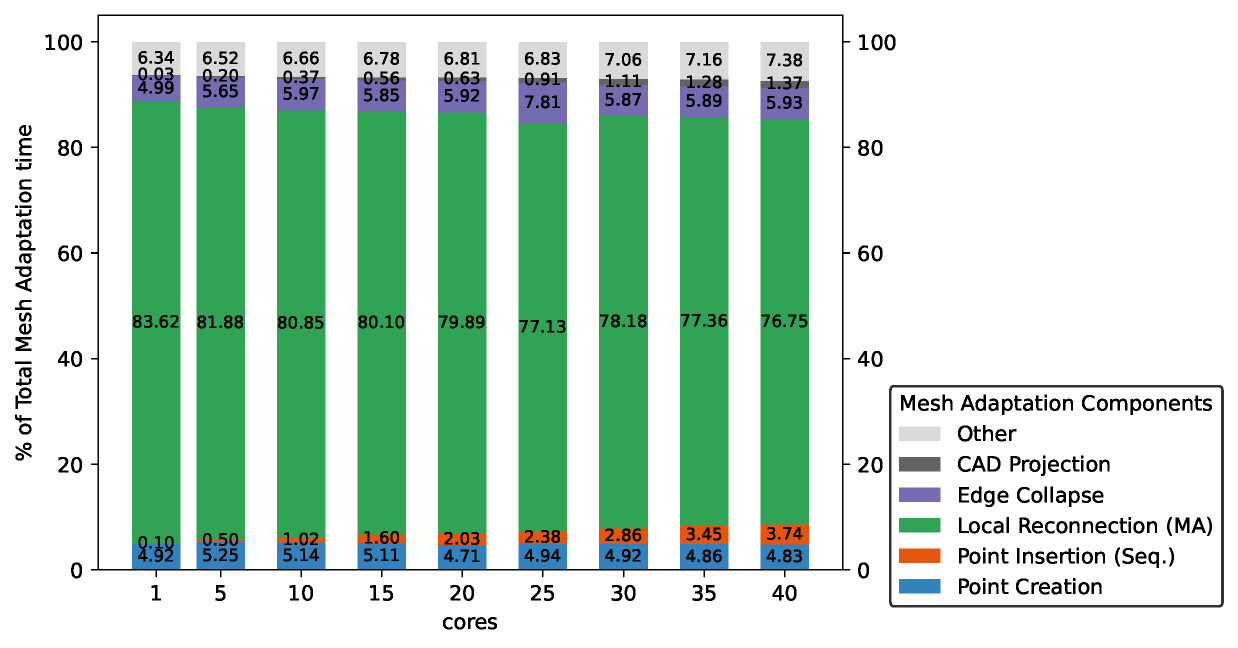}%
	\caption
	{Breakdown of the \emph{Mesh Adaptation} time into the basic operations of
		\cdt (see also Figure~\ref{fig:cad_pipeline}).}
	\label{fig:total-time-breakdown-mr}
\end{figure}

\begin{figure}[h]
	\centering
	\includegraphics[width=\linewidth]{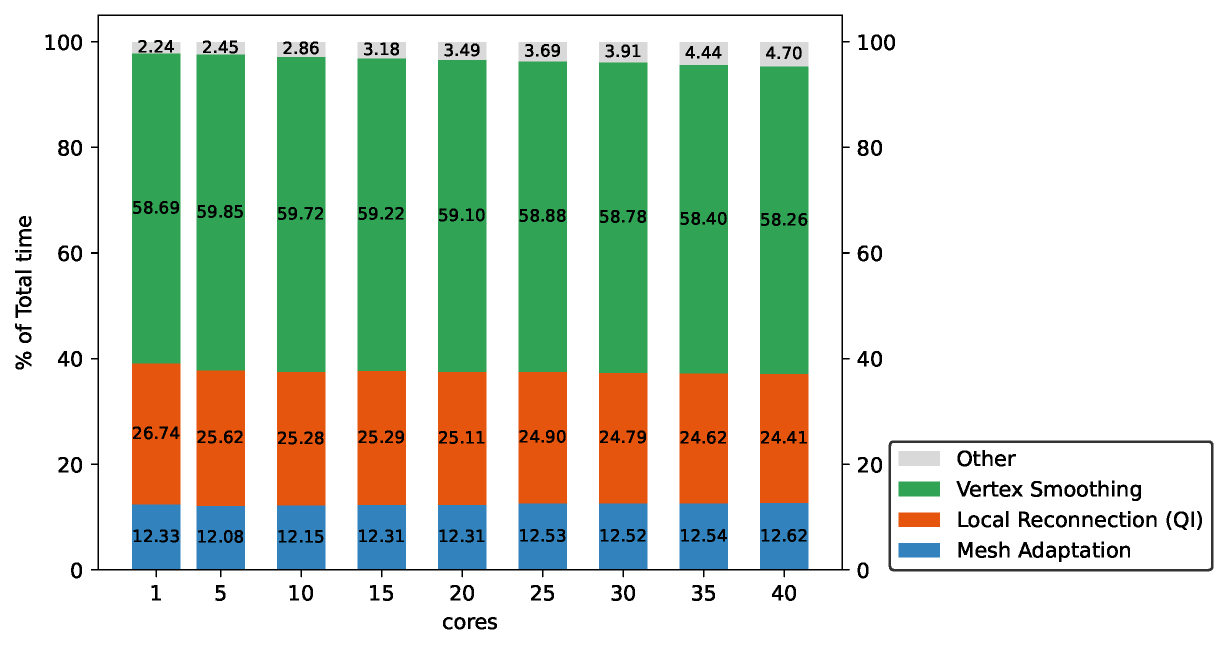}%
	\caption{Breakdown of the total time of
		\cdt (see also Figure~\ref{fig:cad_pipeline}).}
	\label{fig:total-time-breakdown}
\end{figure}




\end{appendices}
\clearpage

\bibliographystyle{ieeetr}
\bibliography{mybibfile}

\end{document}